\documentclass[a4paper,11pt]{article}
\usepackage{jheppub}
\makeatletter
\gdef\@fpheader{}
\makeatother
\usepackage{graphicx} 
\usepackage{amsmath, amssymb}
\usepackage{adjustbox}
\usepackage{MnSymbol,hyperref,float}
\usepackage{slashed,physics,braket,tcolorbox,pgf,tikz,pgfplots,dsfont,multirow,array,relsize}

\usetikzlibrary{decorations.markings}
\usetikzlibrary{decorations.pathreplacing,calligraphy}
\usetikzlibrary{decorations.pathmorphing}

\title{Investigating the Electroweak Phase Transition with a Real Scalar Singlet at a Muon Collider}

\author[1]{Mohamed Aboudonia,}
\author[1]{Csaba Balazs,}
\author[2]{Andreas Papaefstathiou}
\author[3]{and Graham White}

\affiliation[1]{School of Physics and Astronomy, Monash University, Melbourne 3800 Victoria, Australia}
\affiliation[2]{Department of Physics, Kennesaw State University, 830 Polytechnic Lane, Marietta, GA 30060, USA}
\affiliation[3]{School of Physics and Astronomy, University of Southampton, Southampton SO17 1BJ, UK}

\emailAdd{mohamed.aboudonia@monash.edu}
\emailAdd{csaba.balazs@monash.edu}
\emailAdd{apapaefs@kennesaw.edu}
\emailAdd{G.A.White@soton.ac.uk}

\newcommand\numberthis{\addtocounter{equation}{1}\tag{\theequation}}

\newcommand{\eff}{\text{eff}}

\abstract{A strong first-order electroweak phase transition (SFOEWPT) is essential for explaining baryogenesis, and for potentially generating observable gravitational waves. In the present study, we investigate the potential of a high-energy muon collider to examine the occurrence of SFOEWPT within the context of a Standard Model extended by a real scalar singlet (xSM).
We present an analysis of all viable decay modes of a singlet-like scalar particle, in order to constrain the valid parameter space of SFOEWPT, which was extracted numerically at different renormalization scales to account for theoretical uncertainties. This allowed us to determine the sensitivity of a muon collider to the production and decay channels of new heavy scalar singlet-like particles that emerge in the xSM. Our findings demonstrate that a 3 TeV muon collider could directly probe the nature of electroweak symmetry breaking by efficiently detecting new scalar particles associated with a first-order electroweak phase transition through jet-rich final states, thus complementing the indirect constraints from gravitational wave experiments.\\

\vfill

}

\begin{document}

\maketitle

\section{Introduction}

Several aspects of the current structure of the universe remain enigmatic, tracing their origins to the initial moments of its existence, when symmetry breaking resulted in the differentiation of the fundamental forces observed today. Electroweak symmetry breaking (EWSB) occurred approximately $10^{-11}$ s after Big Bang, leading to the generation of masses for most Standard Model (SM) particles, followed by chiral symmetry breaking ($\sim 10^{-3}~\mathrm{s}$), which led to hadron confinement.  Electroweak symmetry was spontaneously broken when the thermal evolution of the universe reached a state where the Higgs field's potential became unstable around its vanishing minimum, at the origin, leading to a transition towards another non-vanishing, ``broken'', minimum that evolved at lower temperatures, a phenomenon referred to as the electroweak phase transition (EWPT)~\cite{kirzhnits1972,coleman1973radiative,dolan1974symmetry,quiros1998,jackiw1974functional,weinberg1973,dolan1974gauge}. This transition could play the role of Sakharov’s third condition for explaining baryogenesis, i.e.\ the generation of the observed matter-antimatter asymmetry, if it was of the first-order type. Additionally, CP-violating processes that constitute Sakharov's second condition would have to occur in thermal disequilibrium, in order to prevent the sphaleron wash-out of the generated asymmetric baryon number (Sakharov's first condition)~\cite{Sakharov:1967dj,dine2003origin,morrissey2012electroweak,bhattacharyya2011pedagogical,mclerran1991baryon,senaha2013overview,cline2006baryogenesis}. Previous studies have shown that this sphaleron condition, which is quantitatively expressed as $\frac{E_{\text{\tiny sp}}(T_c)}{T_c} > 45$, can be translated into a relation between the theoretical parameters: $\frac{v(T_c)}{T_c}\geq1$. This condition can only be satisfied if a first-order EWPT (FOEWPT) occurs~\cite{braibant1993sphalerons, ahriche2007criterion}. A FOEWPT is characterized by the existence of two degenerate minima that are separated by a barrier at some critical temperature $T_c$. In the SM, such behavior can only arise from the cubic term in the Higgs field, generated from thermal loop corrections, which is further suppressed by a mass-screening effect resulting from the IR-divergences from the massless Matsubara modes. Earlier studies have suggested that the SM can still produce a FOEWPT if thermal corrections with the leading contribution of the IR-divergent daisy diagrams are included~\cite{carrington1992effective}. However, subsequent studies demonstrated that this possibility disappears when the subleading terms of the daisy and superdaisy contributions are included, which shifts the SM phase transition more towards the second-order type~\cite{espinosa1993nature,arnold1993effective}. Currently, it is established that a large Higgs boson mass cannot allow for FOEWPT, therefore, at best, the SM EWPT can only be of crossover type when all non-perturbative corrections are included~\cite{kajantie1996there,senaha2020}. Thus, to explain spontaneous symmetry breaking through a FOEWPT, additional degree(s) of freedom must be added to the SM to improve the sphaleron condition, primarily by strengthening the Higgs field's vacuum expectation value through coupling with other scalar degrees of freedom.

Multiple extensions have been proposed in recent decades, including additional real scalars, complex scalars, Higgs doublets, and supersymmetric  extensions~\cite{espinosa2012strong,choi1993real,o2007minimal,ham2005electroweak,ramsey2020electroweak,papaefstathiou2022electro,
branco1998electroweak,barger2009complex,ELLWANGER20101}. 
The real gauge-singlet extension (xSM) is considered to be one of the simplest, yet promising, extensions to the Standard Model (SM) for several reasons. Primarily, it can enhance the sphaleron condition through direct correction to Higgs vacuum expectation value (vev) 
 at the tree level. Additionally, it serves as a simplified model for more complex extensions, such as that of a complex singlet, or the Next-to-Minimal Supersymmetric Standard Model (NMSSM), in which multiple scalar fields exist with a substantial mass hierarchy. This hierarchy effectively isolates the heavier scalars from the electroweak symmetry breaking dynamics, with a contribution that is exponentially suppressed, resulting in only the lighter scalar being the effective scalar particle, which is adequately approximated by xSM~\cite{ELLWANGER20101}. Conversely, this scenario presents significant testing challenges, often referred to as the ``nightmare scenario'', which renders it valuable for evaluating future colliders and assessing their capacity to make definitive statements regarding the nature of electroweak symmetry breaking.  Consequently, a critical examination of  xSM can effectively serve as a concurrent investigation of some of more complex extensions.  For these reasons, this model has been extensively explored in previous works, including studies of its implications for the gravitational waves produced by real vacuum bubble dynamics, and its phenomenological consequences on the precision of the Higgs portal couplings at colliders~\cite{ramsey2020electroweak,carena2020electroweak,papaefstathiou2022electro,no2014probing,huber2008gravitational,curtin2014testing}.

While the nature of electroweak symmetry breaking is currently being explored at the Large Hadron Collider (LHC), it is also advantageous to examine its prospects in future collider facilities~\cite{ramsey2020electroweak}. A high-energy muon collider can provide sufficient energy and integrated luminosity, to obtain information complementary to that of the  LHC~\cite{al2022muon,liu2021probing,forslund2022high}. In particular, the clean experimental environment of the muon collider will facilitate high-precision measurements in the Higgs sector.  These measurements can potentially provide valuable information regarding the possible extensions of SM, which may enable us to elucidate the mechanism of electroweak symmetry breaking.
Furthermore, the direct detection of an additional scalar at a muon collider will facilitate more comprehensive analyses of all potential decay modes due to the enhanced reconstruction capabilities of final state events, including those with invisible final states.
Consequently, it is possible to optimize the search for a scalar in a muon collider by knowing its approximate mass.  In the context of electroweak symmetry breaking, the mass of the scalar augmenting the strength of the phase transition is typically close to the electroweak scale, which is within the operational range of a muon collider. This characteristic renders the muon collider an ideal environment for discovering a new scalar particle associated with electroweak symmetry breaking. In this article, we present an analysis of the minimum sensitivity of a future muon collider to probe a real scalar singlet field that could mediate a strong FOEWPT. We focus on the kinematical region where the new scalar particle's mass exceeds that of the SM Higgs boson mass, thereby allowing for decays into two Higgs bosons, consistent with current LHC constraints. Subsequently, by analyzing the decay channels, we identify the most promising avenues for investigation in a future muon collider.

The remainder of this paper is structured as follows: Section 2 aims to elucidate how the real singlet extension enhances the electroweak phase transition (EWPT) to be of a first-order type. Approximations were employed where appropriate to simplify calculations and highlight the primary features for pedagogical purposes. In addition, this section discusses the constraints utilized to identify the parameter space points that are consistent with FOEWPT. Section 3 addresses the direct and indirect verifications of xSM at colliders. Section 4 presents an analysis of the various channels of the new heavy scalar decay. Finally, Section 5 discusses the results and provides commentary on the subsequent steps of this investigation.

\section{Electroweak Phase Transition with a Scalar Singlet}

The most general renormalizable gauge-singlet scalar potential takes the form,
\begin{align} 
V(S) = \rho S +\frac{1}{2} \mu_s^2 S^2  + \frac{1}{4} \lambda_s  S^4  +  \frac{1}{3} \beta S^3\;.
\end{align}
The singlet scalar $S$ can couple to the Higgs field $H$ using the following terms,
\begin{align}
V_\mathrm{int} = \frac{1}{2} \alpha S  (H^\dagger H)   + \frac{1}{4} \lambda_{hs} S^2  (H^\dagger H)\;.
\end{align}
A $\mathbb{Z}_2$-symmetric potential can be obtained directly by setting the parameters of the odd term to zero (i.e.\ $\alpha, \beta \rightarrow 0$).  In this case, if the mass of the resulting new scalar exceeds half of the Higgs boson mass, it becomes challenging to detect at the colliders. This phenomenon is commonly referred to as the
``nightmare scenario''.
Nevertheless, in this study, we opted to keep these $\mathbb{Z}_2$-asymmeteric terms to maintain a more general representation that encompasses other types of challenging scenarios that induce some hard-to-reach regions of the parameter space, where both the mixing angle and di-Higgs boson decays are minimal. In addition, these specific terms are crucial for facilitating a strong first-order EWPT, see e.g.~\cite{choi1993real}. The singlet term in the scalar field can be eliminated using shift symmetry, which only results in the redefinition of the remaining parameters, as demonstrated in previous studies ~\cite{espinosa2012strong,profumo2007singlet}. Consequently, the most general renormalizable representation of the real gauge-singlet extension of SM (xSM) is,
\begin{align*}
V(H,S) &= -\frac{1}{2} \mu_h^2  (H^\dagger H)  + \frac{1}{4} \lambda _h (H^\dagger H)^2 + \frac{1}{2} \mu_s^2 S^2  + \frac{1}{4} \lambda_s  S^4 + \frac{1}{2} \alpha S  (H^\dagger H)  \\
& + \frac{1}{4} \lambda_{hs} S^2  (H^\dagger H) +  \frac{1}{3} \beta S^3\;. \numberthis \label{reno}
\end{align*}
Electroweak symmetry breaking generates vacuum expectation values (vevs), $v$ and $\omega$, for the Higgs doublet and scalar singlet fields, respectively. The physical states are obtained by expanding around the vevs, $H\rightarrow \frac{1}{\sqrt{2}} (v+h)$ and $S\rightarrow \omega +s$, derived from the minimization conditions:
\begin{align}
\frac{\partial V(h,s)}{\partial h}\Bigg \rvert_{\braket{h} = v,\atop \braket{s} =\omega} = 0,\qquad \qquad \frac{\partial V(h,s)}{\partial s}\Bigg \rvert_{\braket{h} = v,\atop \braket{s} =\omega} = 0, 
\end{align}
which demonstrates that the fields become coupled,
\begin{align}
-\mu_h^2 +\lambda_h v^2 +\alpha \omega+\frac{1}{2} \lambda_{hs}\omega^2 = 0\;,\label{vev1}\\
(2\mu_s^2 + \lambda_{hs} v^2) \omega +  2(\beta +\lambda_s \omega) \omega^2 +\alpha v^2 = 0\;.\label{vev2}
\end{align}
These equations can generate up to eight stationary points, in addition to the typical symmetric point at the origin, ($0,0$). Some of these points may be degenerate, whereas others may not be physically viable. Consequently, the singlet extension in Eq.~\eqref{reno}  directly influences the Higgs vacuum at the tree level, which becomes dependent on the singlet parameters after spontaneous symmetry breaking: 
\begin{align}
v_b = \left(\pm \sqrt{\frac{\mu_h^2 - \left( \alpha +\frac{1}{2} \lambda_{hs} \omega \right)\omega}{\lambda_h}},\,\,\pm \sqrt{\frac{-2(\mu_s^2 +\beta \omega +\lambda_s \omega^2)\omega}{\alpha+\lambda_{hs}\omega} } \right)\;.\label{xSMvev}
\end{align}
If the singlet has a vanishing vev, $\omega =0$, or when $\abs{\alpha}=\frac{1}{2}\lambda_{hs} \omega$, the broken minima in Eq.~\eqref{xSMvev} will converge to the electroweak vev, $v_{\text{\tiny EW}} \approx 246$ GeV, which is also evident from Eq.\eqref{vev1}. Moreover, Eq.~\eqref{vev1} indicates that at high temperatures, where the vevs become temperature-dependent, a large negative value for the cubic term, $\abs{\alpha} >  \frac{1}{2}\lambda_{hs} \omega$, can drive the electroweak phase transition towards the first-order type solely through the tree-level (TL) contribution by improving the sphaleron condition. This is a consequence of a decrease in $\mu_h^2$, and can significantly reduce the critical temperature as $T_c \propto \mu_h$ as will be demonstrated subsequently. This phenomenon may not be immediately apparent from  Eq.~\eqref{xSMvev} as a reduction in $\mu_h^2$ may decrease $v_b$, and consequently may diminish the first-order electroweak phase transition condition, $\frac{v_c}{T_c}$, in totality. However, this is not the case, as will be demonstrated. Consequently, this extensive parameter space facilitates the possibility of obtaining a FOEWPT. Conversely, the tree-level corrections would reduce the mixing angle and the di-Higgs boson decays for some parameter space points, potentially necessitating higher luminosity to examine the $\mathbb{Z}_2$-asymmetric scenario compared with the $\mathbb{Z}_2$-symmetric one~\cite{curtin2014testing,ramsey2020electroweak}. In addition to tree-level corrections, NLO corrections, both at zero temperature from the Coleman-Weinberg one-particle irreducible (1PI) contribution, and at finite temperature, could further catalyze the phase transition towards a FOEWPT. The Coleman-Weinberg correction to the tree-level potential is given by
\begin{align}
V_{\text \tiny CW}^{(1)}(v,\omega) &= \sum_i \frac{n_i}{64\pi^2} m_i^4(v,\omega) \left[\log \frac{m_i^2(v,\omega)}{\Lambda^2} - f_i \right]\;,\label{cw}
\end{align}
where $n_i$ corresponds to the number of degrees of freedom of  particle $i$, $f_i$ represents the residual fraction from dimensional regularization in the $\overline{\text{MS}}$-renormalization scheme and $\Lambda$ is the renormalization scale. The values of the corresponding parameters for each particle contributing to the Coleman-Weinberg potential are listed in Table~\ref{CWparameters}.
\begin{table}
\centering
\begin{tabular}{|>{\centering\arraybackslash}p{15mm}|>{\centering\arraybackslash}p{35mm}|>{\centering\arraybackslash}p{15mm}|>{\centering\arraybackslash}p{15mm}|} 
\hline
       & $m_i(v,\omega) $ &  $n_i$  &  $f_i$  \\[3pt] \hline
$W^\pm$  &  $\frac{1}{2} g_1 v$ & 6 & $\frac{5}{6}$ \\[7pt] 
$Z$  &  $\frac{1}{2} \sqrt{g_1^2+g_2^2}\,\, v$ & 3& $\frac{5}{6}$ \\[7pt] 
$t$  &  $\frac{1}{\sqrt{2}} y_t v$ & -12 & $\frac{3}{2}$ \\[7pt] 
$H$  &  $m_1(v,\omega)$ & 1 & $\frac{3}{2}$ \\[7pt] 
$S$  &  $m_2(v,\omega)$ & 1 & $\frac{3}{2}$ \\[4pt] \hline
\end{tabular}
\caption{Properties of the particles participating in the Coleman-Weinberg potential.}\
\label{CWparameters}
\end{table}
 Generally, all massive particles contribute to this correction; however, owing to the large mass hierarchy among the SM particles, considering only the heavy particles ($W^\pm,\,Z,\,t$) provides a reasonable approximation. Because these particles are not directly coupled to the real singlet, their masses retain their SM values. Nevertheless, the expression for the Higgs boson mass differs from that of SM because of its coupling with the real singlet field. The new Higgs boson mass, in conjunction with the singlet mass, is obtained through the diagonalization of the mass matrix:
\begin{align}
M =
\begin{pmatrix}
\frac{\partial^2 V(h,s)}{\partial h^2} &  \frac{\partial^2 V(h,s)}{\partial h \partial s} \\[7pt]
\frac{\partial^2 V(h,s)}{\partial h \partial s} &   \frac{\partial^2 V(h,s)}{\partial s^2} 
\end{pmatrix}\;, \label{masses}
\end{align}
which returns,
\begin{align*}
m_{1,2}^2 (v,\omega) &= \frac{1}{4} \Bigg[ 2(\mu_s^2-\mu_h^2)+2(\alpha+2\beta)\omega+ (\lambda_{hs}+6\lambda_h)v^2+(\lambda_{hs}+6\lambda_s) \omega^2 \mp \Bigg\{\Big[ 2(\mu_s^2+\mu_h^2)\\
&-2(\alpha-2\beta)\omega+ (\lambda_{hs}-6\lambda_h)v^2-(\lambda_{hs}-6\lambda_s) \omega^2 \Big]^2+16(\alpha+2\beta)v^2\Bigg\}^{\frac{1}{2}} \Bigg], \numberthis \label{zeroTmass}
\end{align*}
The new scalar mass is considered to be the heavier one, $m_2^2(v,\omega)$, to maintain consistency with the SM Higgs boson mass $m^2_h(v) = 2\lambda_hv^2$, which is only satisfied by $m_1^2(v,\omega)$ when all the real singlet parameters vanish. Furthermore, this analysis focuses on the kinematical region where $m_s>m_h$, necessitating that the heavier mass  $m_2^2(v,\omega)$ be the new scalar mass. 
\begin{figure}[htb!]
\centering
\includegraphics[scale=.5]{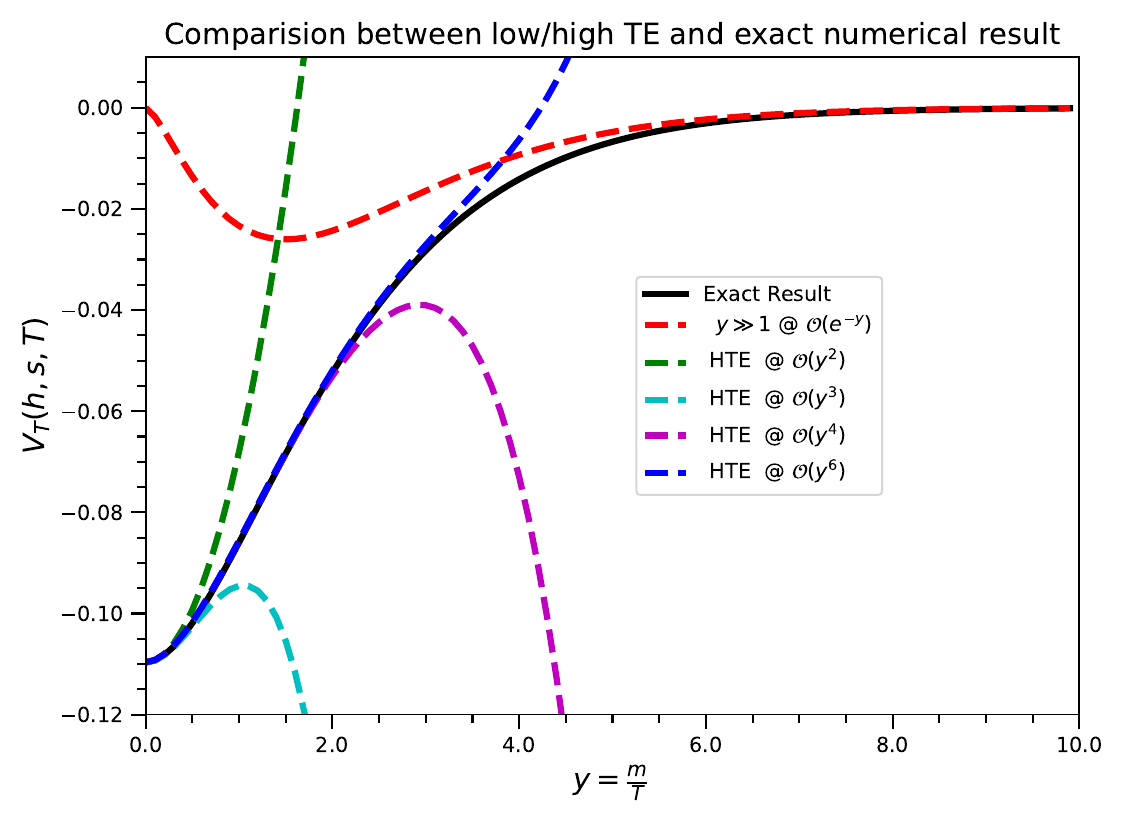}
\caption{The dashed red curve shows the low-temperature expansion of the $V_T(y,T) $ in Eq.~\eqref{fullv}. By definition, it
will only converge to the exact numerical calculation (solid black curve) when $m>T$ and will both coincide at $y\geq 5$. The other colored dashed curves represent the HTE of $V_T(y,T) $ at different order of expansion in $y$.}
\label{numerical}
\end{figure}

\noindent
The finite-temperature corrections are obtained from (refer to, e.g.~\cite{dolan1974symmetry,quiros1998,laine2017basics} for a comprehensive review),
\begin{align}
V_T(y,T) =\sum_i \frac{n_i\,T^4}{2\pi^2} \int_0^\infty dx \,\, x^2 \log \left[1 \mp e^{-\sqrt{x^2 +y^2}} \right]\;,\label{fullv}
\end{align}
where, $\displaystyle{x =\frac{p}{T}, \, y =\frac{m}{T}}$, and the sign depends on the particle type (negative for bosons and positive for fermions). For non-vanishing particle masses, $y\neq 0$, the integral must be evaluated numerically.
Nevertheless, a high-temperature expansion (HTE) to order $\mathcal{O}(y^6)$,
\begin{align*}
V_{T,B}^{\text{\tiny HTE}}(v,\omega,T) &= -\frac{\pi^2T^4}{90} + \mathlarger{\mathlarger{\sum}}_{i=Z,W^\pm,\atop H,S}n_i\Bigg( \frac{m_i^2T^2}{24} -\frac{m_i^3T}{12\pi} -\frac{m_i^4}{2(4\pi)^2} \left[ \ln\left(\frac{m_ie^{-\gamma_E}}{4\pi T} \right)-\frac{3}{4}\right]\\
& + \frac{m_i^6 \xi(3)}{3(4\pi)^4 T^2} +\mathcal{O}(m^8)\Bigg), \numberthis\label{BHTE}\\[8pt]
V_{T,t}^{\text{\tiny HTE}}(v,\omega,T) &= -\frac{21\pi^2T^4}{180} +\frac{m_t^2T^2}{4} +\frac{6m_t^4}{(4\pi)^2} \left[ \ln\left(\frac{m_te^{-\gamma_E}}{4\pi T} \right)-\frac{3}{4}\right] \\
&- \frac{28m_t^6 \xi(3)}{(4\pi)^4 T^2} +\mathcal{O}(m^8) \numberthis\label{FHTE}
\end{align*}
could be a valid approximation up to $m  \lesssim 3T$ as explained in Ref.~\cite{laine2017basics} (See Fig.~\ref{numerical}). In this study, we  set $m_s \in [200:1000]$~GeV, indicating that certain scalar mass values do not conform to this approximation. However, for these values, the thermal corrections can be disregarded because in this case, $m_s>3T$, and the low-temperature expansion will be a valid approximation, superseding the HTE, which is exponentially suppressed. Consequently, this contribution can be disregarded for this pedagogical analysis, particularly given that in $\mathbb{Z}_2$-asymmetric xSM, the FOEWPT is predominantly driven by the tree-level odd terms. Therefore, the calculations can be modified according to the valid mass as follows:
\begin{itemize}
\item  For $m_s \lesssim 3T$, which represents the majority of valid conservative parameter space points (see Fig.~\ref{foewpt_ms}), the thermal corrections can be approximated by the HTE, such that the overall effective potential becomes,
\begin{align}
V_\eff(v,\omega,T) &= V_{TL}(v,\omega) + V_T^{\text{\tiny HTE}}(v,\omega),
\end{align} 
where, $V_T^{\text{\tiny HTE}}(v,\omega)$ is the sum of  Eq.~\eqref{BHTE} and Eq.~\eqref{FHTE} for all bosonic and fermionic contributions, including the real scalar singlet.
\item For $m_s  >  3T$,  which represents  the majority of liberal parameter space points, the overall effective potential becomes,
\begin{align}
V_\eff(v,\omega,T) &= V_{TL}(v,\omega) +\overline{ V}_T^{\text{\tiny HTE}}(v,\omega),
\end{align} 
where, $\overline{ V}_T^{\text{\tiny HTE}}(v,\omega)$  is  given by the sum of Eq.~\eqref{BHTE} and Eq.~\eqref{FHTE} but  with the scalar singlet contribution excluded from Eq.~\eqref{BHTE}, retaining only the contributions of $h,\,Z,\,W^\pm$ particles.
\end{itemize}

\noindent
Notably, the fractional term proportional to $m_i^4$ in Eq.~\eqref{BHTE}, and $m_t^4$ in Eq.~\eqref{FHTE}, cancels out with the corresponding CW correction in Eq.~\eqref{cw}, whereas the logarithmic part is reduced to $\log\left[\frac{e^{2\gamma_{\text{\tiny E}}}T^2}{\Lambda^2}\right]$, where $\frac{e^{2\gamma_{\text{\tiny E}}}T^2}{\Lambda^2}\approx \mathcal{O}(1)$, if the scale is considered equal to the temperature. For other scales, one can qualitatively estimate the relative weight of the CW corrections by examining the possible shift in the Higgs vacuum structure when considering  CW contributions as
\begin{align}
V_\eff(v,\omega) &= V_{TL}(v,\omega) + V_{CW}^{(1)}(v,\omega)\;.
\intertext{Then,}
\frac{1}{v}\nabla{\big( \Delta V_\eff (v,\omega) \big)}&= \frac{1}{v} \frac{\partial V^{(1)}_{CW}}{\partial v}\\
&= \sum_i \frac{n_i \rho_i}{32 \pi^2}\left[ \log\left( \frac{m_i^4}{\Lambda^4}\right) -2f_i +1\right]v^2\;,\label{vcorrection}
\end{align}
where $m_i = \rho_i v$, and $\rho_i$ is  the coefficient of $v$ for each  particle $i$ in Table~\ref{CWparameters}. The logarithmic term in Eq.~\eqref{vcorrection} is attenuated by the corresponding contributions from the thermal corrections and the running of the couplings, which also partially cancel the CW corrections. Hence it can be disregarded in comparison to the other terms in brackets,
\begin{align}
\frac{1}{v} \frac{\partial V^{(1)}_{CW}}{\partial v} \approx \sum_i \frac{n_i \rho_i}{32 \pi^2}\left[ 1 -2f_i \right]v^2, \label{vcorrection2}
\end{align}
This results in an increase in the coefficients of $v^2$ in Eq.~\eqref{vev1} and Eq.~\eqref{vev2}, which is smaller than $0.005$ when only including the effect of $t,\,Z,\,W^\pm$. This becomes even smaller when the effects of $H,S$ are added because they have an overall negative value for the bracket in Eq.~\eqref{vcorrection2}. Consequently, this correction can maximally change Higgs vev by a factor smaller than $1\%$ when compared to the leading $v^2$-coefficients; $\lambda_h \approx 0.13$ and $\alpha $ which could be even larger than $\lambda_h $ for non-vanishing $\omega$, when $\abs{\alpha}> \frac{1}{2}\lambda_{hs} \omega $. Therefore, for an approximate understanding of the thermal evolution of  Higgs vev, one can disregard the CW correction to the effective potential as well.

In addition to thermal corrections, the contributions from the rings diagrams must be considered to secure the cubic mass term\footnote{This term characterizes the bosonic contributions, as it only originates from the vanishing Matsubara mode, see~\cite{leigh1992infraredeffectsbubblepropagation,laine2016}, but the gauge boson terms will not generate any imaginary output for all values of $v$, as their mass can never be negative.} in Eq.~\eqref{BHTE} from having imaginary values for $m_1,m_2$ at certain values of the vevs~\cite{carrington1992effective}, indicating the breakdown of perturbative expansion at high temperatures due to the quartic coupling running with temperature~\cite{arnold1992phase,fendley1987effective,leigh1992infraredeffectsbubblepropagation}. This issue can be resolved by incorporating higher-order IR-divergent contributions, ring (Daisy) diagrams, which, according to Ref.~\cite{carrington1992effective} takes the following form:
\begin{align}
V_{\text{\tiny rings}} = -\frac{T}{12\pi} \Bigg[ \left(m^2(v,\omega) + \frac{\lambda}{4}T^2\right)^{\frac{3}{2}} -m^3(v,\omega) \Bigg].\label{rings}
\end{align}
This corresponds to the shift $m_{1,2}^3 \rightarrow M_{1,2}^3= \left(m_{1,2}^2(v,\omega) + \frac{\lambda}{4}T^2\right)^{\frac{3}{2}}$, where the gauge bosons do not lead to any imaginary parts. Therefore, at very high temperatures, $\lambda T^2 \gg m^2_{1,2}$, the scalar cubic term is reduced to a pure temperature term, independent of the fields. This implies that the cubic term coefficient will be exclusive to the vector gauge boson contributions in this approximation, and will not receive any further corrections in  xSM. Considering the main corrections in Eq.~\eqref{fullv} and Eq.~\eqref{rings}, the effective potential becomes,
\begin{align*}
V_\eff(v,\omega,T) &= V_{TL}(v,\omega) + V_T^{\text{\tiny HTE}}(v,\omega) + V_{\text{\tiny rings}}\\
&= \frac{1}{2} C(T^2-T_0^2) v^2 - E\, T \, v^3+\frac{1}{2}\alpha v^2\omega+\frac{1}{4}\lambda_{hs}v^2\omega^2 \\
&+\frac{1}{4}\lambda_{h}v^4+ \frac{1}{2} D(T^2-T_1^2 )\omega^2+\frac{1}{3} \beta\omega^3 +  \frac{1}{4}\lambda_{s}\omega^4+\mathcal{O}\left(\frac{1}{T^2}\right)\;, \numberthis  \label{xsmlag}
\end{align*}
where,
\begin{align}
C &= \frac{1}{4} \left(\frac{1}{4}(3g_1^2+g_2^2)+2y^2_t +\lambda_h +\frac{\lambda_{hs}}{6} \right)\;,\label{cvalue}\\
E&= \frac{1}{32\pi} \left( 2g_1^2+ (g_1^2+g_2^2)\right)^{\frac{3}{2}}\;,\\
D&= \frac{1}{4}\left(\lambda_s +\frac{\lambda_{hs}}{6} \right),\quad T_0^2 = \frac{\mu^2_h}{C},\quad  T_1^2 = \frac{-\mu_s^2}{D}\;,\label{temp}
\end{align}
and $\mathcal{O}\left(\frac{1}{T^2}\right)$ represents corrections arising from the $m^6$-term in the HTE. The vevs then become temperature dependent,
\begin{align*}
v_b(T) &= \Bigg\{\frac{6ET\pm \sqrt{36E^2T^2 - 8\lambda_h [2C(T^2-T_0^2)+(2\alpha +\lambda_{hs} \omega)\omega]}}{4\lambda_h}\;,\\
& \pm\sqrt{\frac{-2[D(T^2-T_1^2) +\beta\omega +\lambda_s \omega^2]\omega}{\alpha+\lambda_{hs} \omega}}\Bigg\}\;.\numberthis\label{vtemp}
\end{align*}
The above equation represents the temperature-dependent version of Eq.~\eqref{xSMvev}, and it can be rewritten using the relations in  Eq.~\eqref{temp} as, 
\begin{align}
\frac{v_b(T)}{T} = \frac{3E}{2\lambda_h}+\sqrt{\frac{\mu_h^2}{T^2}+\frac{9E^2}{4\lambda^2_h}-\frac{C}{\lambda_h}-\xi(T)}, \label{spha}
\end{align}
where $\displaystyle{\xi(T) = \frac{(2\alpha+\lambda_{hs} \omega)\omega}{2\lambda_h T^2 }}$. At the critical temperature in the SM, the sphaleron condition in Eq.~\ref{spha} returns $\displaystyle{\frac{v_c}{T_c} = \frac{2E}{\lambda_h} \approx 0.15}$, where $\displaystyle{T_c^{\text{\tiny SM}} = \frac{\mu_h}{\sqrt{C\left( 1-\frac{2E^2}{\lambda_h C}\right)}}}$ and $\xi(T)= 0$.
This means that the sphaleron condition in the xSM is primarly improved by the $\xi(T)-$correction term, which can contribute significant negative values to Eq.~\ref{spha} for large negative values of the odd $\mathbb{Z}_2-$asymmetric term $\alpha H^\dagger H S$. An additional supporting correction arises from the decrease in the critical temperature due to the $\lambda_{hs}$ portal coupling correction to the $C$-term, as per Eq.~\eqref{cvalue} and the correction to the $\mu_h^2$  term given in Eq.~\eqref{vev1}. The critical temperature is determined from the degeneracy condition,\footnote{We only consider a one-step phase transition, assuming that the scalar singlet field evolved a non-vanishing vev, which remains the same at $T_c$. In principle, a two-step phase transition mainly accounts for a shift in the effective potential that can affect the gravitational-wave spectrum, while barely affecting the value of the critical temperature~\cite{carena2020electroweak}.}
\begin{align}
V_\eff(0,\omega;T_c) = V_\eff(v_c,\omega;T_c),\qquad
\frac{\partial V_\eff(v,w,T_c)}{\partial v}\Bigg \rvert_{\braket{v} = 0} = \frac{\partial V_\eff(v,w,T_c)}{\partial v}\Bigg \rvert_{\braket{v} = v_c} ,
\end{align}
which yields,
\begin{align}
T_c^{\text{\tiny  xSM}} \approx 6.6 \times 10^{-3} \left(\frac{\mu_h}{\sqrt{C}} \right)^{\text{\tiny xSM}} \Bigg[1-\frac{(\alpha+\frac{1}{2}\lambda_{hs} \omega)\omega}{\mu_h^2} \Bigg]^{\frac{1}{2}}\, T_c^{\text{\tiny SM}}.\label{TcxSM}
\end{align}
Where $\displaystyle{\left(\frac{\sqrt{C}}{\mu_h}\right)^{SM}\approx 6.6 \times 10^{-3}}$. For large negative values for $\alpha$, $\abs{\alpha} >\frac{1}{2} \lambda_{hs} \omega$, the square bracket in Eq.~\eqref{TcxSM} becomes greater than one; however $\mu_h$ simultaneously decreases and the $C$-term increases, which collectively leads to a significant decrease in the critical temperature in xSM, $T_c^{\text{\tiny  xSM}} < T_c^{\text{\tiny  SM}}$.

This approximate analytical investigation of the xSM extension is corroborated by the numerical methods employed to account for the full NLO (CW and finite temperature) corrections to the tree-level potential at different renormalization scales (conservative and liberal categories, see  later) in order to address the theoretical uncertainties in the FOEWPT parameter space points, which are necessary for accurately exploring the muon collider. As illustrated in Fig.~\ref{foewpt_ms}, the improvement of the sphaleron condition by the xSM extension is evident, as is the decrease in critical temperature due to the singlet contribution to the $C$ and $\mu_h^2$-terms, as demonstrated in Fig.~\ref{Tc_mass_C}  for various singlet mass values.

\subsection{First-Order Electroweak Phase Transition Parameter Space within the xSM} 

Catalyzing a FOEWPT through the real singlet extension depends on the parameter space points, particularly on the $\mathbb{Z}_2$-asymmetric parameters and the Higgs-scalar quartic coupling, as shown in Eqs.~\eqref{spha}. The additional scalar degree of freedom introduces five free parameter $\{\mu_s^2,\,\lambda_s,\,\alpha,\,\beta,\,\lambda_{hs}\}$ which need to be determined. In \cite{profumo2007singlet,carena2020electroweak,espinosa2012strong,fernandez2023nu}, a set of diverse constraints, stemming from theoretical and phenomenological sources, were discussed to restrict these parameters,
\begin{itemize}
\item The stability of the effective potential necessiates positive dimensionless quartic couplings, $ (\lambda_h,\lambda_s,\lambda_{hs}) > 0$.
\item The validity of perturbative expansion requires quartic couplings to be smaller than unity, $(\lambda_s, \lambda_{hs}) \in [0:4\pi]$.
\item Electroweak symmetry breaking necessitates a positive determinant of the mass matrix in Eq.~\eqref{masses} as ($v_\text{\tiny EW}, \omega_\text{\tiny EW} $) should constitute the true global minimum, which requires, $4\lambda_h\lambda_s -\lambda_{hs}> 0$.
\item The electroweak precision obeservables (EWPO) were included, where the scalar decay into gauge boson pairs modifies the $S,T,\rho,U$ parameters.
\item Higgs boson branching ratio corrections owing to mixing with the scalar, in addition to the absence of exotic Higgs boson decays, were applied. This shifts the scalar mass towards higher values and constrains the ($\mu_s,\, \alpha,\, \beta$) parameters.
\end{itemize}
\begin{table}
\centering
\begin{tabular}{|>{\centering\arraybackslash}p{20mm}|>{\centering\arraybackslash}p{35mm}|}
\hline
   Parameter    & Range \\[3pt] \hline
$\mu_s$  &  $[-2000,2000]$ GeV\\[7pt] 
$\lambda_s$  &  $[10^{-4},1.5] $ \\[7pt] 
$\lambda_{hs}$  &  $[0,5]$ \\[7pt] 
$\alpha$  &  $[-1000,0]$ GeV \\[7pt] 
$\beta$  &  $[-1800,1800]$ GeV \\[4pt] \hline
\end{tabular}
\caption{Ranges of xSM parameters  obtained from  theoretical and phenomenological constrains used in Ref.~\cite{Papaefstathiou_2021}.}\
\label{parameterspace}
\end{table}
We used the values obtained in Ref.~\cite{Papaefstathiou_2021}, in which nearly identical constraints were applied with minor variations. In this study, valid points satisfying the sphaleron condition $\frac{v_c}{T_c}> 1$ were obtained through a scan using the \texttt{PhaseTracer} package~\cite{athron2020phasetracer}, which tracks vacuum evolution by evaluating the thermal integral of Eq.~\eqref{fullv} numerically by employing the methods described in ~\cite{fowlie2018fast,wainwright2012cosmotransitions}, rather than the HTE approximation discussed in the previous calculations to obtain an approximate understanding of xSM. Furthermore, the ranges of the ``portal'' coupling $\lambda_{hs}$, and singlet quartic coupling $\lambda_s$, were set to be quite loose in \cite{Papaefstathiou_2021}, as listed in Table~\ref{parameterspace}. The rationale behind this approach is that for colliders to exclude the possibility of a strong FOEWPT (SFOEWPT), the candidate parameter space points should be relatively broad.

\begin{figure}[htb!]
\centering
\includegraphics[scale=.6]{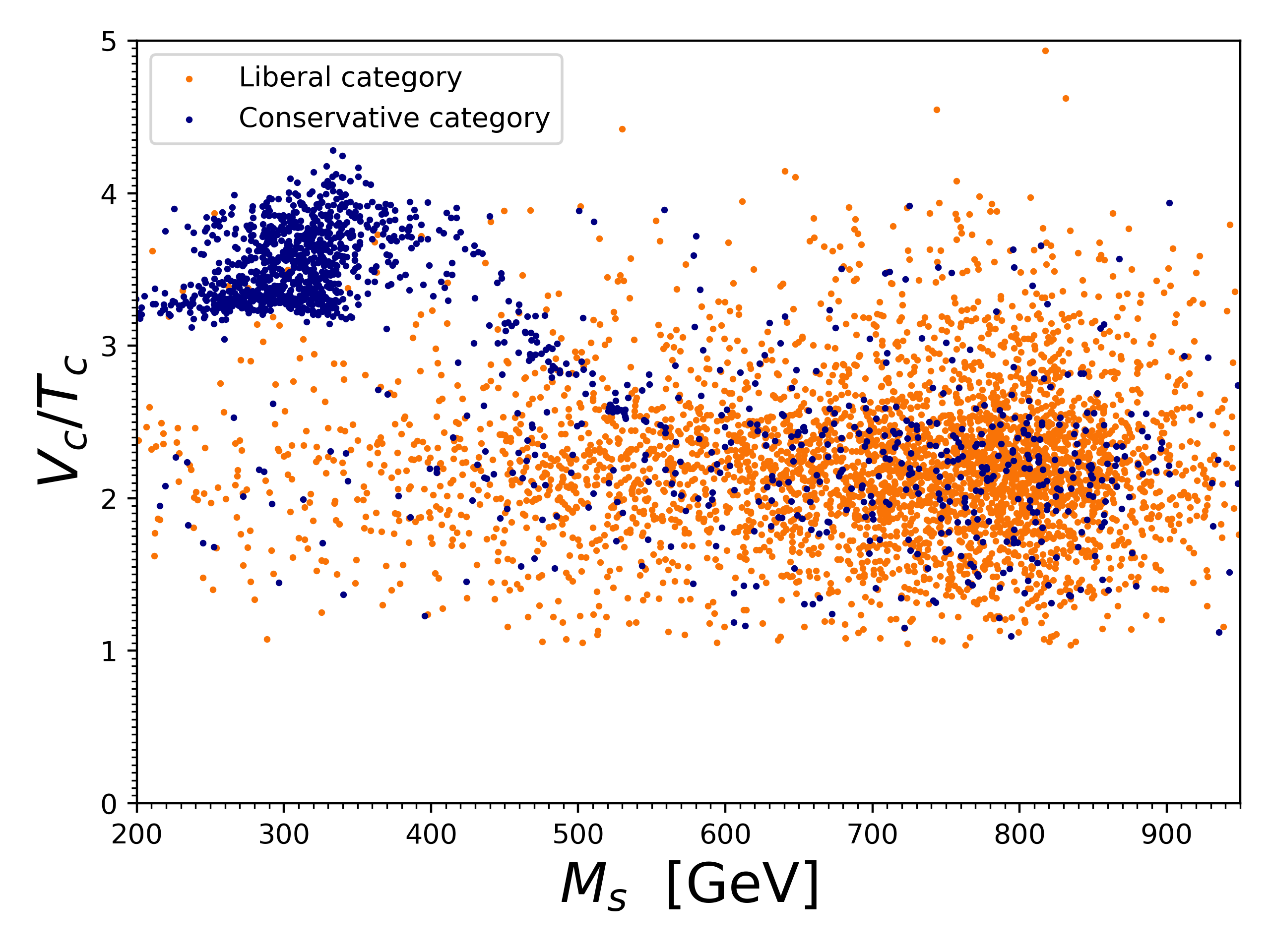}
\caption{Verification of the sphaleron condition against the singlet scalar mass range, as obtained from the xSM parameters.}
\label{foewpt_ms}
\end{figure}
 
The values of the $\lambda_{hs}$ coupling control the shift in the quadratic correction on $C$ from the SM expectation, and consequently improve the sphaleron condition. However,
this raises concerns regarding the validity of the perturbative expansion and its sensitivity to theoretical uncertainties related to the scale and gauge dependencies of the thermal
parameters. This consideration is crucial not only for analyzing specific points, but also for comprehending the parameter space \cite{lewicki2024impacttheoreticaluncertaintiesmodel,gould2023higherorderscosmologicalphase}. According to \cite{Papaefstathiou_2021}, the dominant uncertainty originates from the slow convergence of perturbation theory which is manifested in the dependence on an arbitrary renormalization scale. We follow Ref.~\cite{Papaefstathiou_2021} and define a ``conservative'' point as a point that admits an SFOEWPT for a group of eight scale and gauge variations, with at least one of these variations possessing a Higgs field vev within $246 \pm 30$~GeV, that also corresponds to a transition to the absolute (i.e.\ deepest) minimum of the potential. On the other hand a ``liberal'' point satisfies SFOEWPT for one of the eight scale and gauge variations and the Higgs vev condition for any other. See Ref.~\cite{Papaefstathiou_2021} for further details. 

\begin{figure}[htb!]
\centering
\includegraphics[scale=.46]{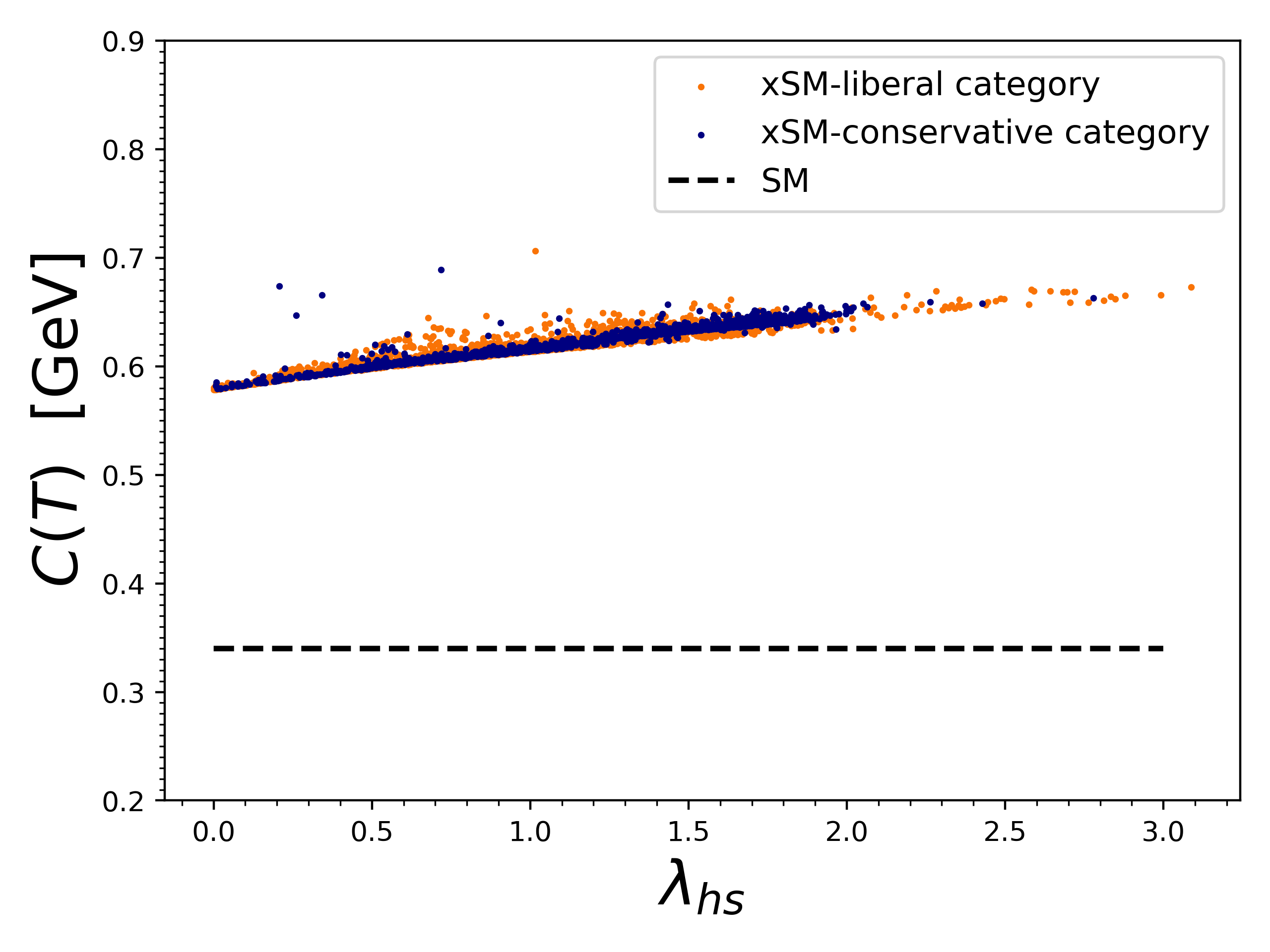}
\includegraphics[scale=.46]{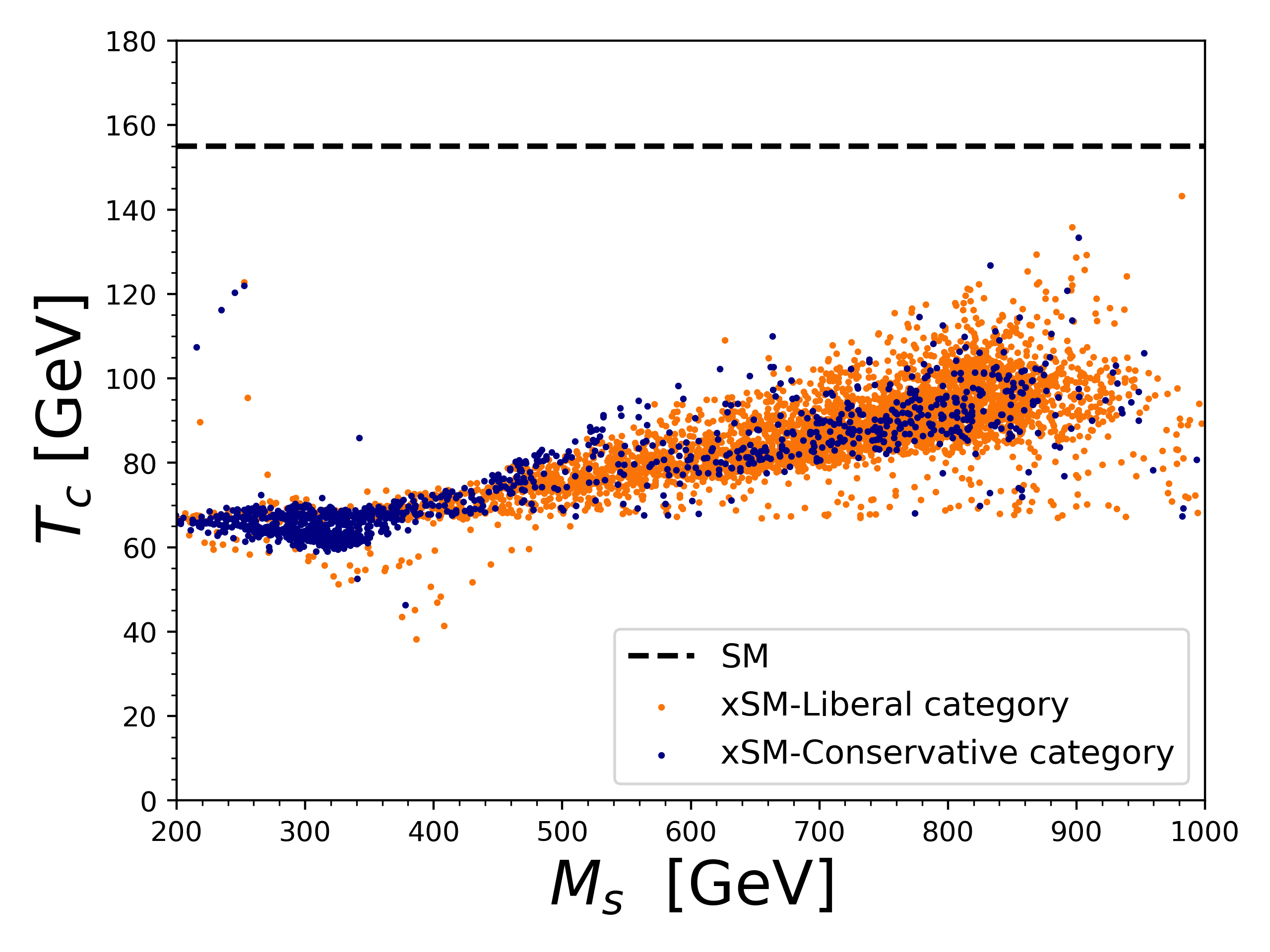}
\caption{The left panel shows the correction to the quadratic thermal term from the $\lambda_{hs}$, while the right panel shows the drop in the critical temperature obtained from the xSM. The observed decrease in the critical temperature and increase in the $C(T)$ term collectively enhance the sphaleron condition. }
\label{Tc_mass_C}
\end{figure}

A substantial number of parameter space points fulfill the SFOEWPT sphaleron condition out of the $\mathcal{O}(10^6)$ points generated according to Table~\ref{parameterspace}. The SFOEWPT condition is plotted against the scalar mass in Fig.~\ref{foewpt_ms} for {\it conservative} and {\it liberal} categories. The observation that  {\it conservative} category values shift more towards lower values of scalar mass, unlike {\it liberal} values, is a good indicator that this category can be verified by current collider searches. For instance, the LHC searches for an SM-like Higgs boson~\cite{cms2013search}, which may lead to the exclusion of the conservative category.
 
\begin{figure}[htb!]
\centering
\includegraphics[scale=.46]{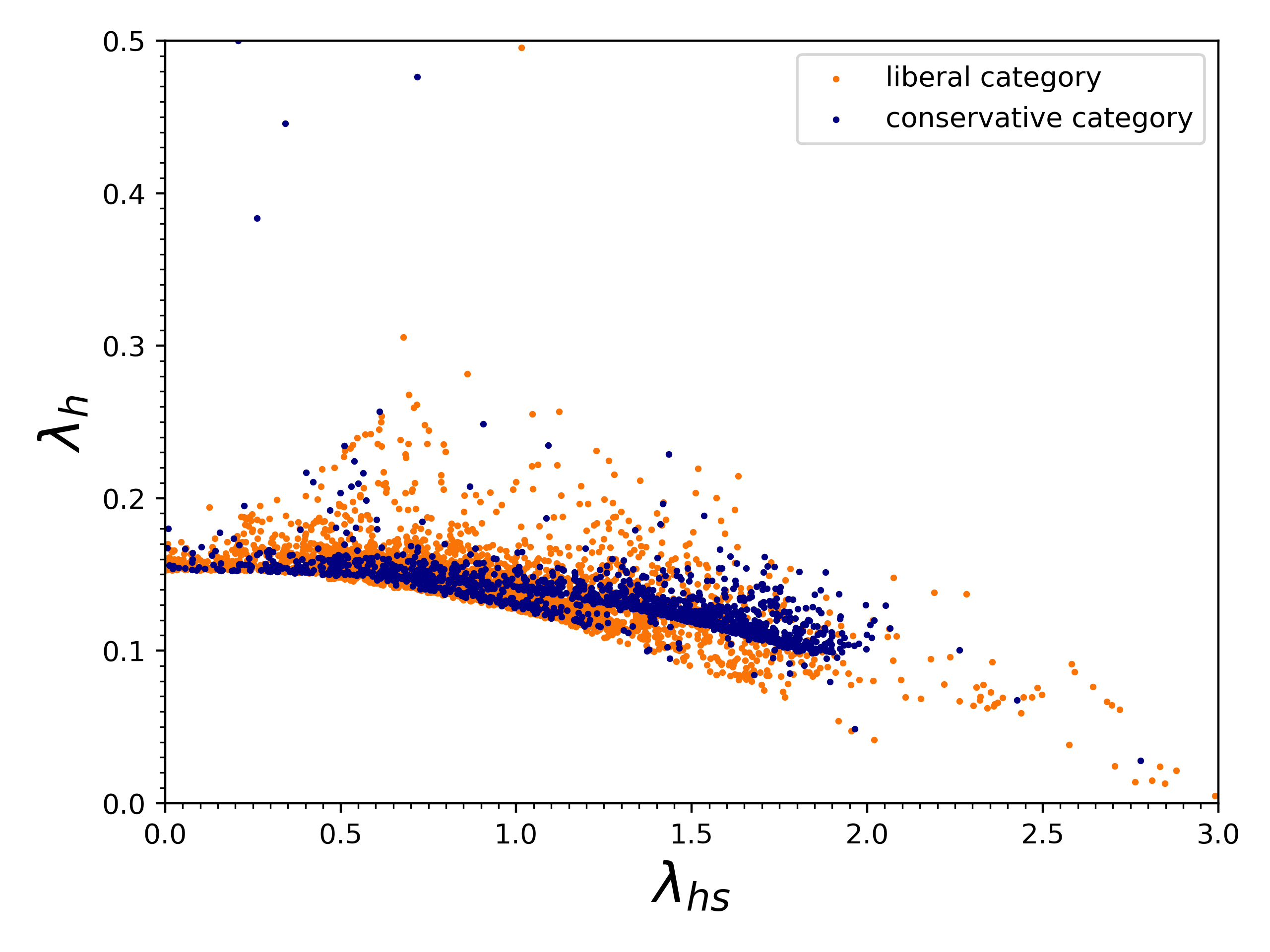}
\includegraphics[scale=.46]{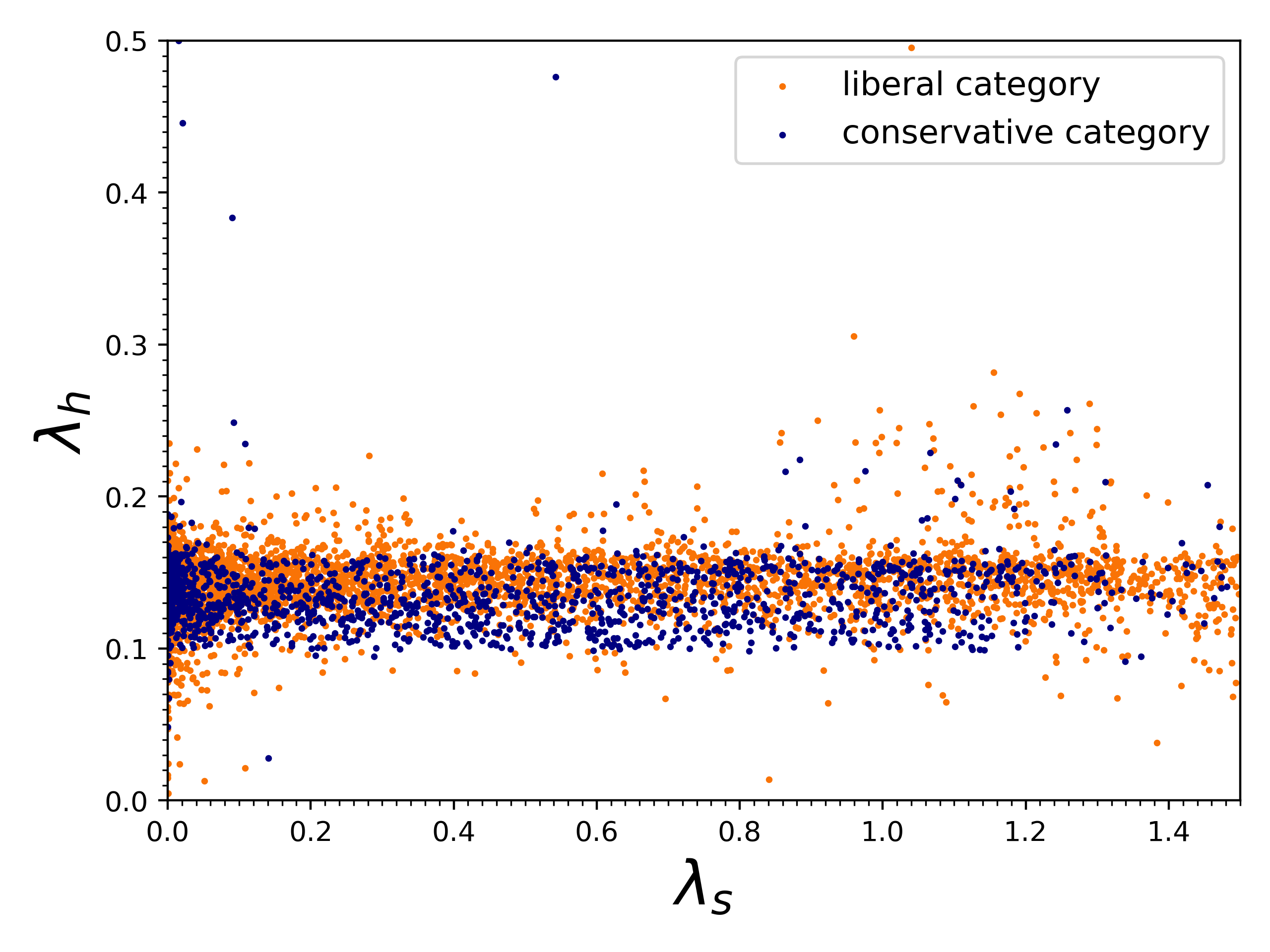}
\caption{The dimensionless quartic couplings are shifted more towards lower values, which is a preferred behaviour, as required by the validity of the perturbative expansion.}
\label{portal_couplings}
\end{figure}

The portal coupling $\lambda_{hs}$ improves the SFOEWPT condition primarily through the correction it contributes to in the quadratic $C$ term, which subsequently reduces the critical temperature. This phenomenon is illustrated in Fig.~\ref{Tc_mass_C} (left panel) which demonstrates the increase in the $C-$term with $\lambda_{hs}$, contrary to the SM expectation. The right panel of Fig.~\ref{Tc_mass_C} shows the tendency toward a lower critical temperature compared to the SM expectation, for both the Conservative and Liberal categories. 
Furthermore, the valid parameter points exhibit a propensity to concentrate towards lower values of the quartic couplings ($\lambda_s,\,\lambda_{hs}$), as shown in Fig.~\ref{portal_couplings}, which indicates favorable behavior regarding the validity of perturbative expansion.

\section{The Electroweak Phase Transition at Colliders}\label{sec:ewptc}

A direct test for SFOEWPT could be derived from the gravitational waves emanating from bubble dynamics. Specifically, sound waves in the plasma and magneto-hydrodynamic turbulence generate gravitational waves that are expected to peak at low frequencies ranging from mHz to
10 Hz. Collider searches provide another promising and complementary approach for verifying the nature of EWPT. This is attributable to the fact that the catalyst of the FOEWPT,
the singlet scalar field in this instance, can be probed either directly through resonant production of a new scalar, if its mass is within the range of current and future searches, or indirectly, through precise measurements of potential deviations in the Higgs boson couplings from the SM expectation.
In contrast to gravitational wave detection, which can only be produced from FOEWPT, collider searches can be sensitive to SFOEWPT as well; however, this is more challenging owing to the weak mixing with the Standard Model particles.

\subsection{Direct Singlet Production at Colliders}

The mass of the new scalar particle in xSM is expected to be of $\mathcal{O}(T_{\text{\tiny EW}})$, which suggests the possibility of resonant production at colliders~\cite{ramsey2020electroweak}. At zero temperature, i.e.\ in the collider scenario, the mass of the singlet-like scalar is given by $m_2(v_{\text{\tiny EW}},\omega_{\text{\tiny EW}})$, Eq.~\eqref{zeroTmass}, which yields a scalar mass in the range $m_s \in [200,1000]$~GeV, when utilizing the range of parameters as shown in Fig.~\ref{foewpt_ms}. This can be  approximated using the curvature of the effective potential in Eq.~\eqref{xsmlag},
\begin{align*}
m_s^2(0) &= \frac{\partial^2 V_\eff}{\partial s^2}\\
&= \frac{1}{2} \lambda_{hs} v^2_{\text{\tiny EW}}  - D(T_1^2) + 2\beta \omega_{\text{\tiny EW}} + 3\lambda_s \omega^2_{\text{\tiny EW}}\\
&= \frac{\lambda_{hs}}{6} \big(3v^2_{\text{\tiny EW}}  -\frac{6}{\lambda_{hs}} (DT_1^2 -2\beta\omega -3\lambda_s \omega^2) \big)\;, \numberthis
\end{align*} 
This coincides with the results of \cite{ramsey2020electroweak}, if we set $\displaystyle{T^2_{\text{\tiny EW}} = \frac{6}{\lambda_{hs}} (DT_1^2 -2\beta \omega -3\lambda_s \omega^2)}$.
This mass range is within the scope of prospective future collider experiments such as a muon collider, where a new scalar can be produced resonantly through one of the main production channels $\mu^+ \mu^- \rightarrow s \rightarrow XX$. The probability of direct detection is more promising for channels with charged fermions in the final states (i.e.\ four leptons, $l^-l^+l^-l^+$, or four quarks, $ b\bar{b}b\bar{b}$), where the invariant mass can be fully reconstructed.

Due to the anticipated low mass of the new scalar, $\lesssim \mathcal{O}$(TeV), both hadron and lepton colliders are valid options for detection. 
Previous studies~\cite{papaefstathiou2022electro,ramsey2020electroweak} have confirmed that a 100 TeV proton-proton collider can substantially explore this parameter space. Recently, there has been a growing interest in muon colliders, which can explore the same parameter space efficiently, at a significantly lower center of mass energy ($\sqrt{s} \in [3-10]$ TeV) \cite{ramsey2020electroweak, liu2021probing}. In a muon collider, most of the available energy is consumed in the hard process, which directly relates the resonance peak to the center of mass energy according to $\frac{\sqrt{s}}{2m_s}\approx 1.7$ as demonstrated in ref.~\cite{ramsey2020electroweak}. This characteristic is unique for lepton colliders, unlike hadron colliders where $E_\text{\tiny CM}$ of the beam differs from that of the parton beams, necessitating integration over the parton density functions. Furthermore, a muon collider will not suffer from the huge QCD radiation resulting from the initial state as in the case of hadron colliders, nor will its energy be dissipated in the large synchrotron radiation as in an electron collider. Notably, the optimal channel for the single production of a new scalar is through vector boson fusion (VBF), mainly $W^+W^-$ fusion, which contributes $\sim 90\% $ of the total cross section~\cite{liu2021probing, forslund2022high}.  

\subsection{Indirect Evidence: Modification of Higgs Boson Couplings}

The most significant terms for indirect detection in the extended potential are those that couple the new scalar to the Higgs field,
\begin{align}
V(h,s) \supset \frac{1}{2} \mu_m S H^\dagger H + \frac{1}{4} \lambda_m S^2 (H^\dagger H)^2\;,\label{mixingp}
\end{align}
\begin{figure}[htb!]
\centering
\includegraphics[scale=.46]{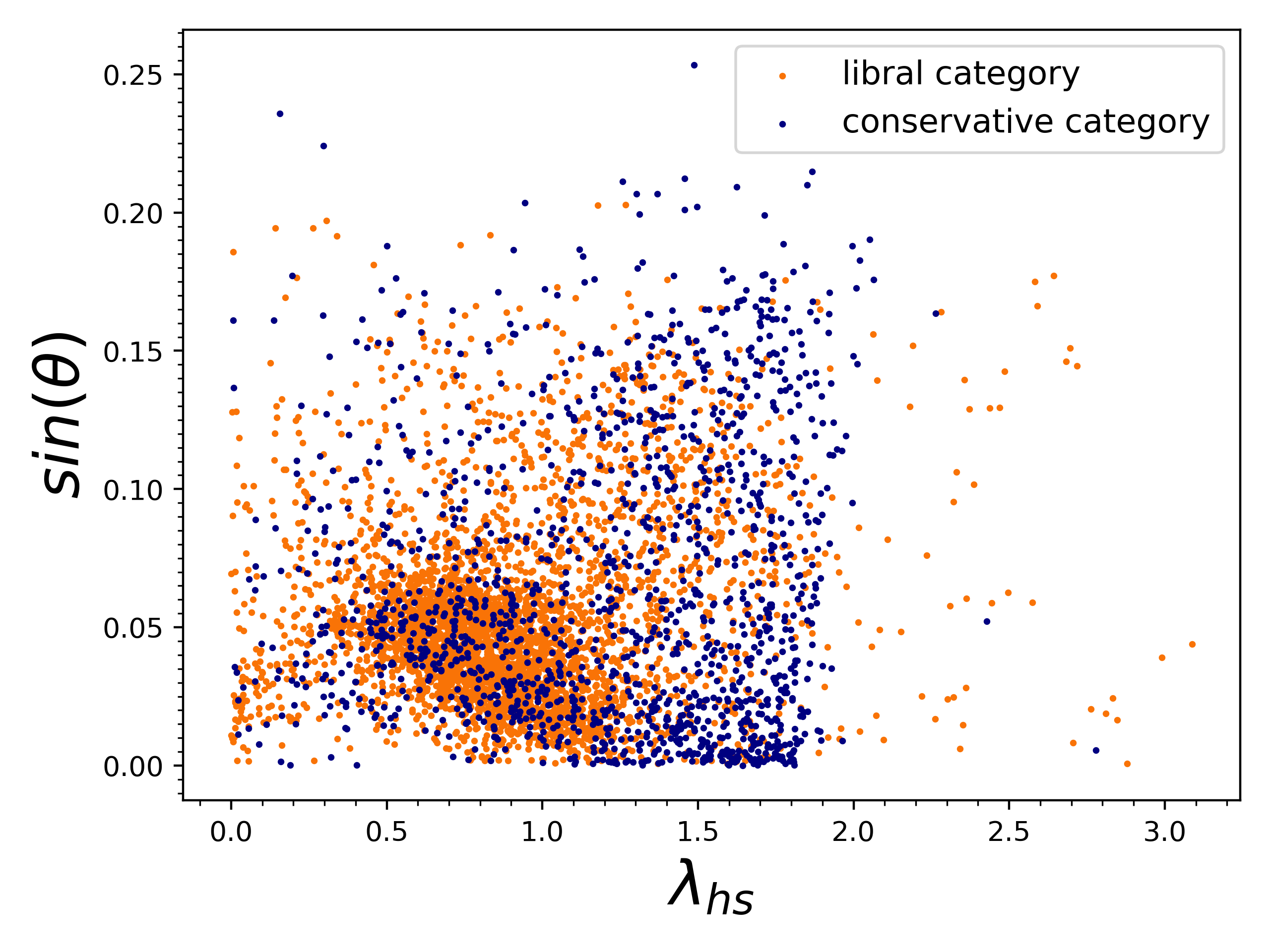}
\includegraphics[scale=.46]{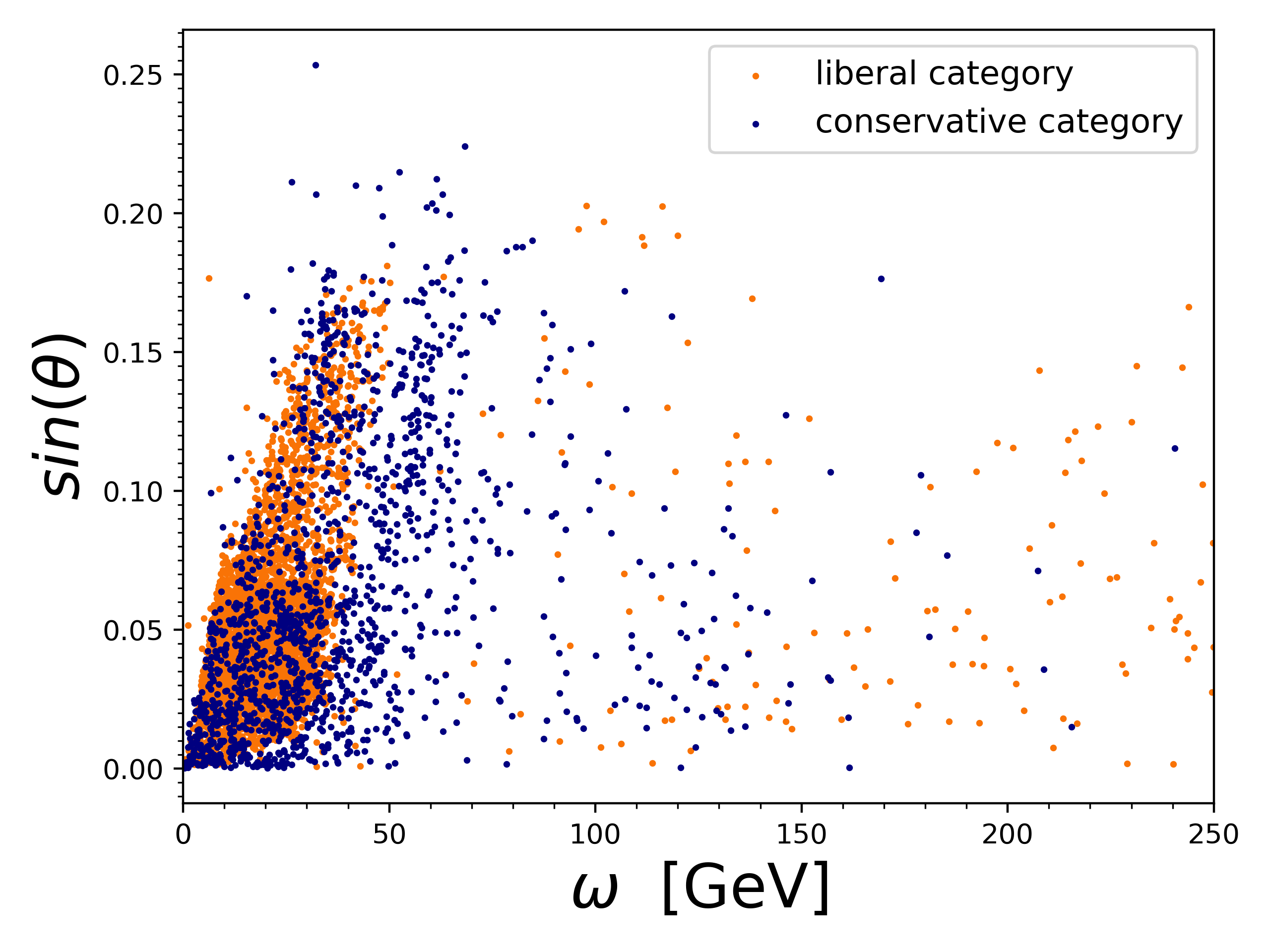}
\caption{The Higgs-singlet mixing angle against the portal coupling $\lambda_{hs}$ and singlet vacuum expectation value,  $\omega$.}
\label{theta}
\end{figure} 
 which consequently form a ``portal'' of the new scalar to the other SM particles. The other self-interaction couplings of the new scalar do play a crucial role in determining the points of EWPT but do not substantially affect phenomenological studies.\footnote{For example, the $\lambda_s$ coupling only appears in the daisy resummation part of the effective potential~\cite{curtin2014testing}.} The terms in Eq.~\eqref{mixingp} inevitably lead to mixing between $H$ and $S$ and consequently the new scalar ``inherits'' interactions with SM particles. The mass eigenstates are obtained using the rotation angle $\theta$, after symmetry breaking as follows:
\begin{align}
\begin{pmatrix} h_1\\[7pt] h_2 \end{pmatrix} = \begin{pmatrix} \cos \theta & \sin \theta\\[7pt] -\sin \theta & \cos \theta\end{pmatrix}\begin{pmatrix} h\\[7pt] s \end{pmatrix}\;,
\end{align}
where the mixing angle is given by,
\begin{align}
\cot \theta = \frac{2M^2_{hs}}{M^2_{hh}-M^2_{ss}+\sqrt{(M^2_{hh}-M^2_{ss})^2+4M^4_{hs}}}\;, \label{mixingangle}
\end{align}
where $M^2_{hh},\,M^2_{ss},\,M^2_{hs}$ are the components of the mass matrix in Eq.~\eqref{masses}. Consequently, the SM-like Higgs boson couplings must be adjusted according to:
\begin{align}
g_{\text{\tiny hXX}}^{\text{\tiny SM}} &\rightarrow  g_{\text{\tiny $h_1$XX}}^{\text{\tiny SM}} \cos \theta - g_{\text{\tiny $h_2$XX}}^{\text{\tiny SM}} \sin \theta.
\end{align}
Here, $g_{\text{\tiny $h_1$XX}}^{\text{\tiny SM}} \cos \theta$ represents the scaling of the original Higgs boson couplings, and $g_{\text{\tiny $h_2$XX}}^{\text{\tiny SM}} \sin \theta$ represents the new scalar couplings to the $XX$-SM-particles.
This opens up a range of possible precision tests such as small deviations in SM-like Higgs boson production rates according to $\sigma^{\text{\tiny BSM}}_h = \cos^2 \theta \times \sigma^{\text{\tiny SM}}_{h_1}$. 
Another significant phenomenological test would be the deviation of Higgs boson trilinear self-coupling from the SM expectation, where after symmetry breaking this term becomes, 
\begin{align}
\lambda_{hhh} &= \frac{1}{4} \left[c_\theta(\lambda_h c_\theta^2 +\lambda_{hs} s_\theta^2)v_{\text{\tiny EW}}  +(\alpha + \lambda_{hs}\omega)c_\theta^2 s_\theta +\frac{4}{3} (\beta +3\lambda_s \omega) s_\theta^3\right]\;,
\end{align}
 where $c_\theta  = \cos \theta$ and $s_\theta = \sin \theta$.
 In \cite{curtin2014testing}, a scan was performed over the possible parameter space which led to FOEWPT demonstrating a significant difference between $\lambda_{hhh}^{\text{\tiny BSM}}$ and  $\lambda_{hhh}^{\text{\tiny SM}}$, that could be as big as $\displaystyle{\frac{ \lambda_{hhh}^{\text{\tiny BSM}}}{ \lambda_{hhh}^{\text{\tiny SM}}} \sim 1.3}$. Conversely, the current precision of the Higgs boson couplings and branching ratios favor a small value of the mixing angle, which in our study was adopted to better constrain the parameter space using the additional phenomenological constraint in Eq.~\eqref{mixingangle}. This constraint indicates that the points that satisfy SFOEWPT favor smaller values of $\lambda_{hs}$ and singlet vev, $\omega$, as depicted in Fig.~\ref{theta}, which is in agreement with the perturbativity constraint. In this study we are focusing on the possible direct detection at a muon collider and save the possible precision tests of the Higgs couplings at muon collider for future work.

\section{Constraints on the xSM at a Muon Collider}
\label{sec:4}

The dominant production mechanism of the new scalar $(S)$ at a muon collider is vector boson fusion (VBF), with $W^+W^-$ fusion (WWF) dominating $ZZ$ fusion (ZZF) as reported in previous studies \cite{buttazzo2018fusing,forslund2022high,liu2021probing,costantini2020vector}.
\begin{figure}[htb!]
\centering
\includegraphics[scale=.5]{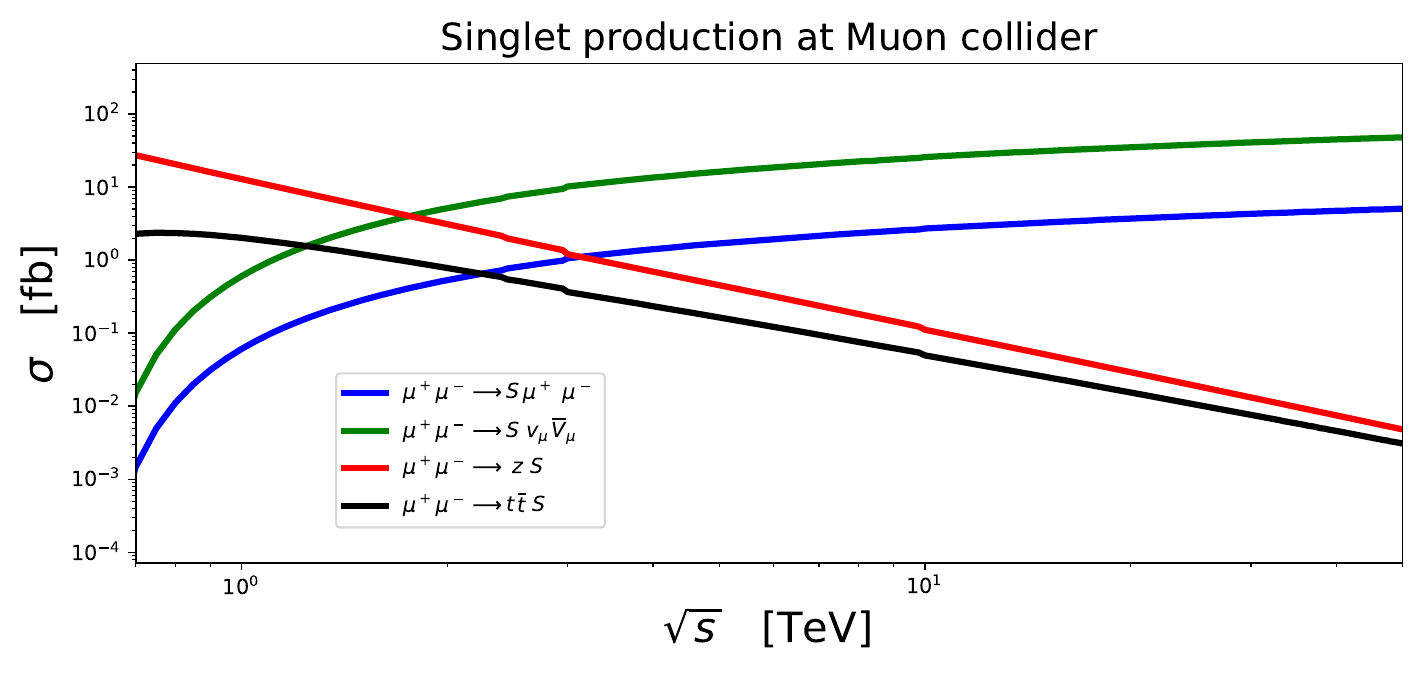}\\
(A)\\
\includegraphics[scale=.5]{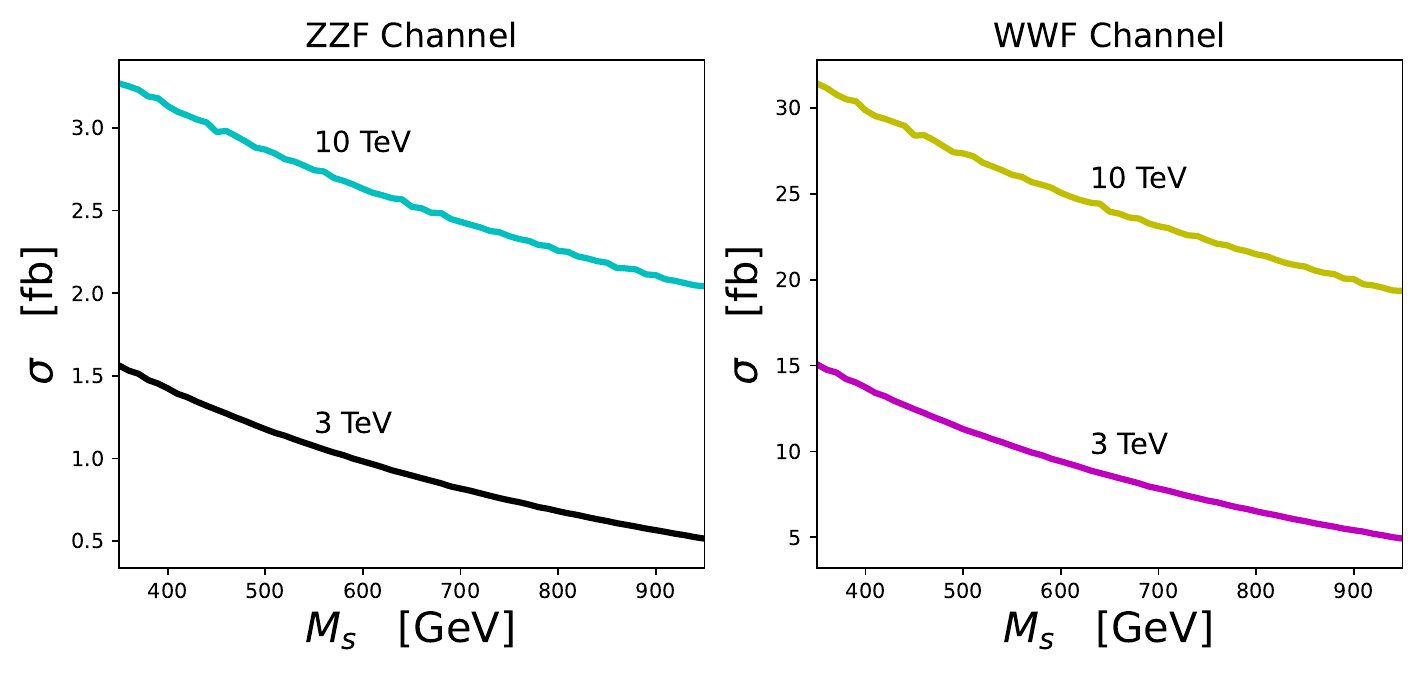}\\
(B)
\caption{The different production mechanisms are illustrated in (A), which also shows that VBF is dominated by $W^+W^-$ fusion (WWF) (green curve), which is further confirmed in (B), showing that WWF accounts nearly for 90\% of VBF. (B) also shows that the $S$-production cross section is only marginally sensitive on its mass $M_S$. The plots have been generated through \texttt{MadGraph5\_aMC@NLO} simulations at parton level for $\sin \theta=1$.}
\label{production}
\end{figure}
In our analysis, we set the muon beam center of mass energy to $\sqrt{s} = 3$ TeV, where VBF became the main production channel, as illustrated in Fig.~\ref{production}. The mass range of the scalar was taken to be $M_S \in [250:1000]$ GeV, stemming from the thermal correction constraint as discussed in the previous section. Consequently, the $S$ decay into two SM-like Higgs bosons becomes possible given a non-vanishing $\alpha$. Based on the potential in Eq.~\eqref{reno}, the new scalar will then primarily decay either directly into two Higgs bosons ($h_1h_1$), or into two vector bosons through the mixing between the two scalars, 
$$\mu^+ \mu^- \rightarrow S \,\mu^+ \mu^-(\nu_\mu \tilde{\nu}_\mu) \rightarrow  XX\, \mu^+ \mu^-(\nu_\mu \tilde{\nu}_\mu),\,\, (XX = h_1h_1, \,W^+ W^-,\,ZZ).$$
The final states obtained from these scalars encompass different topologies, which can be classified into two main categories: pure visible states such as $4l,\,4q,\, 2l2q,\, 2q2\gamma$, and visible-invisible mixed states such as $l\nu_l2q,\, 2l2\nu_l,\, 2q2\nu_l$. For each final state, we considered the backgrounds originating from all possible channels that would produce the same final state as without the resonant production of the $S$ scalar. Therefore, the invariant mass is expected to serve as a highly effective discriminant against the substantially larger background. In our analysis, we initially filter the signal through the invariant mass computed for each specific topology, and subsequently refine the remaining events based on their transverse momentum and pseudorapidity window, obtained by direct comparison of the generated signal to the dominant background. An advantage of a muon collider, that becomes apparent in this analysis, is the ability to track the missing energy that may be present in the final states, and account for it in the invariant mass calculations, as we will see below. In the remaining parts of this section, we enumerate the channels that we have investigated and provide their corresponding expected sensitivities. Subsequently, we project these sensitivity plots onto the FOEWPT xSM parameter space, to assess the potential of the proposed muon collider to provide meaningful constraints.

\subsection{Event Generation}

 Both signal and background Monte Carlo events were generated at the parton level using \texttt{MadGraph5\_aMC@NLO}~\cite{Alwall:2011uj} (\texttt{MG5\_aMC}), while the scalar decays to gauge/Higgs bosons were generated in \texttt{HERWIG 7}~\cite{Bahr:2008pv,Bellm:2017bvx,Gieseke:2011na,Arnold:2012fq,Bellm:2013hwb,Bellm:2019zci,Bewick:2023tfi}, along with the parton showering, underlying event and hadronization. The \texttt{HwSim} plugin for \texttt{HERWIG 7}~\cite{hwsim} was used to generate \texttt{ROOT}~\cite{Brun:1997pa} files for all event samples. Signal events were produced in \texttt{MG5\_aMC} by using the \texttt{loop\_sm\_scalar} model~\cite{Papaefstathiou_2021}.\footnote{We used electron-positron beams for event generation instead of muon-antimuon beams since \texttt{HERWIG 7}, at the time of writing, cannot readily handle muons in the initial state. This does not affect our analysis at all because of lepton universality, and the irrelevance of the lepton mass at high energies.} 
 \begin{figure}[htb!]
    \centering
    \includegraphics[scale=.45]{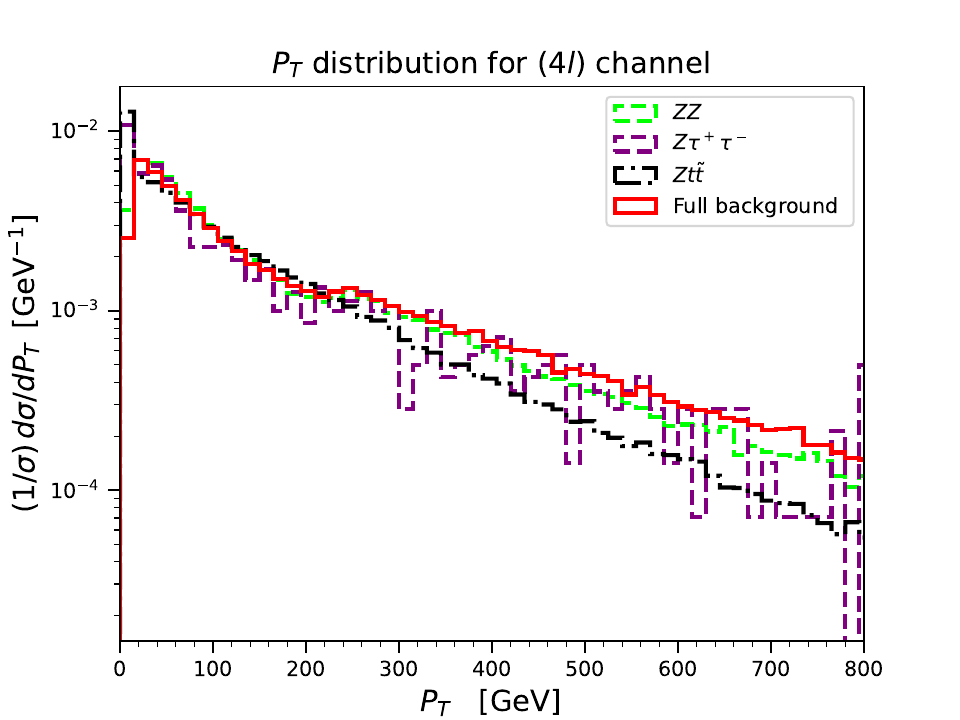}
    \includegraphics[scale=.45]{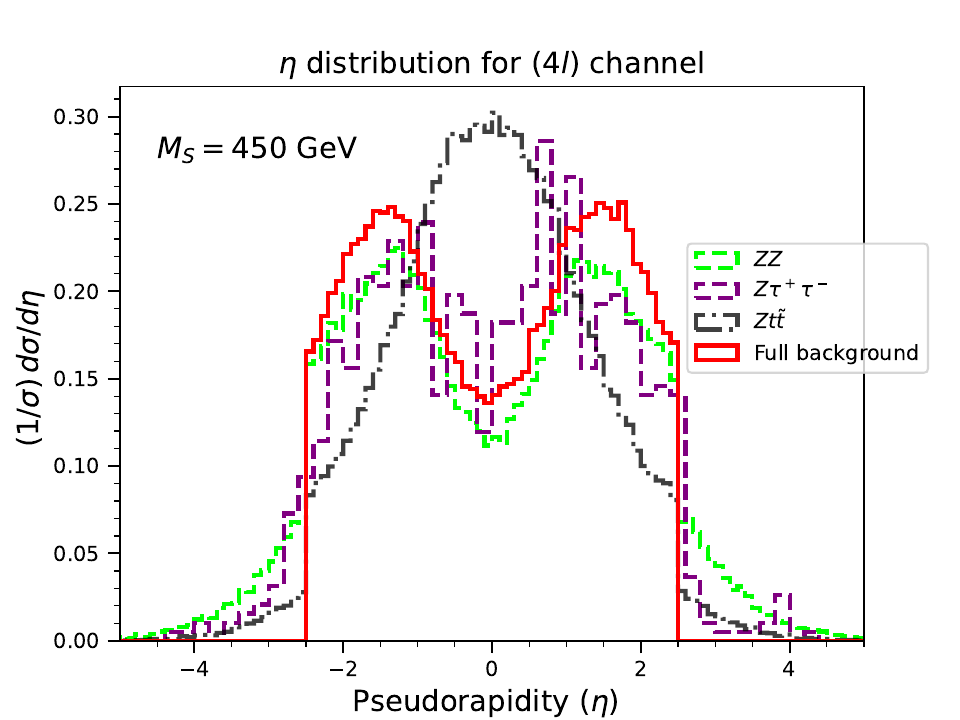}
    \caption{The summed background (red curve) is evidently the sum of all subdominant background resulting from $ ZZ,\, Z\tau^-\tau^+,\, Zt\, \overline{t}$ as seen in transverse momentum distribution (left) and pseudorapidity distribution (right).}
    \label{back_comparision}
\end{figure} 
 Due to background events at a muon collider occurring at much lower rates than at a hadron collider, we took into account all the possible background events for each final state by generating it directly in \texttt{MG5\_aMC}, without importing the \texttt{loop\_sm\_scalar} model. For example, for $\mu^+\mu^-\rightarrow \mu^+\mu^- S \rightarrow \mu^+\mu^- l^+l^+ l^- l^-$, we generated the combined background, 
 $\mu^+\mu^- \rightarrow \mu^+\mu^- l^+l^+ l^- l$.
Combined process generation implicitly contains all the dominant backgrounds resulting from $ZZ,\,Z\tau^+\tau^-,\,Zt\tilde{t},\cdots$. We have examined this claim by comparing the kinematic distributions ($P_T,\, \eta$) of the $4l$-combined background to the dominant $4l$ background sources, such as those coming from $ZZ,\,Z\tau^+\tau^-,\,Zt\tilde{t}$, and confirmed that the combined $4l$- background is indeed almost identical to the sum of the individual dominant backgrounds, as is evident in Fig.~\ref{back_comparision}. Therefore, in our analysis, we compare the signals to the combined backgrounds, instead of just the main ones.

\subsection{$4l$ Final States} 

The four-lepton final state originates from $S \rightarrow ZZ$, and its dominant background is $\mu^+ \mu^- \rightarrow 2l^+ 2l^- \mu^+ \mu^- $ and $\mu^+ \mu^- \rightarrow 2l^+ 2l^- \nu_\mu \bar{\nu}_\mu$. 
The signal leptons are expected to have a higher transverse momentum compared to the background ones, as they originate from $Z$-boson decays, which suggests a constraint on the transverse momentum in terms of $Z$-boson mass as $\sim P_T(l^\pm) \geq \frac{2}{5} M_z$. This assumption was verified by comparing the signal-background transverse momentum distributions for the charged leptons, which showed signal domination in the region $\sim 50\,\mathrm{GeV} \leq P_T(l^\pm) \leq 400\,\mathrm{GeV}$. Consequently, we identified any oppositely charged same-flavour leptons whose $P_T(l^\pm) \geq \frac{2}{5} M_z$, and their invariant mass peaked around the $Z$-boson mass, $0.8 M_Z\leq M(l^+l^-)\leq1.2 M_Z$ as valid pairs resulting from $Z-$boson decay. 
This works as a strong signal-background separator because the leptons produced from the background generated in \texttt{MG5\_aMC@NLO} are sourced from different mediators, as discussed in the previous section. This is in contrast to the signal case where they are mainly produced from $Z$-bosons, $S\rightarrow ZZ$. 
Then, the surviving pairs were further constrained by requiring the invariant mass of the two pairs to be in the vicinity of the scalar mass, $M_s-50 \,\mathrm{GeV} \leq M(l^+l^-l^+l^-)\leq M_s+50 \,\mathrm{GeV} $, which further suppressed the background events, as they were not resonantly produced in this range. 
\begin{figure}[htb!]
\centering
\includegraphics[scale=.4]{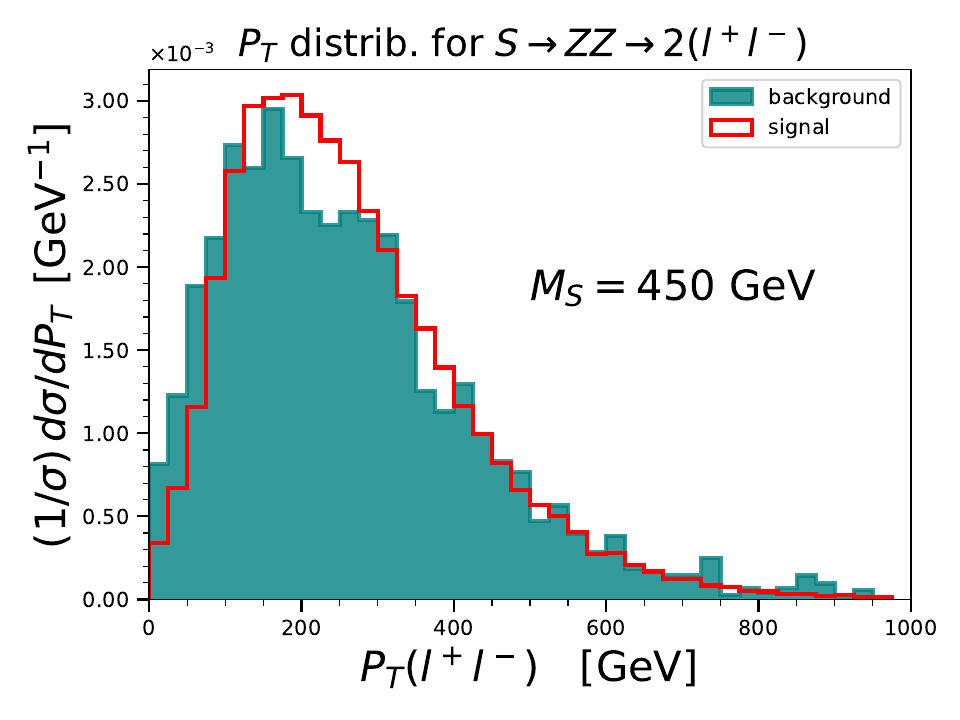}
\includegraphics[scale=.4]{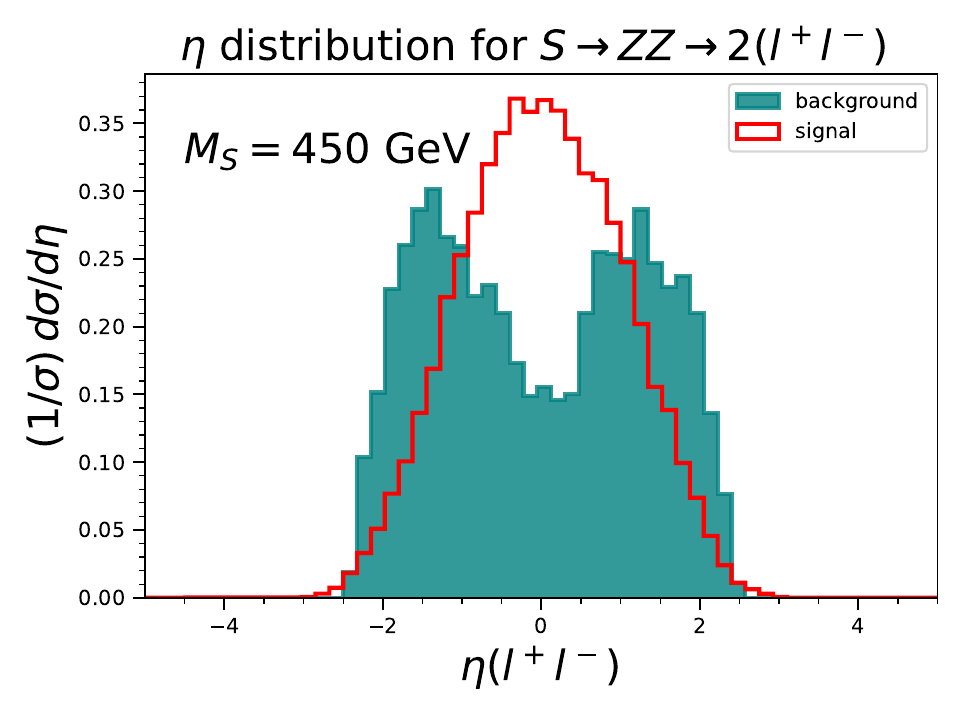}
\\
(A) \hspace{180pt} (B)\\
\includegraphics[scale=.45]{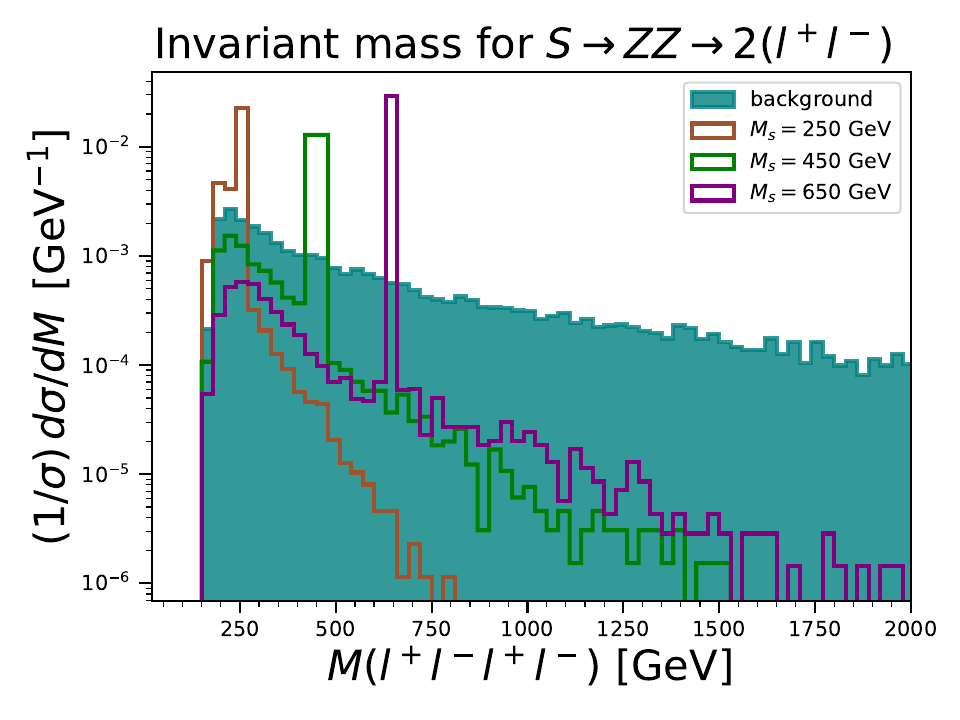}
\\
(C)
\caption{Transverse momentum  and pseudorapidity distributions  for signal events of $M_S=450$ GeV are shown in (A) and (B) respectively. The plot in (C) shows the invariant mass distribution of the four leptons ($M(4l)$) for different values of the $S$ masses. The backgrounds considered are the sum of all backgrounds resulting from $\mu^+ \mu^- \rightarrow \mu^+ \mu^- 2l^+2l^-$.}
\label{4l_constrains}
\end{figure}
A sharp constraint in the invariant mass of the four leptons, $\Delta M(l^+l^-l^+l^-)=50$ GeV, was adopted due to the good expected resolution in the reconstruction of the charged leptons. These combined invariant mass cuts led to significant suppression of background events, with $< 10 \%$ surviving, compared to signal events, where $\sim 50 \%$ survived, as can be seen in Fig.~\ref{4l_constrains} (C). The remaining events were further constrained using the transverse momentum and pseudorapidity of pairs of oppositely charged leptons, see Fig.~\ref{4l_constrains} (A,B).
The selected ranges for $P_T(l^+l^-), \eta(l^+l^-)$ are automated for each $M_s$ value where only regions that contain $N_s \geq 3.5 N_b$ for $P_T(l^+l^-)$ and $N_s \geq 4.5 N_b$ for $\eta(l^+l^-)$ were selected.\footnote{$N_s,N_b$ are the number of signal  and background events respectively.} These ranges change according to the $M_s$ value, as can be seen in Table~\ref{4lcuts}. The efficiencies obtained via this analysis were then used to estimate the expected exclusion cross section using,
\begin{align}
\mathcal{S} = \frac{S}{\sqrt{B+ (\alpha B)^2}}, \label{significance}
\end{align}
where $S= \varepsilon_s L \sigma_s$ is the signal number of events, and $B= \varepsilon_b L \sigma_b$ is the background number of events. The exclusion limit $\mathcal{S}$ is set to 2, corresponding to a 95\% confidence level (C.L.) exclusion.\\

\begin{table}[tp]
\centering
\begin{tabular}{|>{\centering\arraybackslash}p{30mm}|>{\centering\arraybackslash}p{15mm}|>{\centering\arraybackslash}p{15mm}|>{\centering\arraybackslash}p{15mm}|>{\centering\arraybackslash}p{15mm}|} 
\hline
 $M_S$ [GeV]   &  250  & 500  &  750  & 1000\\[3pt] \hline \rule{0pt}{1.5\normalbaselineskip}
$P_T(l^+l^-)$ [GeV] & 86 : 956  & 37 : 877  & 18 : 938 &  76 : 916  \\[7pt] 
$\abs{\eta(l^+l^-)}$  &  $< 1.15$ & $< 1.35$ & $< 1.75$ & $< 1.75$ \\[4pt] \hline
\end{tabular}
\caption{Samples of transverse momentum and pseudorapidity cuts for $4l$ channel, automated for different values of the scalar mass.}
\label{4lcuts}
\end{table}

\begin{figure}[htb!]
\centering
\includegraphics[scale=.45]{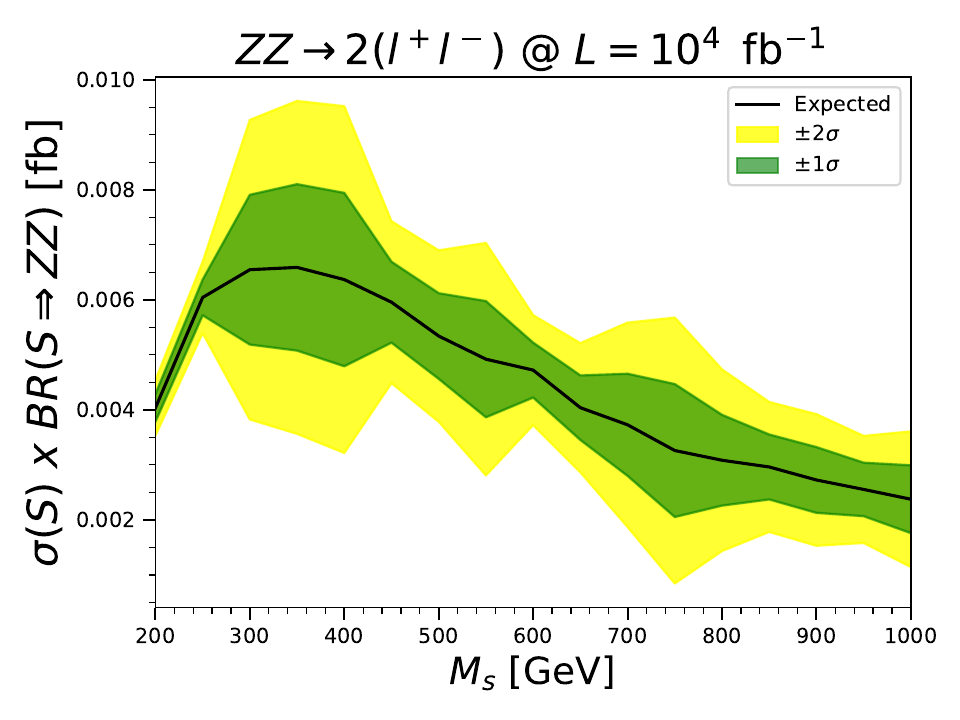}
\caption{Cross section exclusion curve for $\sigma(S) \times \mathrm{BR}(S\rightarrow ZZ)$ with $ZZ\rightarrow 4l$. The green and yellow intervals correspond to the $1\sigma$ and $2\sigma$ confidence levels.}
\label{4lsensitivity}
\end{figure}

\noindent
The four-lepton invariant mass cut, $M(l^+l^-l^+l^-)$, is the most crucial constraint in this analysis, as backgrounds lack resonant production, so we calculated the uncertainty in our calculations by repeating the analysis for different invariant mass ranges, $\Delta M(l^+l^-l^+l^-)=100$~GeV where both $\varepsilon_s, \, \varepsilon_b$ would deviate from the previous one leading to uncertainty in the signal cross section given by,
\begin{align}
U(\sigma_s) = \frac{\sigma_s(\varepsilon_s,\varepsilon_b)}{2}\sqrt{\left(\frac{\delta \varepsilon_s}{\varepsilon_s}\right)^2+ \left(\frac{\delta \varepsilon_b}{\varepsilon_b}\right)^2}\;. \label{uncerti}
\end{align}
Finally, the resulting sensitivity plot for a luminosity of $10 ^4 ~\mathrm{fb}^{-1}$ is shown in Fig.~\ref{4lsensitivity}, at $68\%$ and $95\%$ C.L.

\subsection{$4q$ Final States}

The $4q$ final state can result from three different pathways: $S\longrightarrow ZZ,\,W^+W^-,\,hh$, and could also provide strong constraints.
\begin{figure}[htb!]
\centering
\includegraphics[scale=.4]{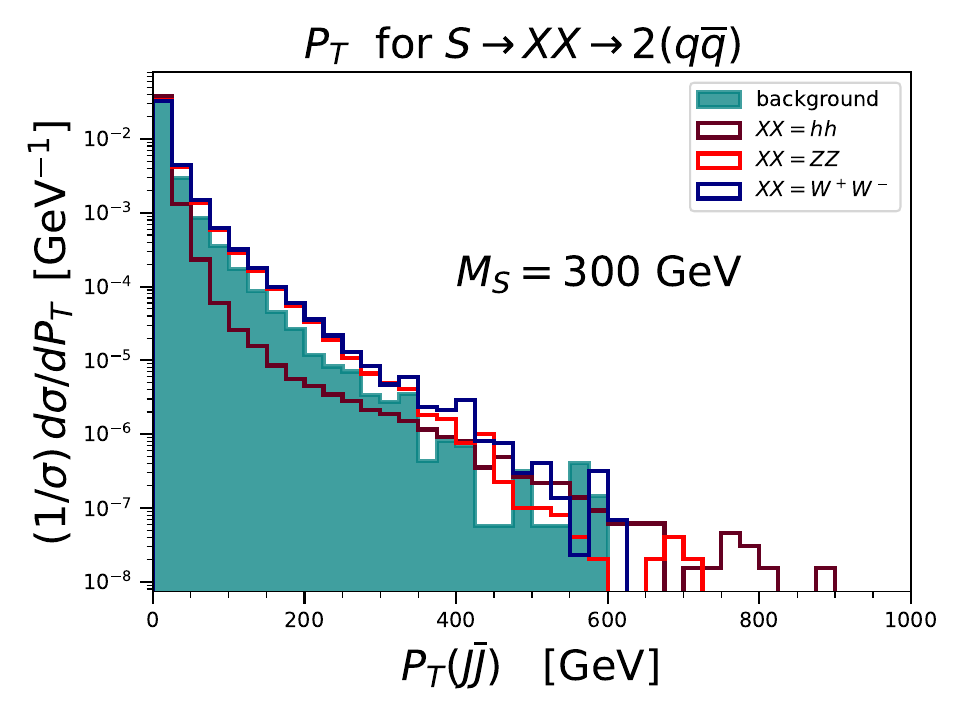 }
\includegraphics[scale=.4]{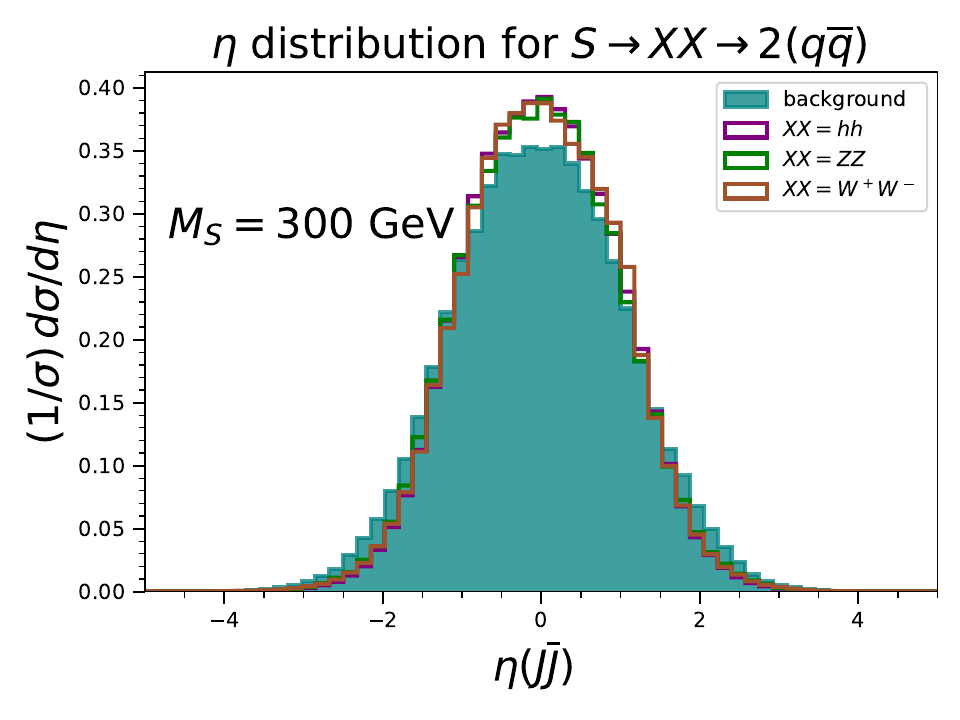 }\\
 (A) \hspace{200pt}  (B)\\
\includegraphics[scale=.46]{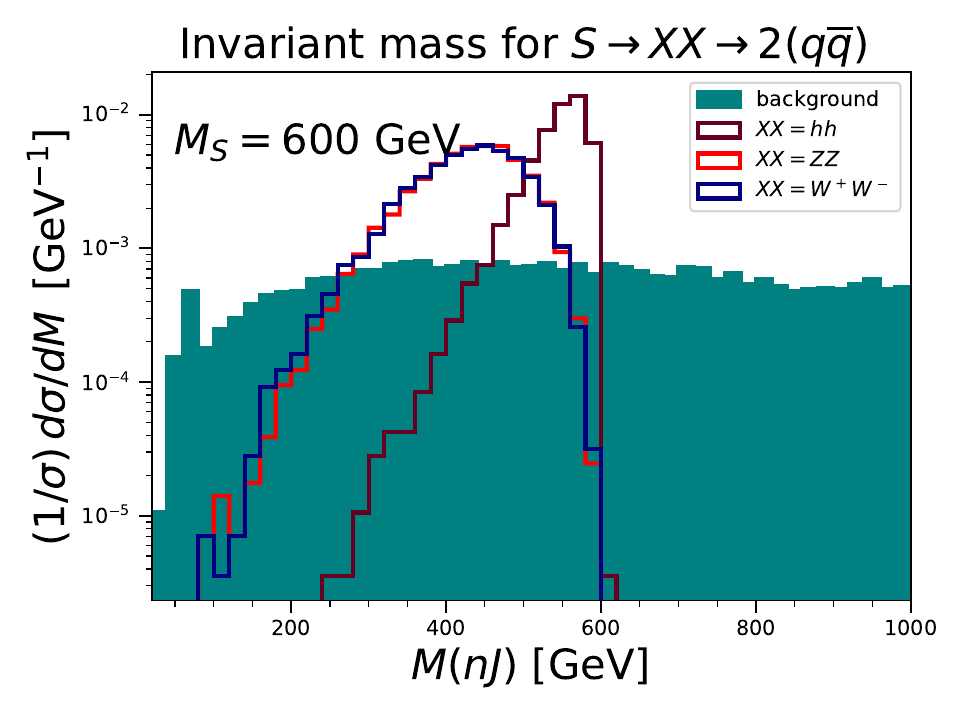}
 \\
 (C)
\caption{Plots in (A) and (B) show the transverse momentum and pseudorapidity distributions for the paired jets in final states for the signal at $M_S= 300$ GeV coming from $S\rightarrow hh,\, ZZ,\, W^+W^-$, as well as for the background. Subplot (C) shows the invariant mass $M(nJ)$ for jets originated from $h,\,Z,\,W^\pm$ decays respectively at $M_S=600$ GeV.}
\label{4j_kinematics}
\end{figure}
The main background is similar to those discussed in the previous channel, with $l\rightarrow q$. In principle, the jets originating from  ($ZZ,\,\,W^+W^-,\,\, hh$) decays  will have almost identical kinematics due to the approximate symmetry in the $Z,W^\pm,h$ masses especially for $M_S\gg M_Z,M_{W^\pm},M_h$. 
Nevertheless, the differences in the branching ratios of Higgs boson and weak gauge boson decays into quarks will lead to a deviation in the kinematical behavior of the two cases, especially for the transverse momentum and invariant mass distributions, as can be seen in Fig.~\ref{4j_kinematics} (A,C). These differences can be further investigated by tracking the jet's origin, and this can play a vital role in exploring the $\mathbb{Z}_2$ nature of xSM. Hence, this specific channel, which has not been explored sufficiently to date, to the best of our knowledge, can further constrain the parameter space of FOEWPT. This proves to be an additional advantage of a muon collider, in which such a channel could be explored with much less effort than at hadron colliders due to the lower QCD backgrounds. A detailed study of these differences is left for future work. Jets were clustered using \texttt{FastJet} (v3.3.2)~\cite{Cacciari:2011ma}, where the anti-kT algorithm~\cite{Catani:1993hr, Ellis:1993tq, Cacciari:2008gp} with a radius parameter $R = 0.4$ was chosen to be the default jet clustering algorithm. The calculation of the invariant mass for this channel is more challenging than in the previous one, mainly due to the high tendency of quarks to radiate via QCD. This means that we cannot simply calculate the invariant mass as previously using $\displaystyle{M^2(4j) = \big( \sum_{i=1}^4 p(J_i)\big)^2}$.  Instead, for each event, we defined  $\displaystyle{P = \sum_{i=1}^n p(J_i)}$, that sums all the jets to a single four vector, from which we construct $M^2(nj)= P^2$. This, still, will not add up to the parent $S$ particle mass, as some of the energy will have escaped in the form of radiation, but this could be neglected as a similar effect will also be found in background events. The main effect of this final-state radiation is a shift in $M(nj)$ towards lower values, as evident in Fig.~\ref{4j_kinematics} (C), which was taken into account in the analysis by setting a loose $M(nj)$ interval cut. An event was accepted if its jets invariant mass was in the range $0.55\,M_S  \leq M(nj) \leq 1.05\, M_S$ for the $ZZ,WW$ channels and $0.8\,M_S  \leq M(nj) \leq 1.05\, M_S$ for the $hh$ channel. Based on Fig.~\ref{4j_kinematics} (C), the $hh$ channel accepted range was adopted to be narrower compared to that of the $ZZ,WW$ channels. Most likely, this is because jets coming from a Higgs boson's decay are produced more collinearly, and hence lose less energy in the form of radiation. This fact is supported by examining in Fig.~\ref{4j_kinematics} (A), that shows the tendency of jet pairs coming from the Higgs boson's decay to accumulate towards lower $P_T$ values in comparison to those coming from $WW,ZZ$. 
\begin{table}[tp]
    \centering
    \begin{tabular}{|c|c|c|c|c|c|}
    \hline
     \multicolumn{1}{|c}{ $M_S \,\,$[GeV]} &   & 300 & 500 & 750 & 1000\\\hline \rule{0pt}{1.25\normalbaselineskip}
       \multirow{2}{*}{$ZZ\rightarrow 2(q\overline{q}) $} & $P_T$ [GeV] & $6:376$ & $5:445$ & $5:265 $ &  $5 :205$ \\[5pt] 
       & $\abs{\eta}$ & $<1.75$ & $< 1.75$ & $< 1.25$ & $< 1.05$ \\ \hline \rule{0pt}{1.25\normalbaselineskip}
       \multirow{2}{*}{$W^+W^-\rightarrow 2(q\overline{q}) $} & $P_T$ [GeV] & $5 :375 $ & $5 :305$ & $5 :175$ &  $5 :185$ \\ [5pt]
       & $\abs{\eta}$ & $<2.25$ & $< 1.65$ & $<1.35$ & $< 1.05$ \\ \hline \rule{0pt}{1.25\normalbaselineskip}
       \multirow{2}{*}{$hh\rightarrow 2(q\overline{q}) $} & $P_T $ [GeV] & $-5:55$ & $5 : 25$ & $5 : 25$ &  $5 : 25$ \\[5pt] 
       & $\abs{\eta}$ & $-< 3.95$ & $< 2.15$ & $< 1.15$ & $< 1.065$ \\ \hline
    \end{tabular}
    \caption{Samples of transverse momentum and pseudorapidity cuts for the different $4q$ channels, automated at each value of the scalar mass.}
    \label{4qcuts}
\end{table}
Since jets in both cases (signal and backgrounds) arise either due to weak gauge bosons or the Higgs boson, which have approximately similar masses, then applying a di-jet invariant mass cut around the mediator's mass will not help much in suppressing the background, except for the case of jets coming from $W^+W^-$, which yielded better efficiency for signal separation when applying $0.4M_W\leq M(J\overline{J}) \leq  1.5 M_W$, with $P_T(J,\overline{J}) \geq \frac{1}{5} M_W$. The events passing the $M(nJ)$ cut were then used to build all the possible jet pairs, and the event is accepted at the end if it contains at least two jet pairs. 

The surviving events were then further filtered using transverse momentum and pseudorapidity constraints via direct comparison of signal events to the background events, as illustrated in Fig.~\ref{4j_kinematics} (A,B). Bins of transverse momentum, $P_T$, were selected such that $N_s \geq 1.02 N_b$ for the $ZZ,hh$ channels and $N_s \geq 1.4 N_b$ for the $WW$ channel. The $\eta$ bins were selected by requiring $N_s \geq 1.4 N_b$ for both $ZZ,hh$ channels, and $N_s \geq 1.5 N_b$ for the $WW$ channel. 

\begin{figure}[htb!]
\centering
\includegraphics[scale=.4]{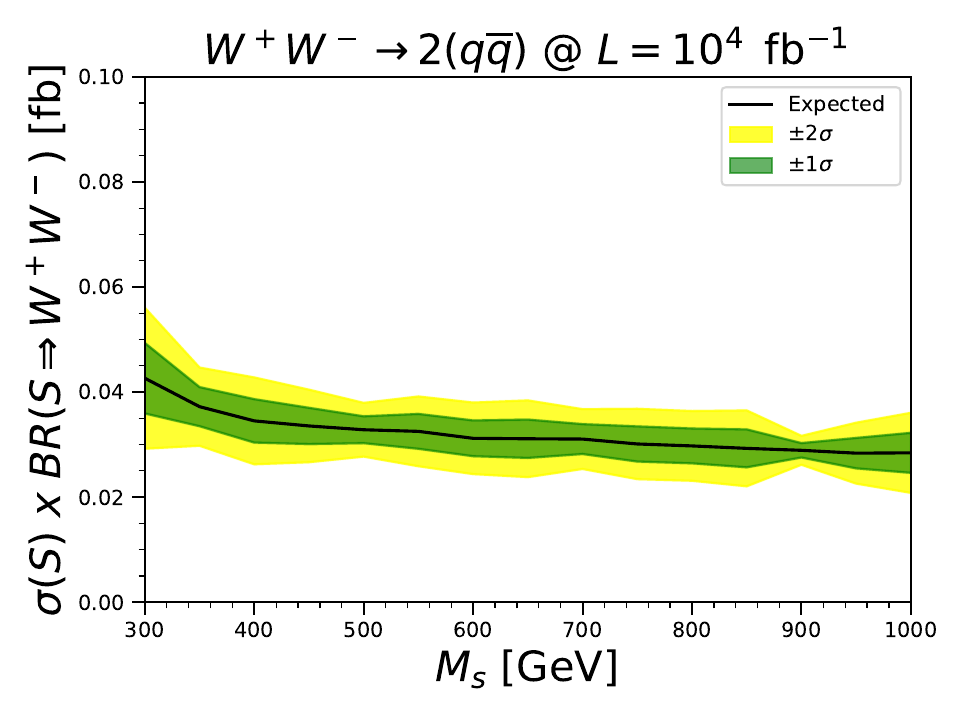}
\includegraphics[scale=.4]{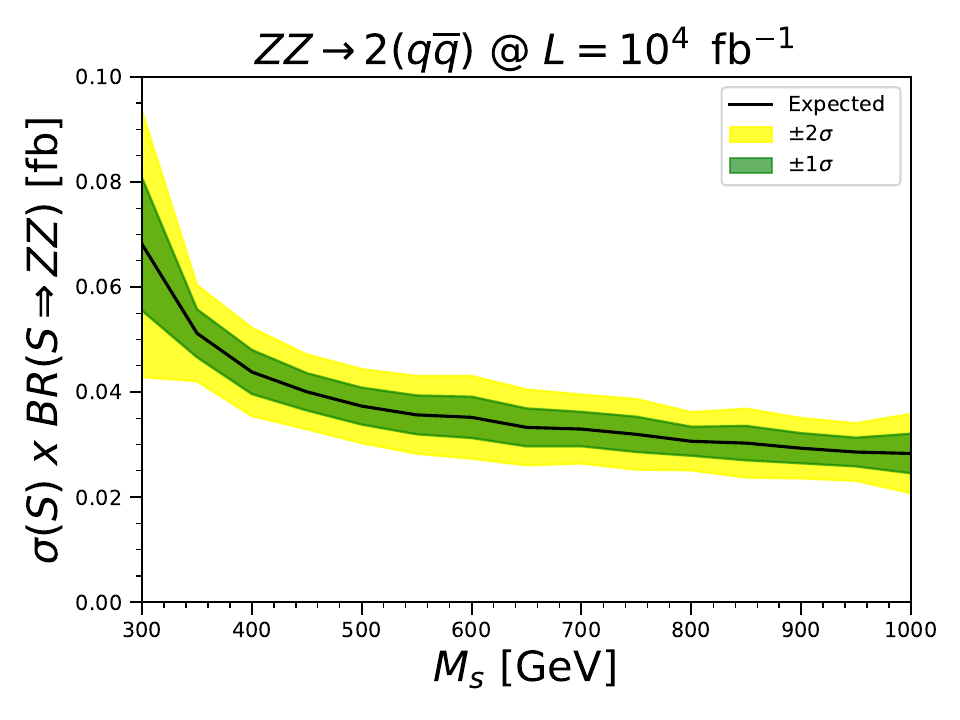}\\
(A)\hspace{180pt} (B)\\
\includegraphics[scale=.4]{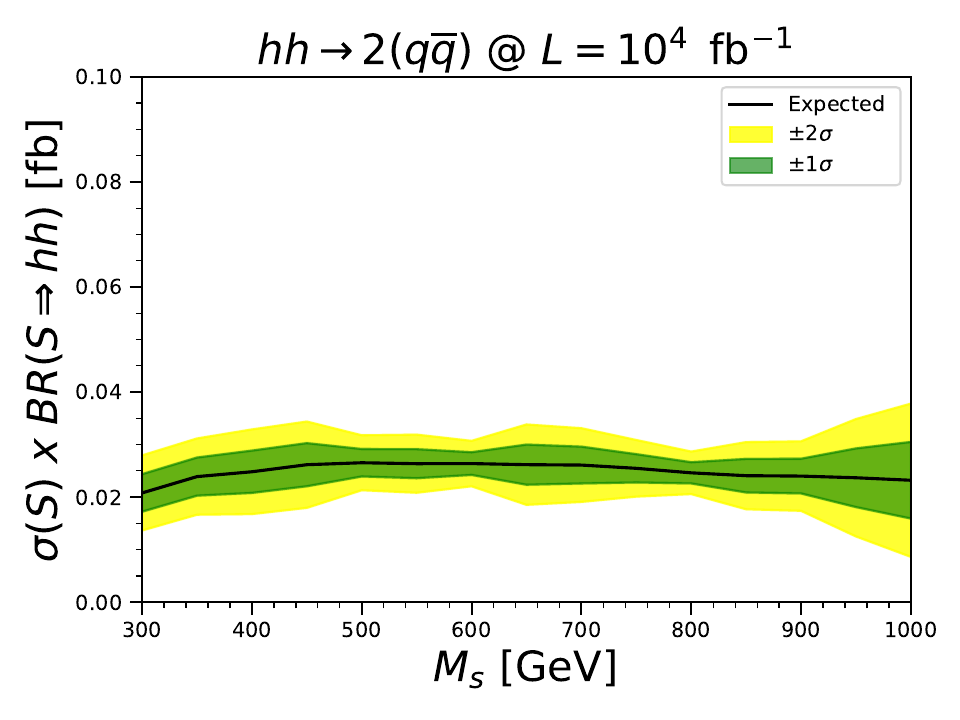}
\\
(C)
\caption{The sensitivity plots for (A) $S\longrightarrow (WW,\,\,ZZ,\,\, hh)\rightarrow 4J$,  (B) $S\longrightarrow ZZ\rightarrow 4J$, and (C) $S\longrightarrow W^+W^-\rightarrow 4J$, shown with the $68\%$ and $95\%$ C.L. intervals. }
\label{4j_sensitivity}
\end{figure}

These combined cuts led to an overall signal efficiency above $35\%$ on average for all signal channels and below $10\%$ on average for the backgrounds. The cuts were automated based on the $M_s$ value, as shown in Table~\ref{4qcuts}. Furthermore, the uncertainties in the signal and background efficiencies ($\varepsilon_s,\varepsilon_b$) were computed using Eq.~\ref{uncerti}, by repeating the previous analysis for different $M(nJ)$ ranges, as this represents the most critical cut in this analysis. For the second analysis, we chose a wider range for the invariant mass, $0.45M_s \leq M(nJ) \leq 1.1 M_s$ for the $ZZ,WW$ channels and $0.75M_s \leq M(nJ) \leq 1.1 M_s$ for the $hh$ channel, which led to the uncertainty in the exclusion curves shown in Fig.~\ref{4j_sensitivity} for a luminosity of $10^4$~fb$^{-1}$.

\subsection{$2l2q$ Final States}
The $2l2q$ channel is also exclusively produced by $S\rightarrow ZZ$. All possible backgrounds were generated in \texttt{MG5\_aMC} via $\mu^+ \mu^- \rightarrow \mu^+ \mu^- l^+l^- q\overline{q}$. As in the previous channels, the most crucial distinction between signal and background is the invariant mass, which separates the resonant production from the background. 

\begin{figure}[htb!]
\centering
\includegraphics[scale=.4]{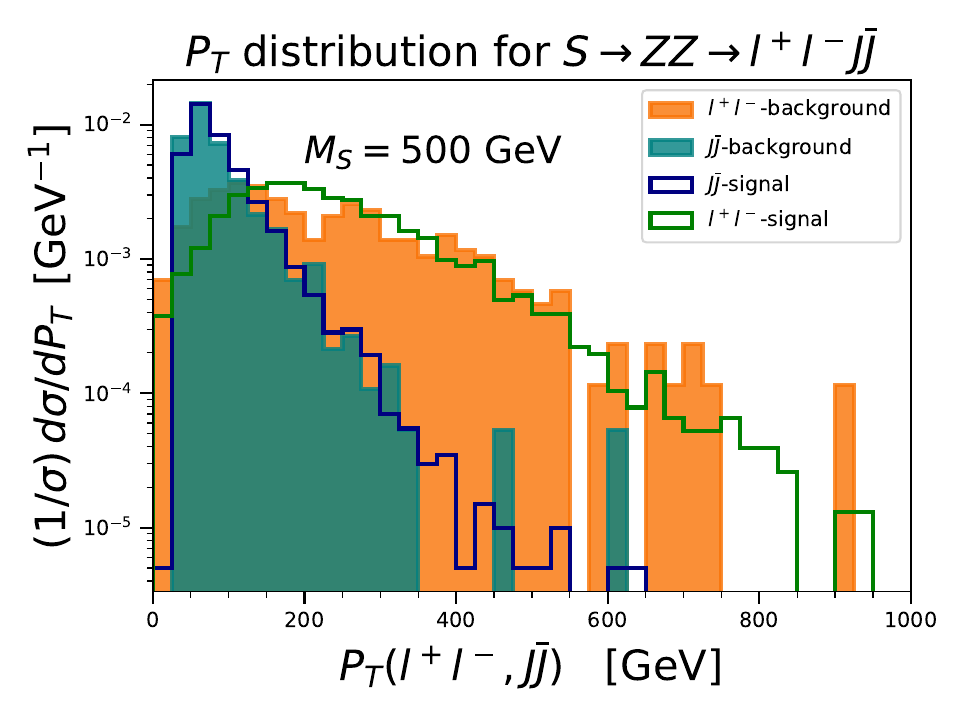}
\includegraphics[scale=.4]{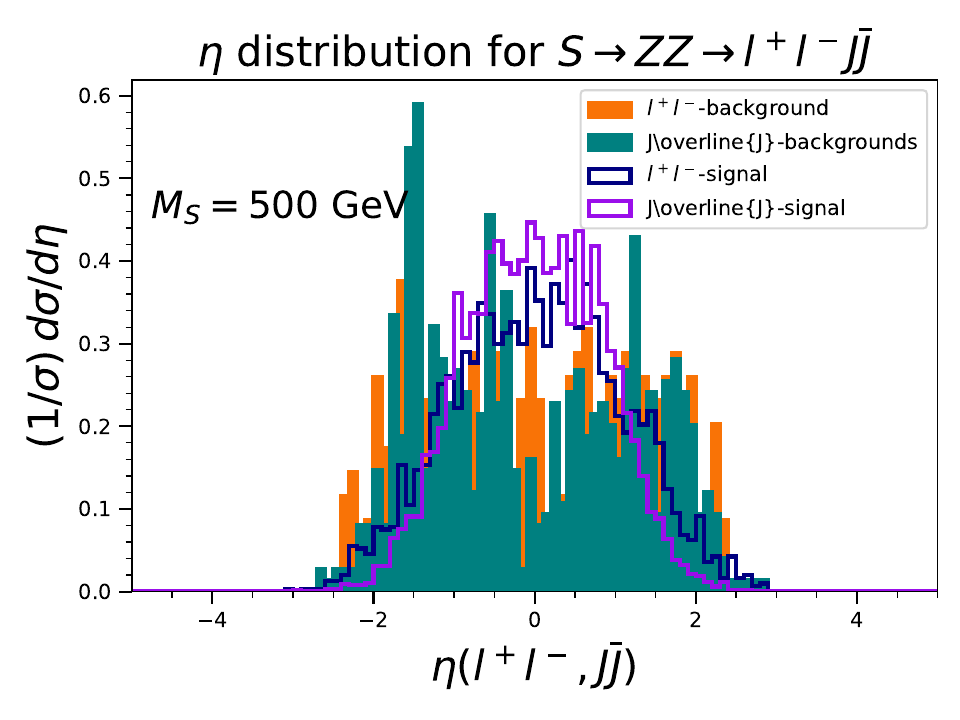}\\
(A)\hspace{180pt} (B)\\
\includegraphics[scale=.46]{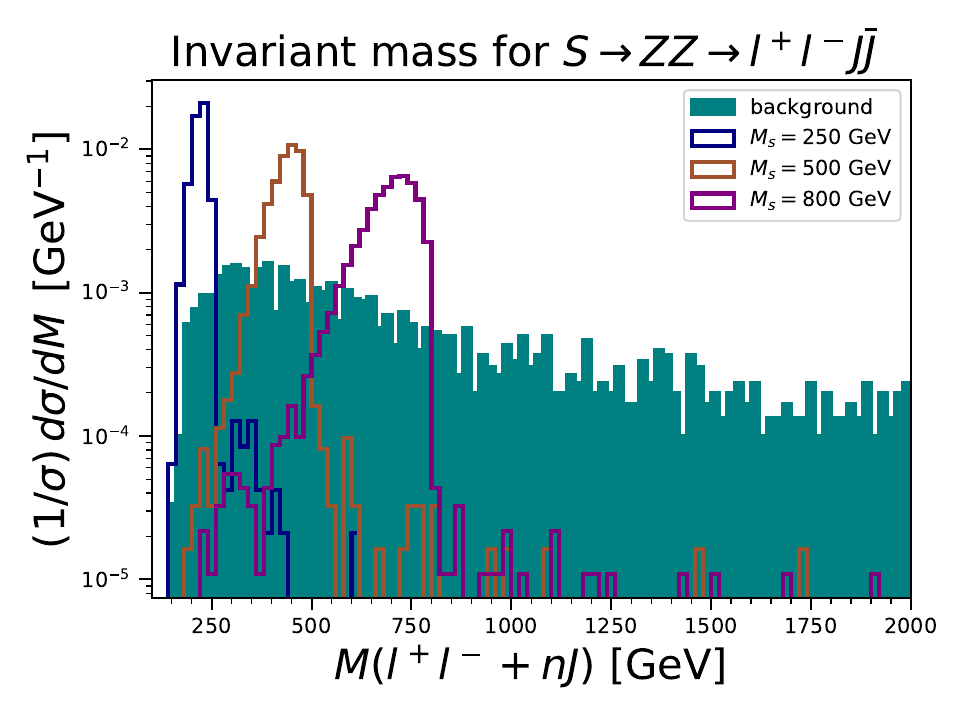}\\
(C)
\caption{Plots in (A) and (B) show the transverse momentum and pseudorapidity comparisons between signal and background for $M_S = 500$~GeV. (C) shows the invariant mass $M(l^+l^-+nJ)$ obtained from Eq.~\eqref{2l2qmij} for $M_S = 250, 500,800$~GeV. }
\label{2l2q_mij}
\end{figure}

On the leptonic side, we required $P_T(l^\pm)\geq \frac{1}{5} M_Z$, based on a direct comparison of the transverse momentum of single leptons, to guarantee that we did not lose any possible leptons resulting from $Z$ decays. Then, we paired the same-flavour oppositely charged leptons if their invariant mass satisfied $0.75 M_Z \leq M(l^+l^-)\leq 1.25 M_Z$. Similarly, on the jet side, we asked for $0.4M_Z \leq M(nJ)\leq 1.5M_Z$, and then paired jets according to their transverse momentum $P_T(J)\geq \frac{1}{5} M_Z$, which is also confirmed from single $P_T(J)$ comparison between signal and background. We then look for  resonant production around the scalar mass by computing the invariant mass of all possible combinations of the accepted ($l^+l^-$) pairs, together with all the accompanying jets ($nJ$) in each event using the formula,
\begin{align}
M^2(l^+l^-+nJ) &= \left(\sum_{i=1}^3 \left(p(l^+_i)+p(l^-_i) \right)+\sum_{k=1}^n  p(J_k)\right)^2\\
&= \left(P(J)+\sum_{i=1}^3 p(l^+_i)+p(l^-_i) \right)^2 \label{2l2qmij}\;,
\end{align}
such that $P(J) = \sum_k p(J_k)$ is the previously defined four vectors that add up all the jets present in a single event. 
This approach returned the correct invariant mass, which shifted towards lower mass values because of the unavoidable final-state radiation, especially toward higher $M_s$ values, as depicted in Fig.~\ref{2l2q_mij} (C). The same figure also suggests a selection range of $M_s -150 \,~\mathrm{GeV} \leq M(l^+l^-+nJ) \leq M_s+10 \,~\mathrm{GeV}$. The event is then selected if it contains at least one combination that passes the $ M(l^+l^-+nJ) $ cut and has at least one $l^+l^-$ pair and one ($J\overline{J}$) pair passing the previous $M_Z(l^+l^-,nJ),P_T(l^+l^-,J\overline{J})$ constraints. Following the invariant mass cuts, the remaining events were subjected to cuts on $P_T(l^+l^-,J\overline{J}), \,\eta(l^+l^-,J\overline{J})$, through direct comparison between the signal and background for each specific $M_s$ value, as shown in Fig.~\ref{2l2q_mij} (A,B) for $M_s = 500$~GeV. The $P_T$ bins for this channel were selected by requiring $N_s \geq 2 N_b$, and $N_s \geq 1.8 N_b$ for the pseudorapidity ($\eta$) bins.
\begin{table}[tp]
    \centering
    \begin{tabular}{|c|c|c|c|c|c|}
    \hline
     \multicolumn{1}{|c}{ $M_S \,\,$[GeV]} &   & 250 & 500 & 750 & 1000\\\hline \rule{0pt}{1.25\normalbaselineskip}
       \multirow{2}{*}{Leptons} & $P_T$ [GeV]& $ 6 : 346 $ & $ 19:689 $ & $46:686$ &  $67:657$ \\ [5pt]
       & $\abs{\eta}$ & $  < 2.25$ & $ < 1.65$ & $  < 2.35$ & $ < 2.35$ \\ \hline \rule{0pt}{1.25\normalbaselineskip}
       \multirow{2}{*}{Jets} & $P_T$ [GeV]& $ 28:568 $ & $37:607$ & $34:584$ &  $41:591$ \\ [5pt]
       & $\abs{\eta}$ & $< 2.15$ & $ < 1.35$ & $< 1.75$ & $< 1.85$ \\ \hline
    \end{tabular}
    \caption{Samples of transverse momentum and pseudorapidity cuts for the $2l2q$ channel, automated for each value of the scalar mass.}
    \label{2l2qcuts}
\end{table}

\noindent
These bins fluctuate from $M_s$ value to another, as shown in Table~\ref{2l2qcuts}. These cuts led to significant suppression of the background, where $\varepsilon_b \sim 3\%$ on average and $\varepsilon_s \sim 35\%$ for the signal.  

\begin{figure}[htb!]
\centering
\includegraphics[scale=.45]{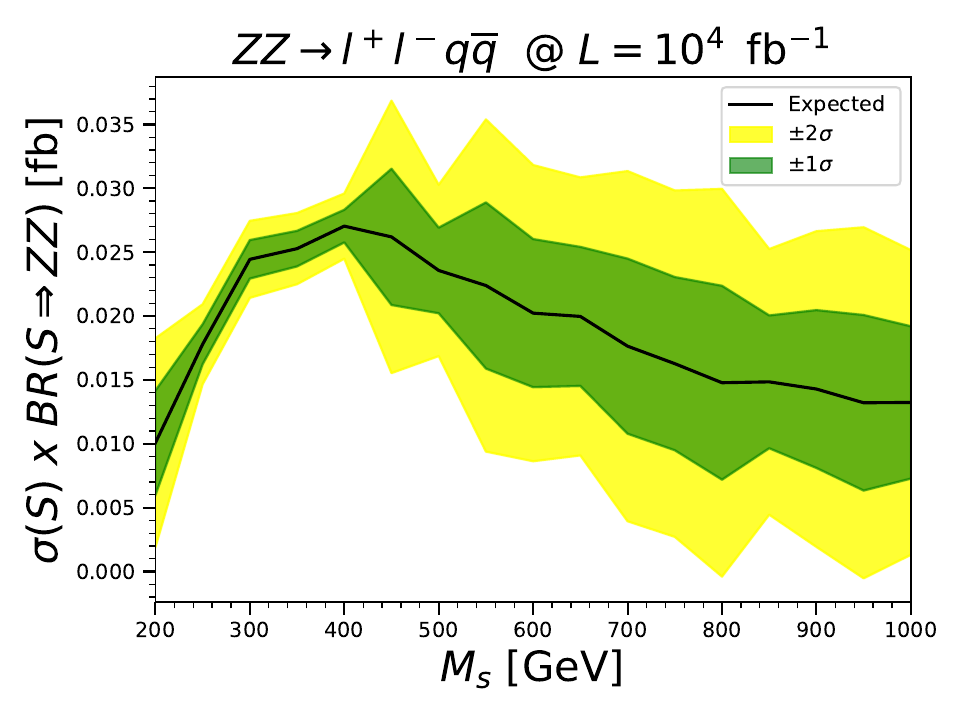}
\caption{ Sensitivity curves at $L=10^4$~fb$^{-1}$ and $\sqrt{s}=3$ TeV for $S\longrightarrow l^+l^- 2J$ channel at a muon collider with $68\%$ and $95\%$ CL intervals. }
\label{2l2q_sensitivity}
\end{figure}

The fluctuations in $P_T,\eta$ were then considered according to Eq.~\ref{uncerti}, by repeating the previous analysis for different invariant mass ranges $M_s -250$~GeV~$\leq M(l^+l^-+nJ) \leq M_s+20$~GeV which led to the sensitivity curve for this channel, plotted in Fig.~\ref{2l2q_sensitivity} at $L=10^4$~fb$^{-1}$.

\subsection{$2\gamma2q$ Final States}

This final state arises from the $S\rightarrow h_1 h_1$ decay channel in addition to all the different, nonresonant background channels that lead to the same final state at the muon collider ($\mu^+\mu^- \longrightarrow \mu^+\mu^-  2\gamma2J $), which was generated in \texttt{MG5\_aMC}. This channel is very challenging due to the difficulty in reconstructing both photons and jets in the final state, since a considerable number of photons would be present from the muon and quark QED radiation.

\begin{figure}[htb!]
\centering
\includegraphics[scale=.4]{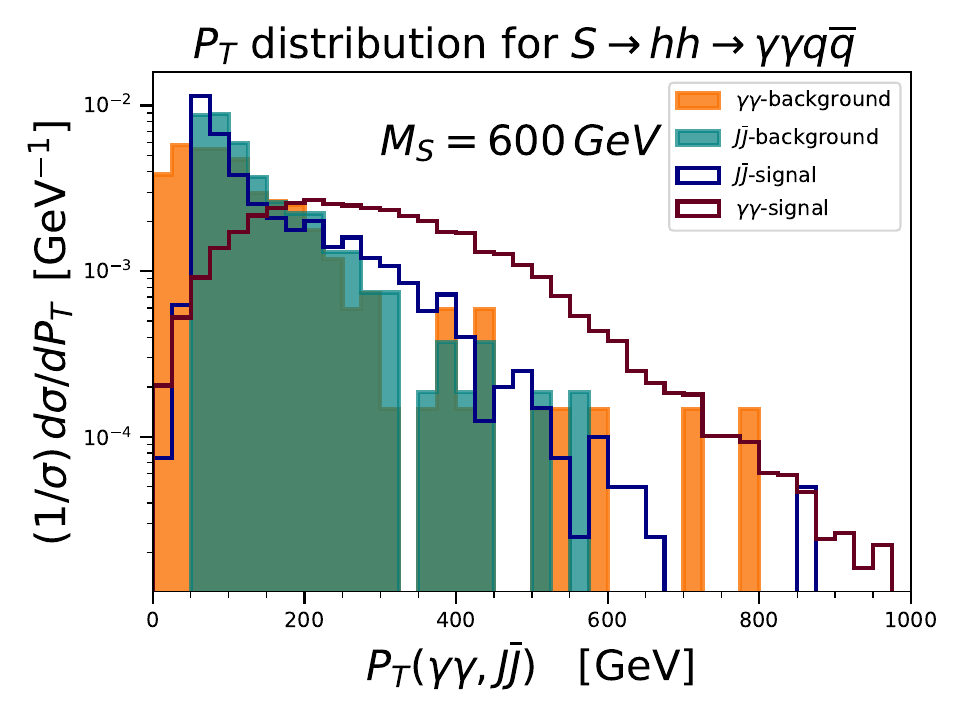}
\includegraphics[scale=.4]{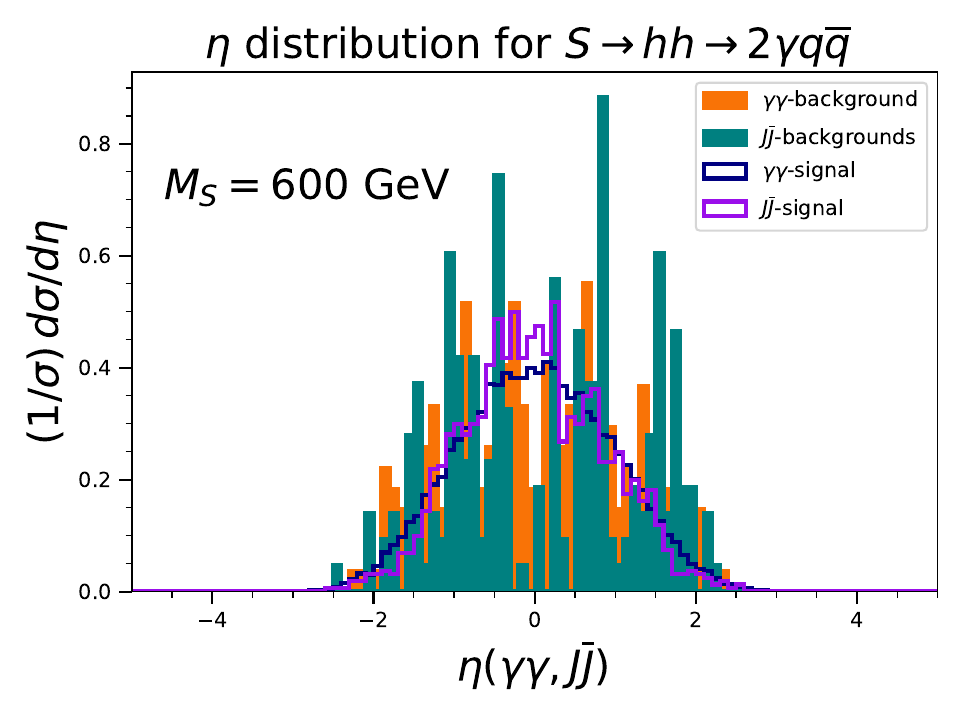}\\
(A)\hspace{180pt} (B)\\
\includegraphics[scale=.46]{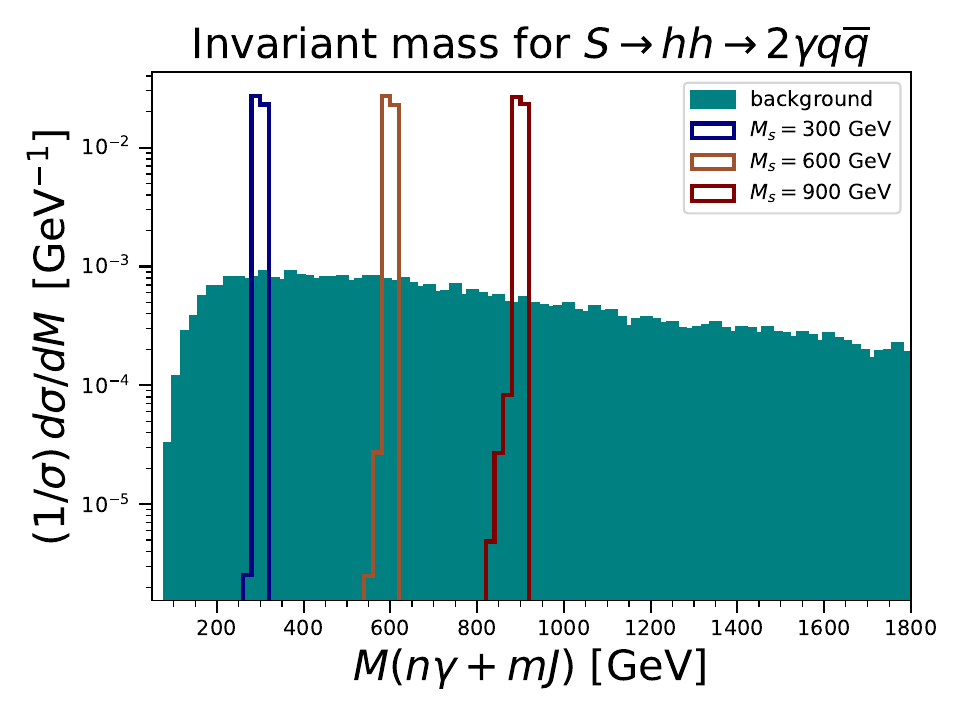}\\
(C)
\caption{Subplots (A) and (B) show the transverse momentum and pseudorapidity comparisons between signal and background for $M_S = 600$ GeV, while (C) shows the invariant mass obtained from Eq.~\eqref{2a2qmij} for different $M_s$ values.}
\label{2a2q_observables}
\end{figure}

\begin{figure}[htb!]
\centering
\includegraphics[scale=.45]{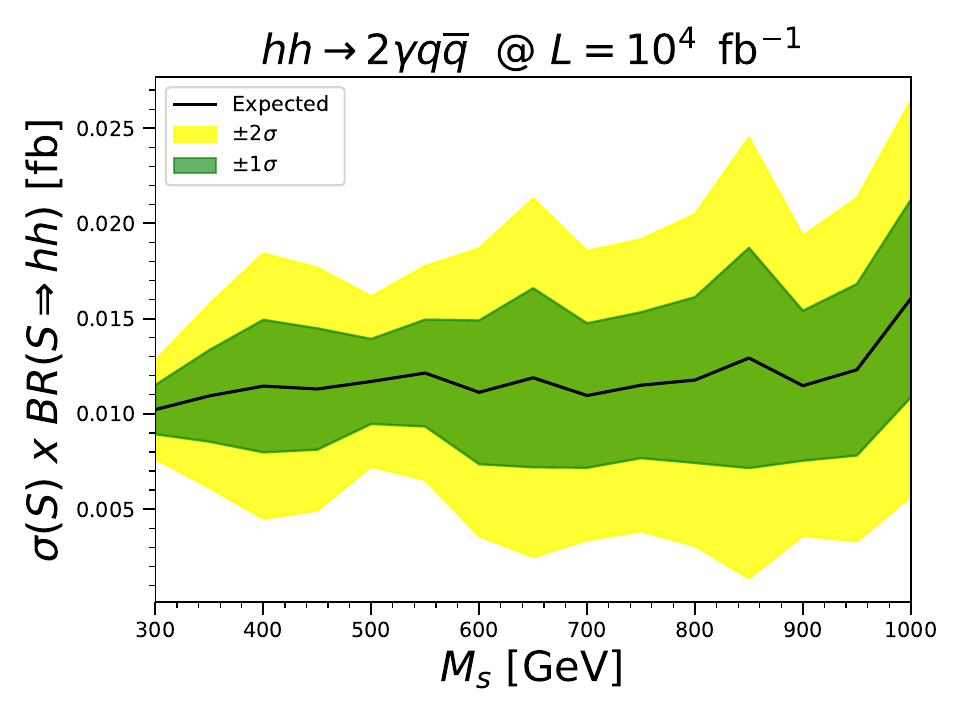}
\caption{The sensitivity curve at $L=10^4$~fb$^{-1}$ and $\sqrt{s}=3$ TeV, for the $S\rightarrow 2\gamma 2J$ channel at a muon collider with $68\%$ and $95\%$ CLs interval.}
\label{2a2q_sensitivity}
\end{figure}

 As before, the main difference between the signal and background is the absence of resonant production in the background, which suggests invariant mass cuts on the final state as an effective probe to identify the signal. Photons resulting from the Higgs boson decay could be separated from photons resulting from other decay modes, or initial/final-state radiation, by requiring $P_T(\gamma)\geq \frac{2}{5} M_h$, which was confirmed through direct comparison of the transverse momentum distributions. Then, to pair photons, we computed the invariant mass of all the possible combinations of the photons that passed the $P_T$ cut, and only kept the pairs that had $0.8 M_h\leq M(\gamma\gamma)\leq 1.2 M_h$. We then computed the invariant mass of the selected photon pairs and all jets contained in the event, $M(\gamma\gamma+nJ)$, given by,\footnote{This is a modification of Eq.~\eqref{2l2qmij}, where we take into account all the possible combinations of the accepted $\gamma\gamma$ pairs with the entire $n$-jets available in the event.} 
\begin{align}
M^2(\gamma\gamma + nJ) &= \left(p(\gamma\gamma)+\sum_{k=1}^n  p(J_k)\right)^2,\\
&=  P^2(\gamma \gamma+nJ) \label{2a2qmij},
\end{align}
where $p(\gamma\gamma) = p(\gamma_1) +p(\gamma_2)$, represents the summed vector of the photon pairs, which returns very precise peaks around the parent $M_s$ value, as shown in Fig.~\ref{2a2q_sensitivity} (C).  
\begin{table}[tp]
    \centering
    \begin{tabular}{|c|c|c|c|c|c|}
    \hline
     \multicolumn{1}{|c}{ $M_S \,\,$[GeV]} &   & 300 & 500 & 750 & 1000\\\hline \rule{0pt}{1.25\normalbaselineskip}
       \multirow{2}{*}{Photons} & $P_T$ [GeV]& $5:355$ & $6:476$ & $7:457$ &  $6:446$ \\ [5pt]
       & $\abs{\eta}$ & $ < 2.25$ & $  <2.15 $ & $  < 2.25$ & $  < 2.15$ \\ \hline \rule{0pt}{1.25\normalbaselineskip}
       \multirow{2}{*}{Jets} & $P_T$ [GeV]& $7:747$ & $14:684$ & $7:877$ &  $26:826$ \\ [5pt]
       & $\abs{\eta}$ & $< 1.45$ & $ < 1.35 $ & $< 2.35$ & $< 1.55$ \\ \hline
    \end{tabular}
    \caption{Samples of transverse momentum and pseudorapidity cuts for the $2l2q$ channel, automated for each value of the scalar mass.}
    \label{2a2qcuts}
\end{table}
The event was accepted if it had $M_s-80\,~\mathrm{GeV}\leq M(\gamma\gamma+nJ)\leq M_s+80\,~\mathrm{GeV}$, and contained at least one accepted $\gamma \gamma$ pair. Subsequently, jet pairs were constructed for all possible combinations of the remaining jets after the $P_T(J) \geq \frac{2}{5}$ constraint was applied. 

The accepted events were further subjected to transverse momentum and pseudorapidity constraints, obtained through a direct comparison between the remaining signal and background events, as shown in Fig.~\ref{2a2q_observables} (A,B), for $M_s = 600$ GeV. We automated these cuts to make them sensitive to the $M_s$ value, as shown in Table~\ref{2a2qcuts}. The $P_T$ bins were obtained by requiring $N_s \geq 1.1 N_b$, while $N_s \geq 1.01 N_b$ was adopted for pseudorapidity ($\eta$).  
These observables are, in turn, dependent on the invariant mass, $M(\gamma\gamma+nJ)$, cut applied, so we repeated the above analysis for a different range, $M_s-150\, \mathrm{GeV}\leq M(\gamma\gamma+nJ)\leq M_s+100\, \mathrm{GeV}$, and computed the resulting uncertainty in the signal and background efficiencies ($\varepsilon_s,\varepsilon_b$) using Eq.~\eqref{uncerti}. 
This set of cuts significantly suppressed the background, where the background efficiency was $\varepsilon_b \sim 1\%$, compared to the signal efficiency of $\varepsilon_s \sim 50\%$ on average. These results were reflected in the characteristic sensitivity curve for the $2\gamma2q$-channel as plotted in Fig.~\ref{2a2q_sensitivity} at a luminosity of $10^4$~fb$^{-1}$.

\subsection{$2l2\nu_l$ Final states}

The $2l2\nu_l$ final state can originate either from the $S$ decay into weak gauge bosons $S\rightarrow ZZ,\,W^+W^-$, or other backgrounds that can be generated at a muon collider through $\mu^+\mu^- \rightarrow \mu^+\mu^-  (\nu_\mu\overline{\nu}_\mu) 2l2\nu_l$, and all were considered in our analysis. The invariant mass of such final states is not immediately calculable, because of the missing energy carried away by neutrinos. However, unlike in hadron colliders, the 
 energy of the muon beam used in scalar production can be precisely estimated, giving us an opportunity to reconstruct the parent particle's mass, emphasizing one of the advantages of muon colliders over hadron colliders.

\begin{figure}[htb!]
\centering
\includegraphics[scale=.4]{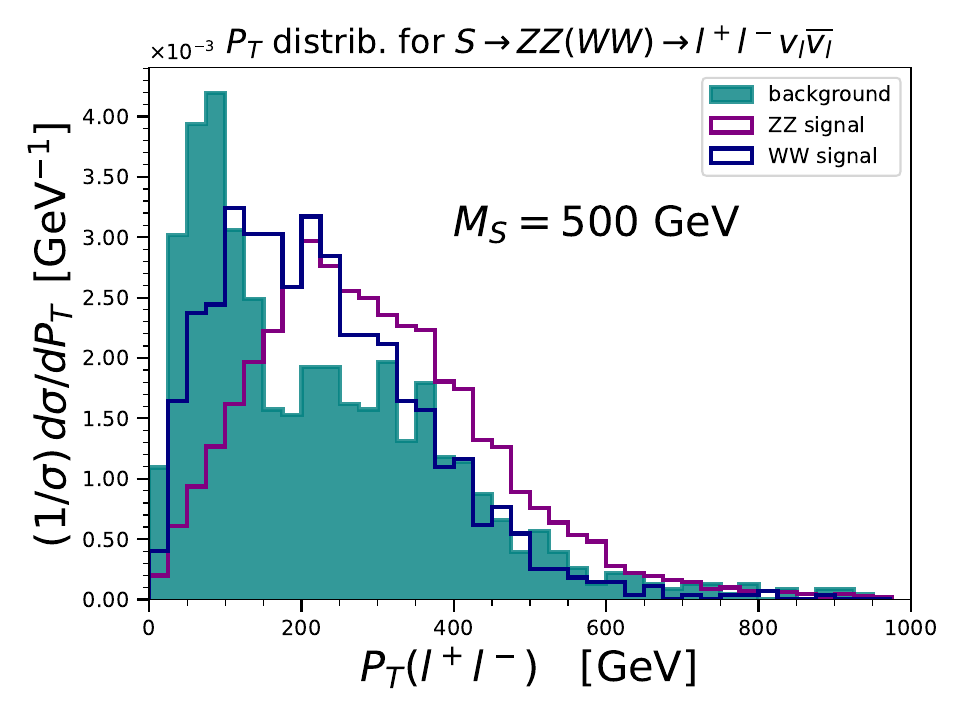}
\includegraphics[scale=.4]{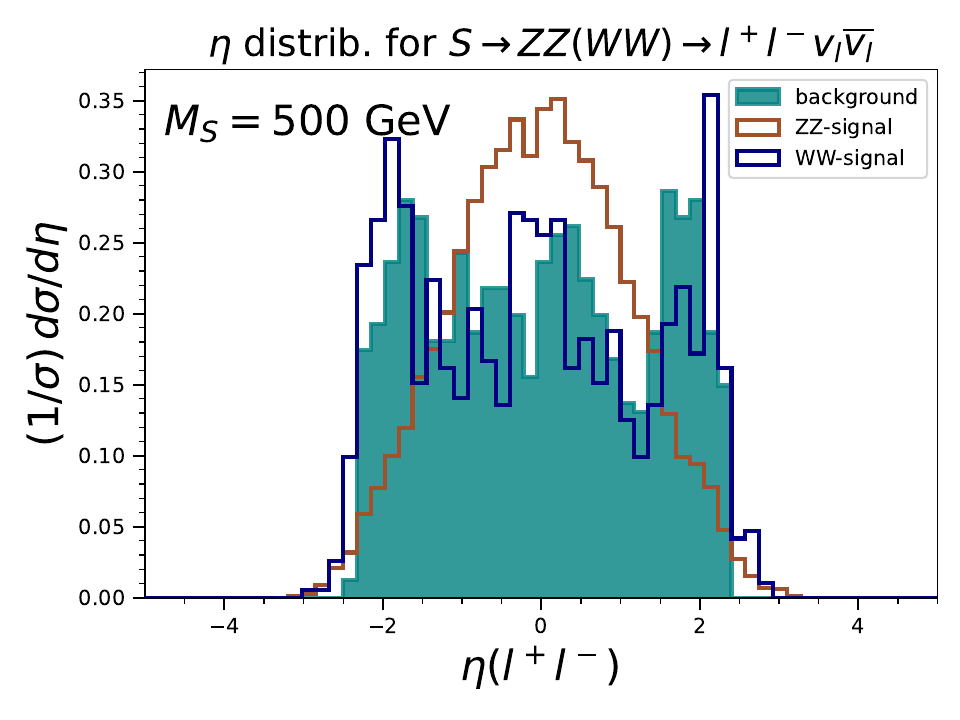}\\
(A)\hspace{180pt} (B)\\
\includegraphics[scale=.46]{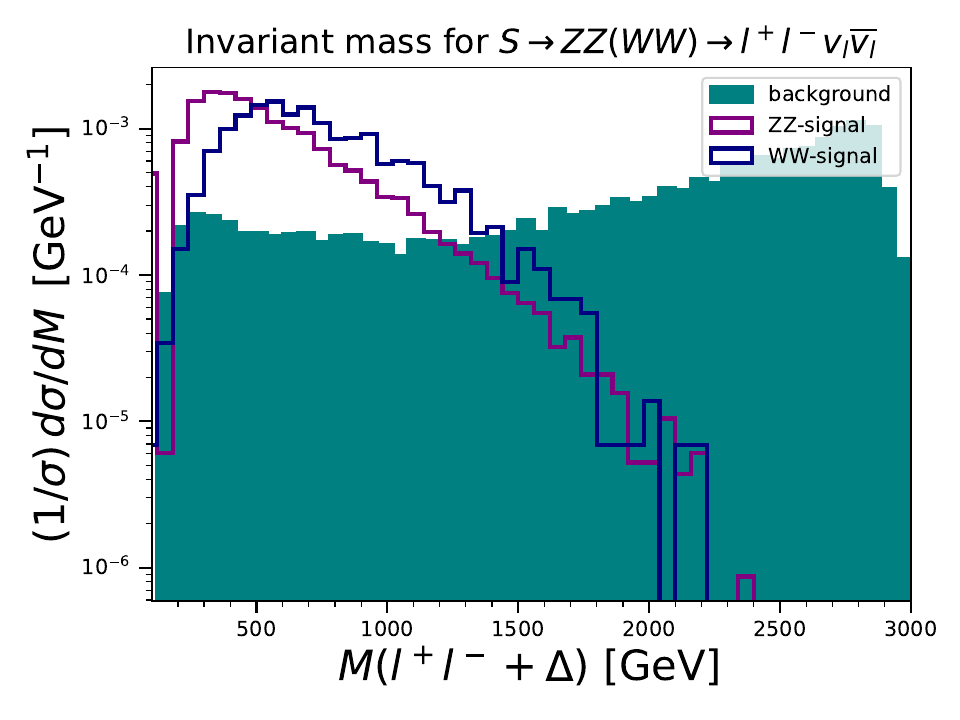}\\
(C)
\caption{Subplots (A), (B) show the transverse momentum and pseudorapidity comparisons between signal and background for $Z$ and $W$ decays respectively, for $M_S = 500$ GeV. Subplot (C) shows the invariant mass from $2l2\nu_l$-channel for $M_S = 600$ GeV.}
\label{2l2vl_observables}
\end{figure}  

The kinematics of a typical ZZF to $2l2\nu_l$ state can be derived from the Feynman diagram in Fig.~\ref{ZZF-2l2vl} as follows:
\begin{align}
\sqrt{s} &= (q_1+q_2+q_3+q_4)+(k_1+k_2).
\end{align}
We define $\Delta=(k_1+k_2)$, as the invariant mass of the $X$ particle. $\Delta$ can then be easily evaluated from the muon beam energy and detected charged leptons. In other words, $\Delta$ would be a function of all four visible leptonic vectors, $\Delta(s,q_1,q_2,q_3,q_4)$. Then $\Delta \neq 0$ signifies the existence of missing energy, and the scalar mass is given by
\begin{align}
M_s &= M(l^+l^-) + \Delta(s,q_1,q_2,q_3,q_4),\label{invisible_mij}
\end{align}
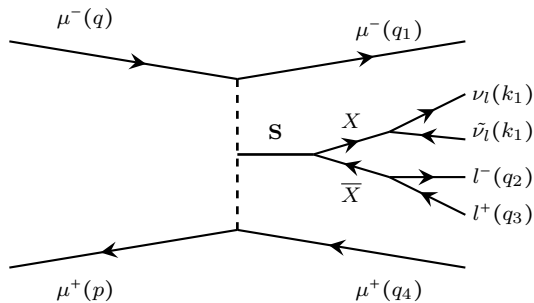
\begin{figure}[thb!]
\centering
\begin{tikzpicture}[baseline=0cm]
\draw[thick,decoration={markings, mark=at position 0.6 with {\arrow[scale=1.5]{stealth}}},
        postaction={decorate}] (0,.5) -- (3,0);   
\draw[thick,decoration={markings, mark=at position 0.6 with {\arrow[scale=1.5]{stealth}}},
        postaction={decorate}] (3,0) -- (6,.5);  
\draw[thick,decoration={markings, mark=at position 0.6 with {\arrow[scale=1.5]{stealth}}},
        postaction={decorate}]  (3,-2) -- (0,-2.5);   
\draw[thick,decoration={markings, mark=at position 0.6 with {\arrow[scale=1.5]{stealth}}},
        postaction={decorate}]  (6,-2.5) -- (3,-2);  
\draw[dashed,line width = 1pt] (3,0) -- (3,-2);
\draw[thick,line width = 1pt] (3,-1) -- (4,-1);
\draw[thick,decoration={markings, mark=at position 0.6 with {\arrow[scale=1.5]{stealth}}},
        postaction={decorate}] (4,-1) -- (5,-.7);  
\draw[thick,decoration={markings, mark=at position 0.6 with {\arrow[scale=1.5]{stealth}}},
        postaction={decorate}]  (5,-1.3) -- (4,-1);  
\draw[thick,decoration={markings, mark=at position 0.6 with {\arrow[scale=1.5]{stealth}}},
        postaction={decorate}] (5,-.7) -- (6,-.2);  
\draw[thick,decoration={markings, mark=at position 0.6 with {\arrow[scale=1.5]{stealth}}},
        postaction={decorate}] (6,-.8) --  (5,-.7);  
\draw[thick,decoration={markings, mark=at position 0.6 with {\arrow[scale=1.5]{stealth}}},
        postaction={decorate}] (5,-1.3) -- (6,-1.3);  
\draw[thick,decoration={markings, mark=at position 0.6 with {\arrow[scale=1.5]{stealth}}},
        postaction={decorate}] (6,-1.8) -- (5,-1.3);  
\begin{scriptsize}
\draw[] (1,.8) node{$\mu^-(q)$};
\draw[] (5,.65) node{$\mu^-(q_1)$};
\draw[] (1,-2.8) node{$\mu^+(p)$};
\draw[] (5,-2.8) node{$\mu^+(q_4)$};
\draw[] (3.5,-.7) node{\bf S};
\draw[] (4.5,-.55) node{$X$};
\draw[] (4.5,-1.5) node{$\overline{X}$};
\draw[] (6.5,-.2) node{$\nu_l(k_1)$};
\draw[] (6.5,-.7) node{$\tilde{\nu_l}(k_1)$};
\draw[] (6.5,-1.3) node{$l^-(q_2)$};
\draw[] (6.5,-1.8) node{$l^+(q_3)$};
\end{scriptsize}      
\end{tikzpicture}\label{feynmannu}
\caption{Feynman diagram of ZZF to $2l2\nu_l$.}
\label{ZZF-2l2vl}
\end{figure}
where $M(l^+l^-)$ is the invariant mass of the two oppositely charged leptons. In the case of ZZF, the $l^+l^-$ pair will originate from a single $Z$-boson, so only pairs with $0.8M_Z\leq M(l^+l^-)\leq 1.2M_Z$ were accepted. For WWF, $l^+$ and $l^-$ will originate from different particles, that is, $W^+,W^-$ respectively, and therefore $M(l^+l^-)$ is not expected to peak around the $M_W$ mass. Despite this, we found that the signal from WWF tends to outnumber the background in the region $ 250\,~\mathrm{GeV}\leq M(l^+l^-)\leq 600\,~\mathrm{GeV}$. In both cases, ZZF and WWF, only leptons with $P_T(l^\pm)\geq \frac{1}{5}M_X$ were included in pair formation. In this way, it is most likely that leptons included are those resulting from mediator ($X$) decay. However, a fraction of forward muons may satisfy these requirements. Therefore, a peak is expected to occur at $m_{ij} \sim \sqrt{s}$. However, this is not of interest to us, since we are already limiting $M_S \in [200,1000]$~GeV, and consequently such peaks will automatically be cut out (see Fig.~\ref{2l2vl_observables} (C)). We accepted events that had $0.6 M_X\leq M(l^+l^-+\Delta)\leq 1.2M_X$ due to the broadening of the resulting invariant mass distributions observed in Fig.~\ref{2l2vl_observables} (C). 

\begin{table}[tp]
    \centering
    \begin{tabular}{|c|c|c|c|c|c|}
    \hline
     \multicolumn{1}{|c}{ $M_S \,\,$[GeV]} &   & 250 & 500 & 750 & 1000\\\hline \rule{0pt}{1.25\normalbaselineskip}
       \multirow{2}{*}{$ZZ\rightarrow 2l2\nu_l$} & $P_T$ [GeV] & $7:627$ & $7:797$ & $13:803$ &  $67:827$ \\[5pt] 
       & $\abs{\eta}$ & $<2.35$ & $<2.35$ & $<2.35$ & $  < 2.35$ \\ \hline \rule{0pt}{1.25\normalbaselineskip}
       \multirow{2}{*}{$W^+W^- \rightarrow 2l2\nu_l$} & $P_T$ [GeV] & $26:726$ & $126:506$ & $85:595$ &  $45:725$ \\[5pt] 
       & $\abs{\eta}$ & $<2.35$ & $<2.35$ & $<2.35$ & $  < 2.35$ \\ \hline
    \end{tabular}
    \caption{Samples of transverse momentum and pseudorapidity cuts for the $2l2\nu_l$ channel automated for each value of the scalar mass.}
    \label{2l2vlcuts}
\end{table}

The accepted events up to this point were further constrained by applying cuts on $P_T(l^+l^-),\eta(l^+l^-)$, from a signal-background comparison, automated for each $M_s$ value as shown in Fig.~\ref{2l2vl_observables} (A,B) for $M_s=500$~GeV. The results are shown in Table~\ref{2l2vlcuts} for a sample of different $M_s$ values. The $P_T$ bins were selected by requiring $N_s \geq 1.2N_b$ and $N_s \geq 1.1 N_b$ for the $\eta$ bins for the $ZZ,WW$ channels. Similarly to the previous channels, the uncertainty in $\varepsilon_s,\varepsilon_b$ was computed from Eq.~\eqref{uncerti}, by repeating the previous analysis for different invariant mass range, $0.5 M_X\leq M(l^+l^-+\Delta)\leq 1.3M_X$. 

\begin{figure}[htb!]
\centering
\includegraphics[scale=.42]{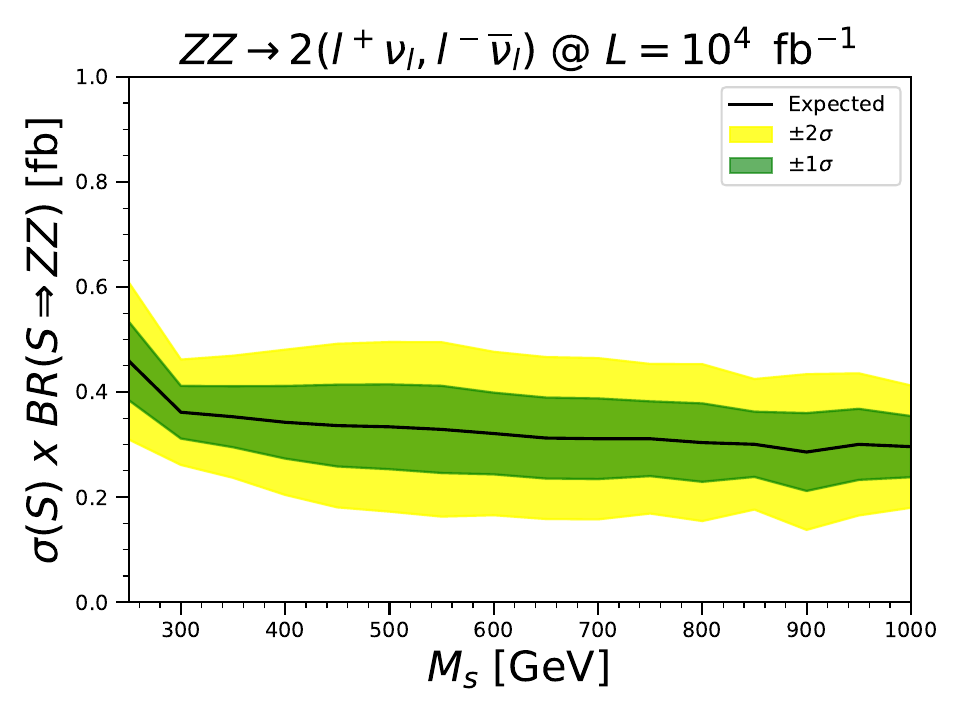}
\includegraphics[scale=.42]{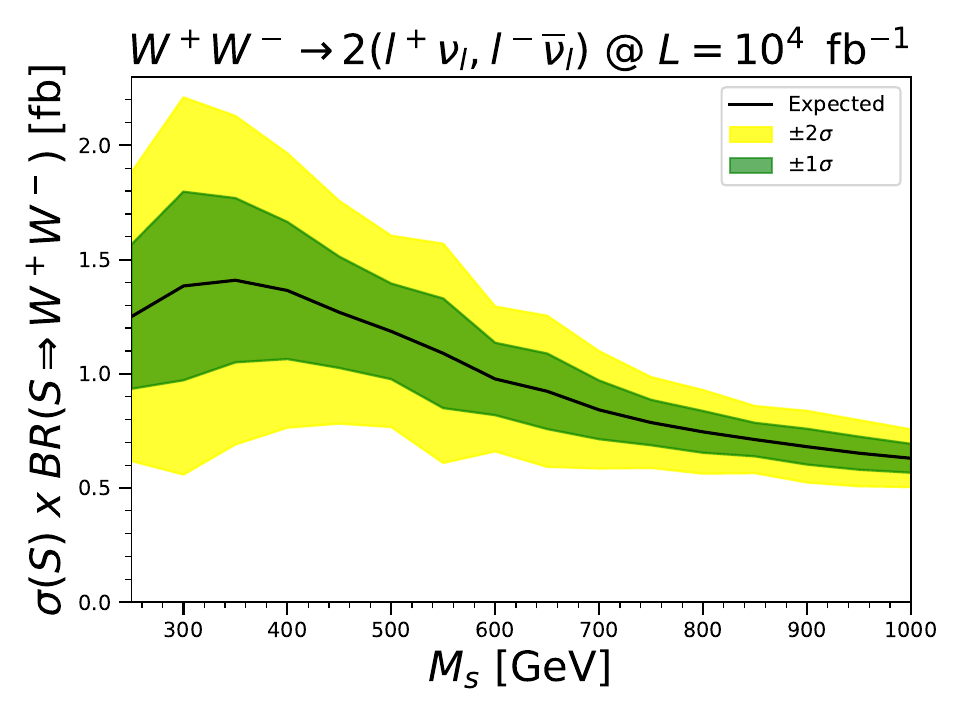}\\
(A) \hspace{180pt} (B)
\caption{(A) and (B) show the sensitivity curves at $L=10^4~\mathrm{fb}^{-1}$ and $\sqrt{s}=3$ TeV for $S\longrightarrow ZZ \rightarrow 2l2\nu_l$ and $S\longrightarrow W^+W^- \rightarrow 2l2\nu_l$ channels respectively at a muon collider with $68\%$ and $95\%$ CLs intervals. }
\label{2l2vl_sensitivity}
\end{figure}

The full analysis yielded a background efficiency of $\varepsilon_b \sim 4\%$ against a signal efficiency of $\varepsilon_s \sim 40\%$ for $ZZ\rightarrow2l2\nu_l$. Instead, for the $WW\rightarrow2l2\nu_l$ final state, the results were not as good, because the leptons did not originate from the same particle, which led to a less efficient signal-background separation. This is reflected in the sensitivity plots in Fig.~\ref{2l2vl_sensitivity}, which show a more uncertain curve for the $WW$ channel compared to the $ZZ$ channel.

\subsection{$2q2\nu_l$ Final states}

Unlike in the $2l2\nu_l$ case, this final state can originate from the $S$ decay into $Z$-bosons, $S\rightarrow ZZ,$ together with all possible backgrounds resulting from $\mu^+\mu^- \rightarrow \mu^+\mu^- (\nu_\mu\overline{\nu}_\mu) 2q2\nu_l$. 
Similarly to the previous channel, it would be impossible to directly compute the invariant mass owing to the missing energy carried away by the neutrinos. However, we can employ the same formula as  Eq.~\eqref{invisible_mij}  to evaluate it, with a slight replacement of $l^+l^-\rightarrow nJ$, where all the jets contained in the event are sourced from a single $Z$ boson, and $\Delta$ becomes a function of the forward muons ($q_1,q_2$) and all the jets ($p_1,p_2,\cdots p_n$). 

\begin{table}[tp]
\centering
\begin{tabular}{|>{\centering\arraybackslash}p{30mm}|>{\centering\arraybackslash}p{15mm}|>{\centering\arraybackslash}p{15mm}|>{\centering\arraybackslash}p{15mm}|>{\centering\arraybackslash}p{15mm}|} 
\hline
 $M_S$ [GeV]   &  250  & 500  &  750  & 1000\\[3pt] \hline \rule{0pt}{1.5\normalbaselineskip}
$P_T(J\overline{J})$ [GeV] & 9 : 979  & 5 : 1185  & 5 : 1105 &  5 : 1205  \\[7pt] 
$\abs{\eta(J\overline{J})}$  &  $< 0.55$ & $< 2.45$ & $< 2.65$ & $< 2.75$ \\[4pt] \hline
\end{tabular}
\caption{Samples of transverse momentum and pseudorapidity cuts for the $2q2\nu_l$ channel, automated for each value of the scalar mass.}
\label{2q2vlcuts}
\end{table}

\begin{figure}[htb!]
\centering
\includegraphics[scale=.4]{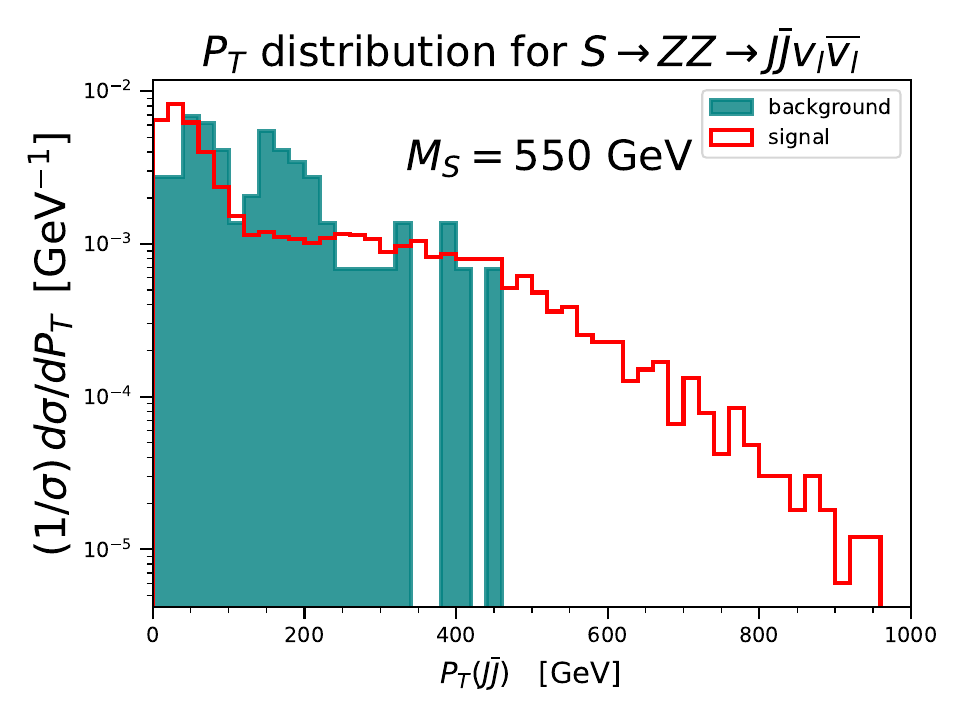}
\includegraphics[scale=.4]{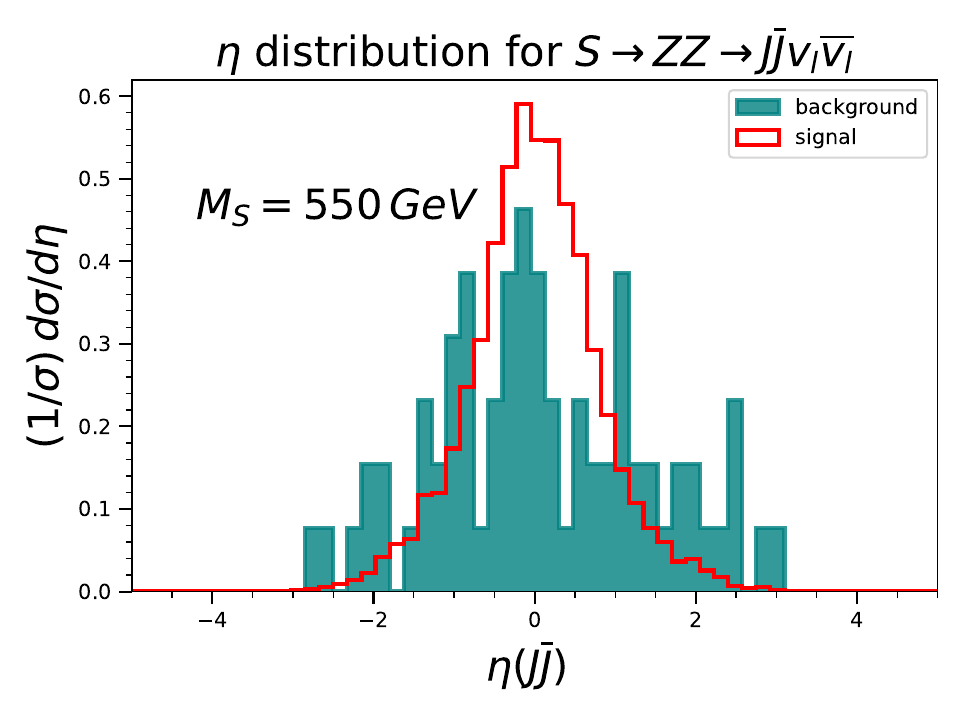}\\
(A)\hspace{180pt} (B)\\
\includegraphics[scale=.46]{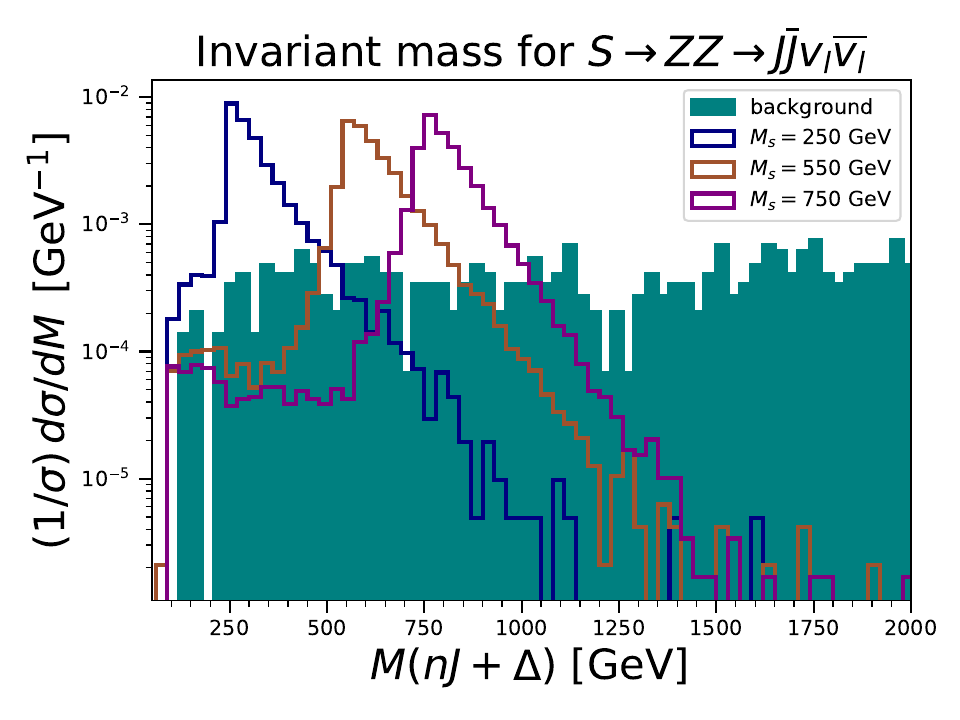}\\
(C)
\caption{Subplots (A), (B) show the transverse momentum and pseudorapidity comparisons between signal and background for $M_S = 550$ GeV. Subplot (C) shows the invariant mass obtained for $M_S = 250,550.750$ GeV.}
\label{2q2vl_observables}
\end{figure}  
\noindent
Therefore, the invariant mass in this case is given by
\begin{align}
M(nJ+\Delta) = M(nJ) + \Delta(s,q_1,q_2,p_1,p_2,\cdots p_n)\;.
\end{align}
This formula correctly returns a resonant peak around each $M_s$ value,  as shown in Fig.~\ref{2q2vl_observables} (C) for $M_s = 250,550,750$~GeV. Figure~\ref{2q2vl_observables} (C) also suggests an invariant mass range of $.9 M_s\leq M(nJ+\Delta) \leq 1.5 M_s$ to separate the signal from the background. 

In order to accept an event, we further require that it must contain at least one jet pair, $J\overline{J}$, with $0.5M_Z\leq M(J\overline{J})\leq 1.4 M_Z)$ for $P_T(J,\overline{J})\geq \frac{1}{5} M_Z$. We then applied cuts on $P_T(J\overline{J}),\eta(J\overline{J})$, obtained through a direct signal-background comparison, as shown in Fig.~\ref{2q2vl_observables} (A,B) for $M_s=550$ GeV. These secondary cuts were automated for each $M_s$ value, as illustrated in Table~\ref{2q2vlcuts} for a sample of selected $M_s$ values. The $P_T$ bins were selected by requiring $N_s \geq 1.2N_b$ and $N_s \geq 1.4 N_b$ for the $\eta$ bins. Similarly to all previous channels, the uncertainty in $\varepsilon_s,\varepsilon_b$ was computed using Eq.\eqref{uncerti} by repeating the preceding analysis for a different invariant mass range, $0.85 M_X\leq M(l^+l^-+\Delta)\leq 1.6M_X$. 

\begin{figure}[htb!]
\centering
\includegraphics[scale=.45]{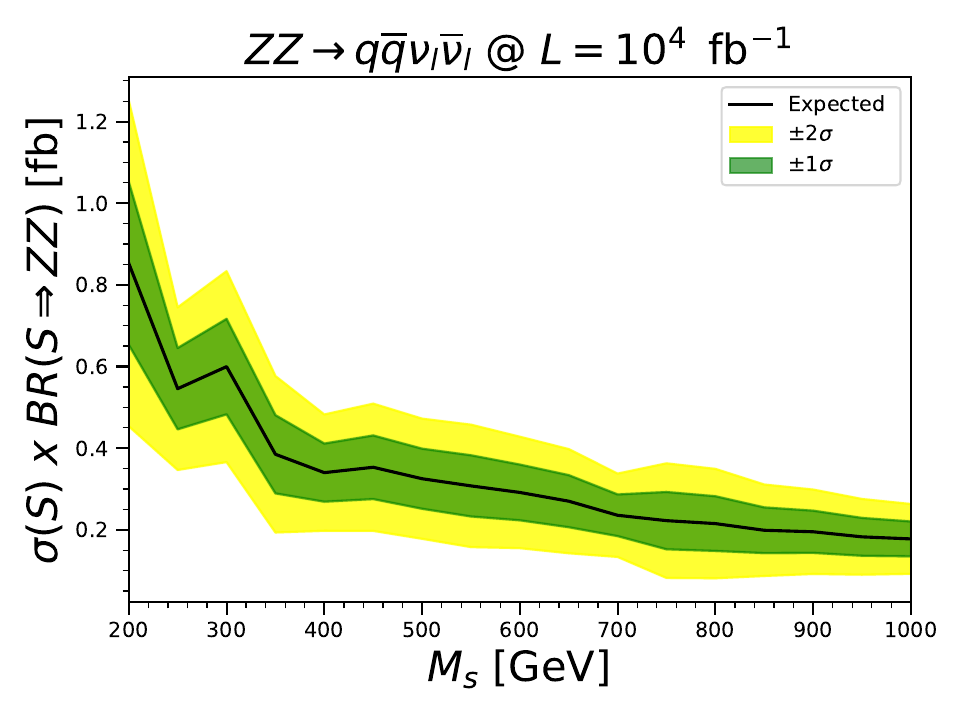}
(A) \hspace{180pt} (B)
\caption{The sensitivity curves at $L=10^4 \,\,\mathrm{fb}^{-1}$ and $\sqrt{s}=3$ TeV for $S\rightarrow ZZ \rightarrow 2q2\nu_l$ at a muon collider with $68\%$ and $95\%$ CLs intervals.\\[1pt]}
\label{2q2vl_sensitivity}
\end{figure}

\noindent
This analysis scheme greatly suppressed the background, yielding a background efficiency of $\varepsilon_b \sim 1\%$ on average, while the signal efficiency $\varepsilon_s$ exceeded $25\%$ on average. The corresponding sensitivity plot for this final state is shown in Fig.~\ref{2q2vl_sensitivity} for a luminosity of $L=10^4~\mathrm{fb}^{-1}$.

\subsection{$l\nu_l2q$ Final states}

This final state is another gauge boson-initiated channel that arises only from $S\longrightarrow W^+W^-\rightarrow l^+\nu_l (l^- \tilde{\nu_l}) q\overline{q}$. All backgrounds contributing to this channel that could be generated at the muon collider from $\mu^+\mu^- \longrightarrow mu^+\mu^- l^+\nu_l (l^- \tilde{\nu_l}) q\overline{q}$ were considered. The invariant mass of the fully visible final states can also be evaluated using Eq.~\ref{invisible_mij}, but now `$\Delta$' will be a function of an extra lepton vector (i.e. $p_3$) which results from $W$-boson decay,

\begin{align}
M_s&= M(l^\pm+nJ) + \Delta(s,q_1,q_2,q_3,p_1,p_2,\cdots p_n ),
\end{align}
where `$q_1,q_2,q_3$' are the four vectors of the two forward muons and extra lepton resulting from one $W^\pm$-boson decay. `$p_1,p_2,\cdots p_n $' are the entire n-jets four vectors in the event.  
\begin{figure}[htb!]
\centering
\includegraphics[scale=.4]{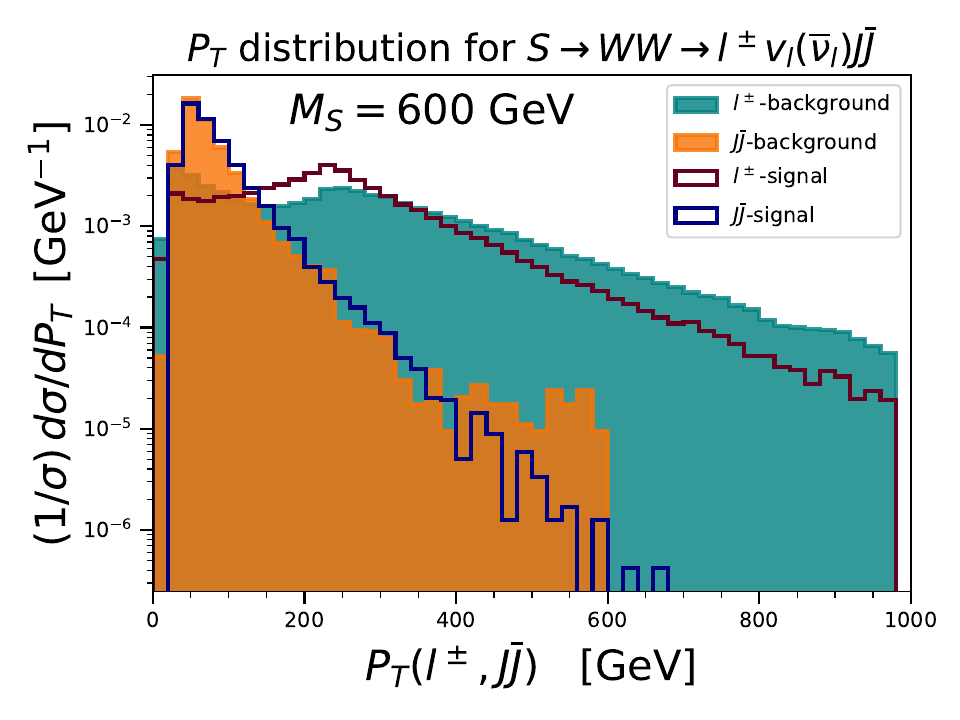}
\includegraphics[scale=.4]{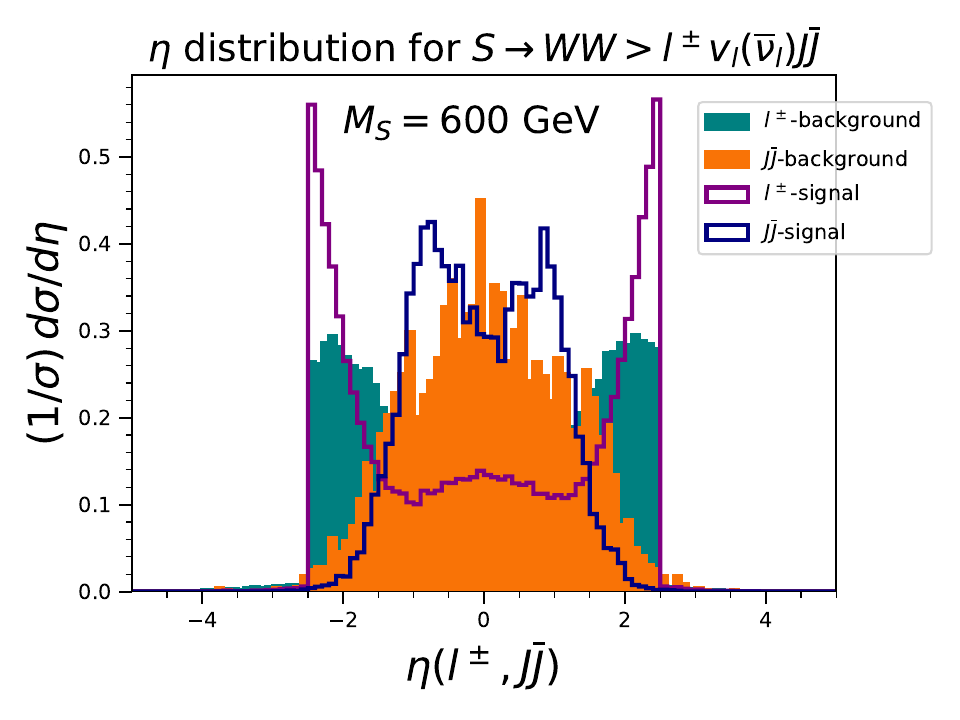}\\
(A)\hspace{180pt} (B)\\
\includegraphics[scale=.4]{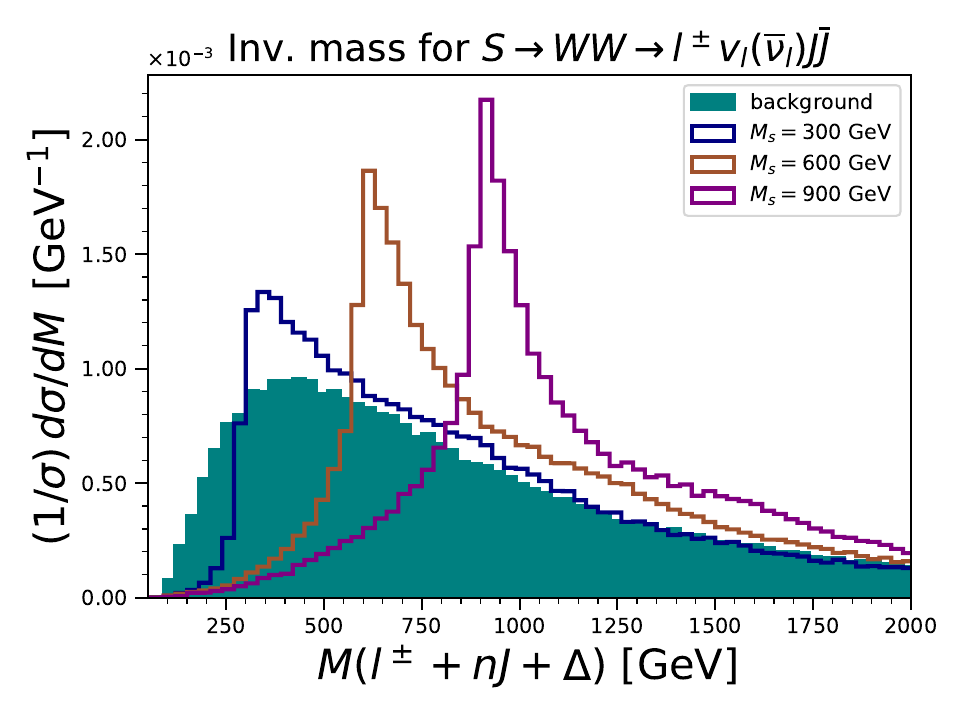}\\
(C)
\caption{(A) and (B) show the transverse momentum and pseudorapidity distributions of paired jets and a single charged lepton obtained from $l\nu_l2q$-channel for $M_S=600$~GeV. (C)  shows the invariant mass distributions of $l\nu_l2q$-channel for $M_s= 300,600,900$ GeV. }
\label{lvl2q_observables}
\end{figure} 

\begin{table}[tp]
    \centering
    \begin{tabular}{|c|c|c|c|c|c|}
    \hline
     \multicolumn{1}{|c}{ $M_S \,\,$[GeV]} &   & 250 & 500 & 750 & 1000\\\hline \rule{0pt}{1.35\normalbaselineskip}
       \multirow{4}{*}{$W^+W^- \rightarrow l^\pm\nu_l(\overline{\nu}_l) q \overline{q}$} & $P_T(l^\pm)$ [GeV]& $111:361$ & $111:311$ & $91:331$ &  $81:371$ \\[5pt] 
       & $\abs{\eta(l^\pm)}$ & $2:2.45$ & $2.1:2.45$& $<2.45$ & $<2.45$ \\[5pt]
       & $P_T(J\overline{J})$ [GeV]& $39:509$ & $409:469$ & $32:612$ &  $35:725$ \\ [5pt]
       & $\abs{\eta(J\overline{J})}$ & $<2.25$ & $ <1.95$ & $<1.65$ & $  < 1.65$ \\ [7pt]\hline 
    \end{tabular}
    \caption{Samples of transverse momentum and pseudorapidity cuts for the $2q2\nu_l$ and $l\nu_l 2q$ channels, automated for different values of the scalar mass.}
    \label{lvlqcuts}
\end{table}

$M(l^\pm+nJ)$ runs over all possible combinations of a single charged lepton in the final state, with $n$ jets available in the event. This includes forward high-energy scattered muons, so a secondary peak around $M_S\approx \sqrt{s}$ should be expected, and could be ignored for the reasons discussed earlier. In fact, this will not even be seen in this case because the jets largely outnumber the forward muons, and hence the secondary peak at $M_S=\sqrt{s}$ will be smeared away, which would be further enhanced by only considering charged leptons with $P_T(l^\pm) \geq \frac{1}{5}M_W$ that shall suppress the forward muon contributions. This approach generated well-defined invariant mass distributions for all the examined $M_s$ values, as can be seen in Fig.~\ref{lvl2q_observables} (C), which suggests a narrower invariant mass range compared to the previous channel, $0.85 M_s\leq M(l^\pm+ nJ+\Delta) \leq 1.5 M_s$. If the event passes this primary cut, we then check whether the jet invariant mass peaks around the $W$-boson mass, $0.75 M_W \leq M(nJ) \leq 1.25 M_W$. If it does, then we construct all possible jet pairs with $P_T\geq \frac{1}{5} M_W$. 
At least one pair of jets, $J\overline{J}$, is required in order to pass this event to the secondary analysis. In this step, we check for the observables $P_T(J\overline{J}), \eta(J\overline{J})$, the regions where the signal outnumbers the background, keeping only the bins where $N_s \geq 1.1N_b$ for $P_T(J\overline{J})$ and $N_s \geq 1.2N_b$ for $\eta(J\overline{J})$. This scanning procedure is automated for each $M_s$ value as shown in Fig.~\ref{lvl2q_observables} (A,B) for $M_s=600$ GeV, and for other different scalar mass values in Table\ref{lvlqcuts}. We finally account for the fluctuations in the measured efficiencies by repeating the previous analysis for a different invariant mass range, $0.8 M_s\leq M(l^\pm+ nJ+\Delta) \leq 1.65 M_s$, and then compute $U(\sigma_s)$ using Eq.~\eqref{uncerti}. 

\begin{figure}[htb!]
\centering
\includegraphics[scale=.45]{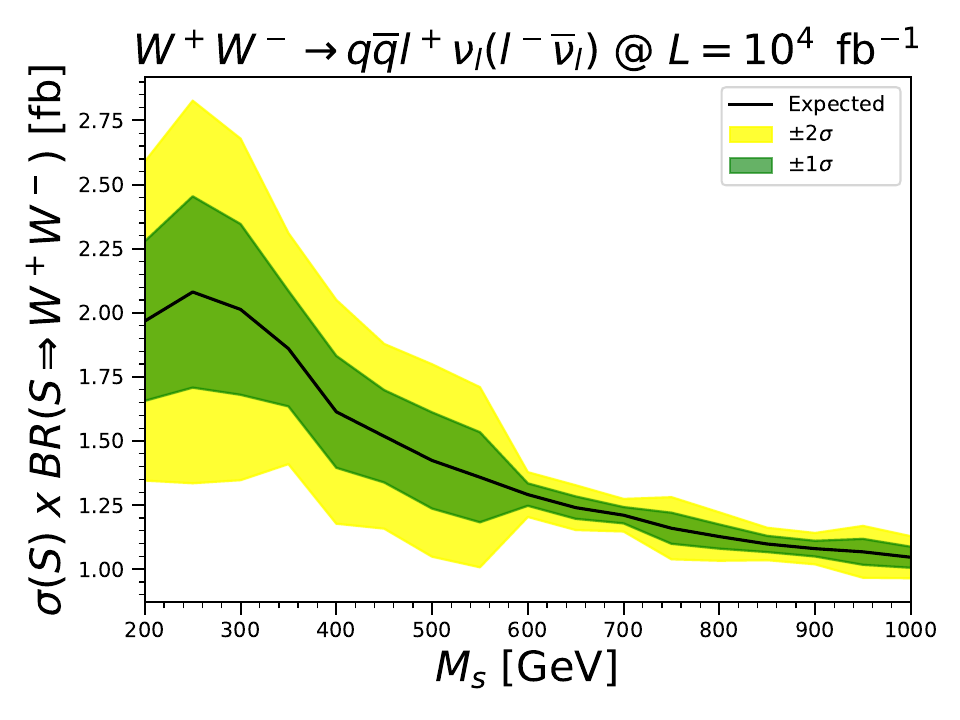}
\caption{Sensitivity curves for $ S\rightarrow l \nu_l2q$  at a muon collider with $L=10^4$~fb$^{-1}$ and $\sqrt{s}=3$ with $68\%$ and $95\%$ C.L. intervals.}
\label{lvl2q_sensitivity}
\end{figure}

Our analysis successfully suppressed backgrounds in comparison to signal events, with $\varepsilon_b\sim 10:40 \%$, and $\varepsilon_s \sim 25:88 \%$. This channel shows a sensitivity in $\varepsilon_s, \varepsilon_b$ to the scalar mass value, as both tend to grow for higher $M_s$ values, a feature that was not observed in the other channels. This feature is reflected in the sensitivity plot in Fig.~\ref{lvl2q_sensitivity}, where the uncertainty in the signal cross section converges towards higher values of the scalar mass as $\varepsilon_s$ becomes larger than $50\%$, which guarantees that the uncertainties will be smaller.

\section{Results}

\begin{figure}[htb!]
\centering
\includegraphics[scale=.46]{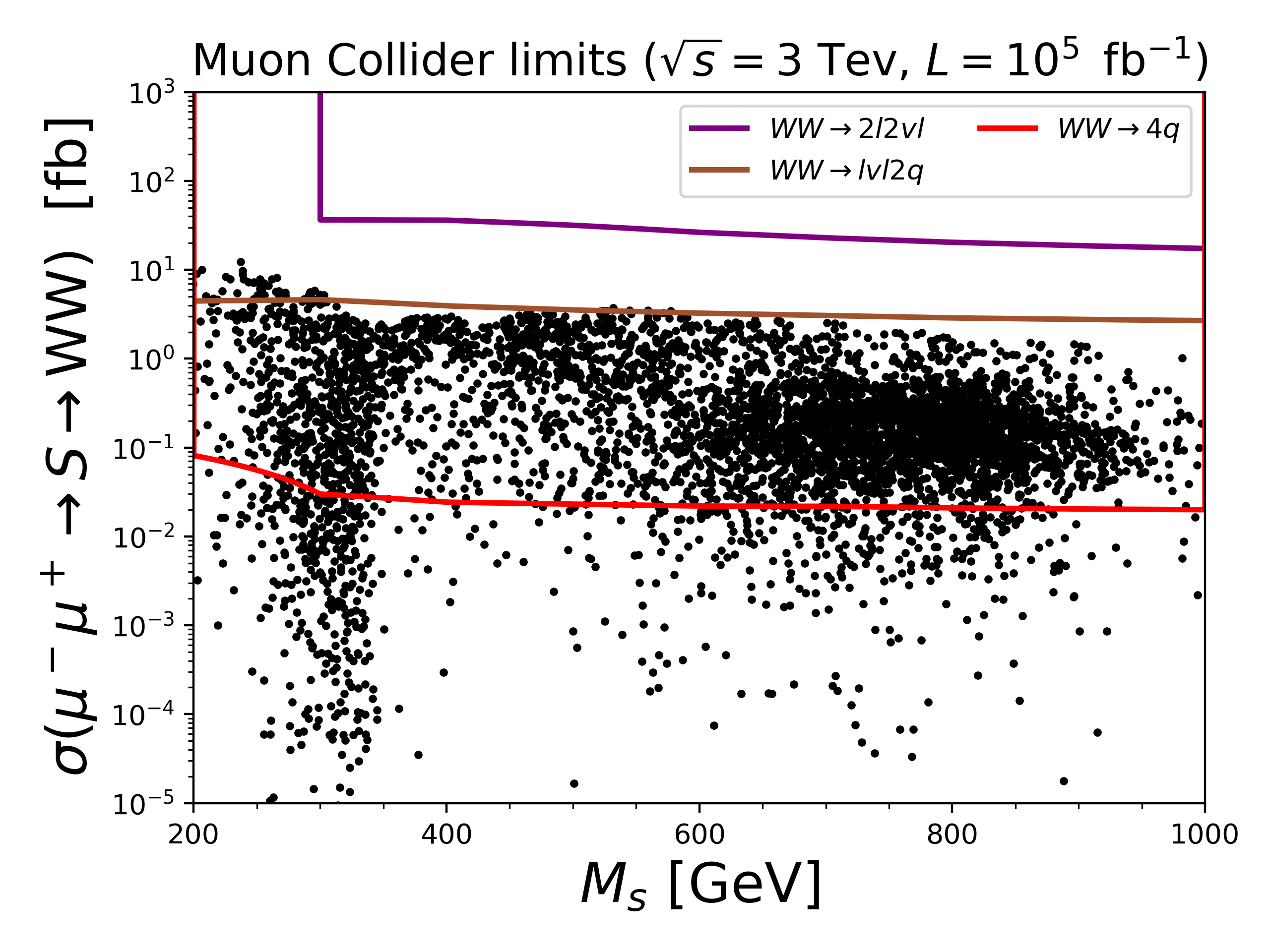}
\includegraphics[scale=.46]{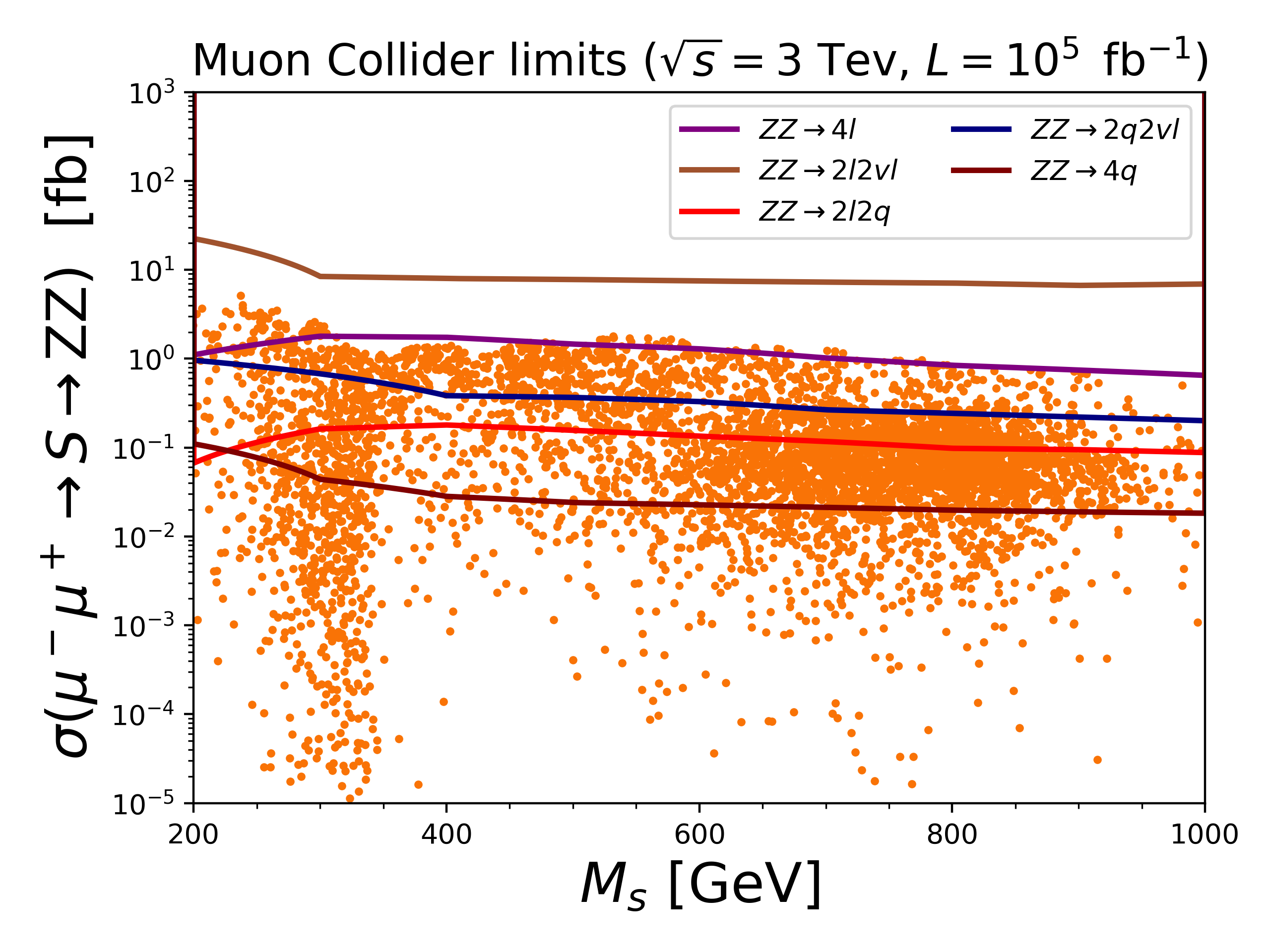}\\
(A) \hspace{200pt} (B)\\
\includegraphics[scale=.46]{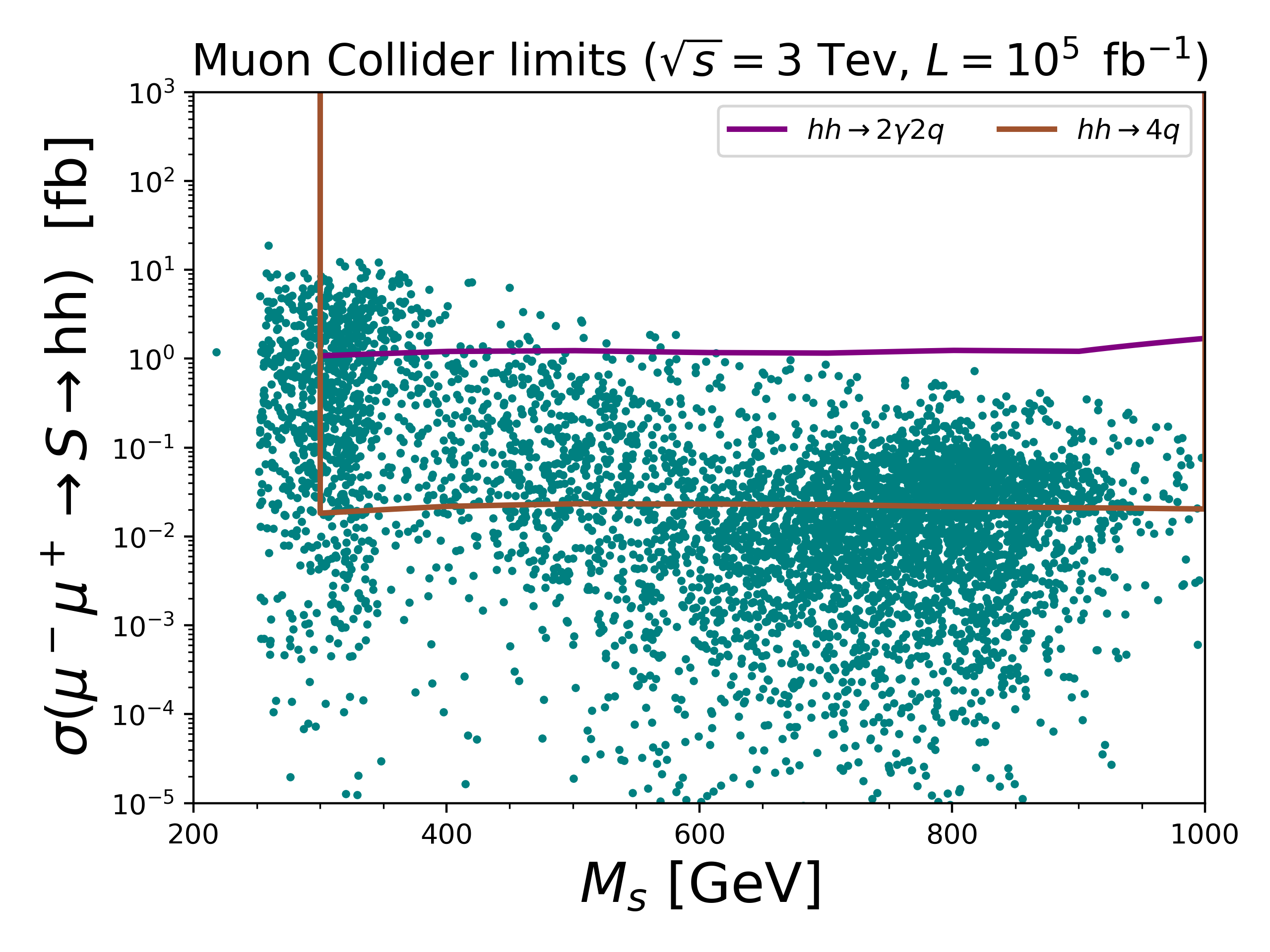}\\
(C)
\caption{Exclusion curves for all channels from $S\longrightarrow WW,\,ZZ,\,hh$ respectively at a muon collider with $\sqrt{s} = 3$ TeV and $L=10^5\, ~\mathrm{fb}^{-1}$.}
\label{exclusion_curves}
\end{figure}

Using the phenomenological analyses described above, we explored the parameter space points that fulfill the SFOEWPT conditions using the derived sensitivity plots, as illustrated in  Fig.~\ref{exclusion_curves}. 
It is evident that channels containing quarks in the final state can explore a larger portion of the xSM SFOEWPT parameter space. This is primarily attributed to the higher branching ratios of the Higgs bosons and vector bosons to the jets. The relative absence of substantial QCD backgrounds at a muon collider, in comparison to hadron colliders, enables the reconstruction of jets via a comparatively straightforward approach. This factor facilitates efficient background suppression, even for channels with invisible final states. This same rationale supports the pursuit of potential precision tests of the Higgs boson self-couplings at a muon collider, and the possible confirmation or exclusion of the SFOEWPT scenario. We reserve such an investigation for future studies, along with the examination of potential kinematical differences between jets originating from $S\rightarrow hh$ and $S\rightarrow ZZ (WW)$, which may be related to the $\mathbb{Z}_2$ nature of the extended potential of the xSM.  
 
\begin{figure}[htb!]
\centering
\includegraphics[scale=.46]{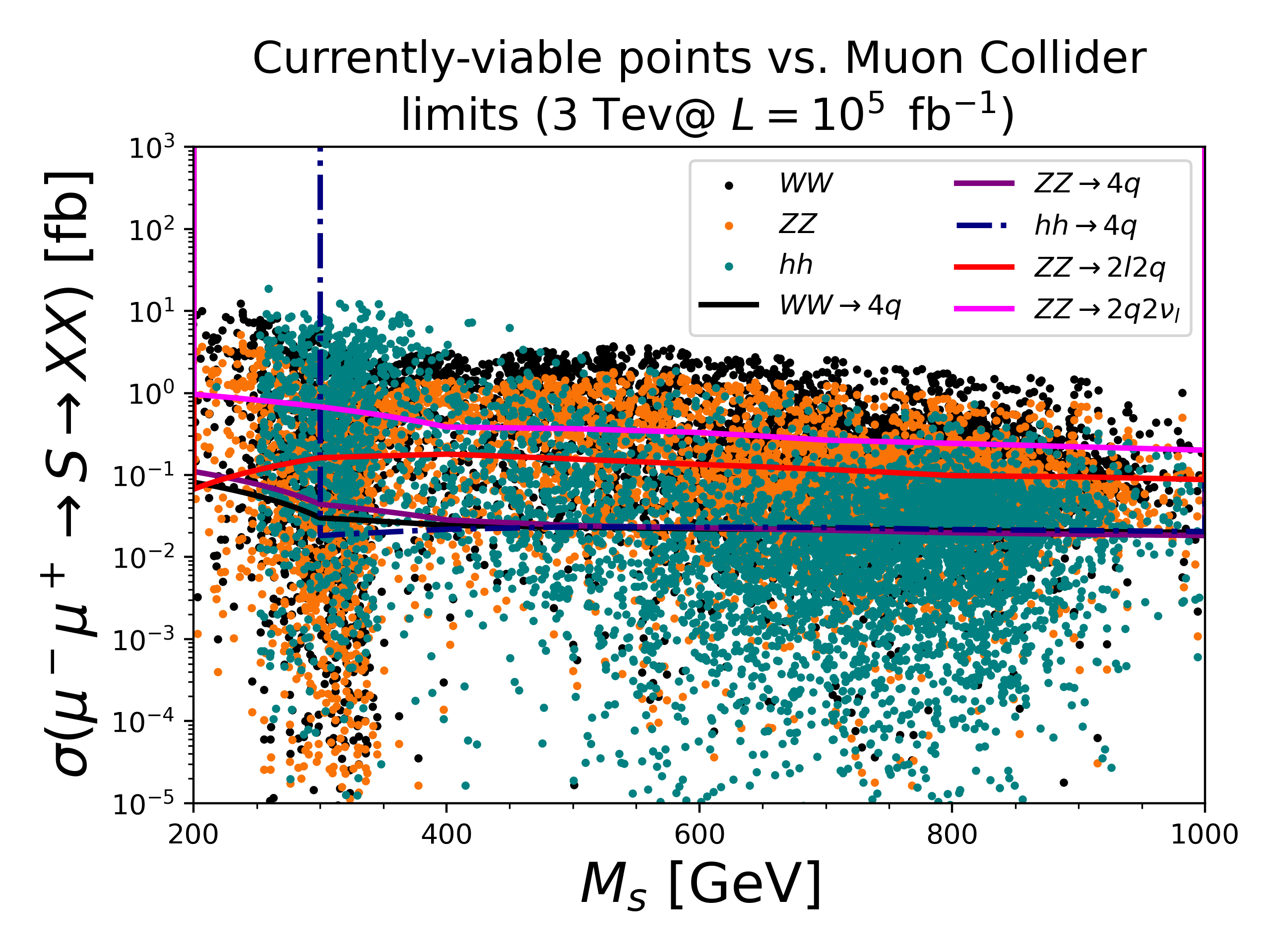}
\includegraphics[scale=.46]{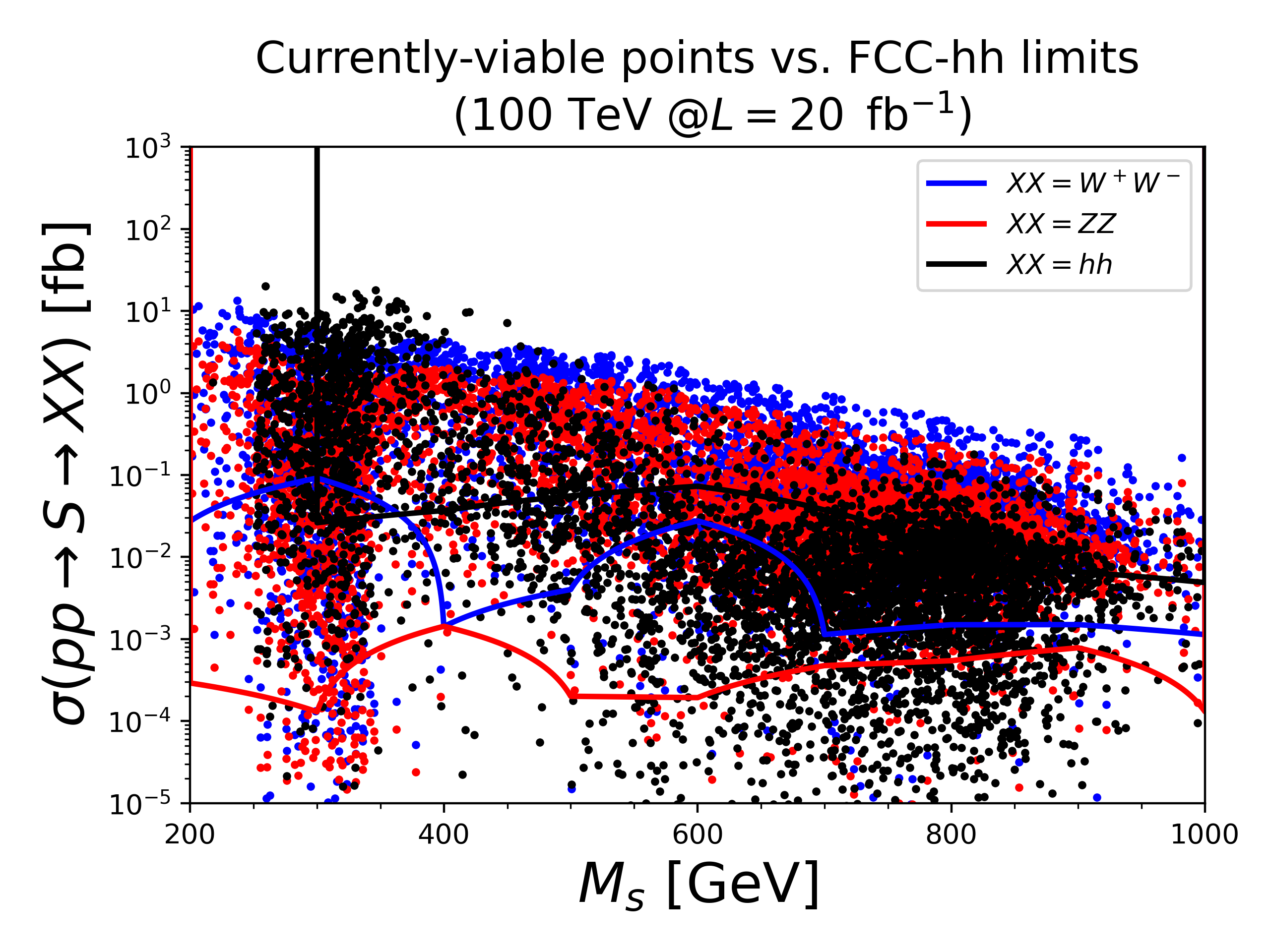}\\
(A) \hspace{200pt} (B)\\
\includegraphics[scale=0.5]{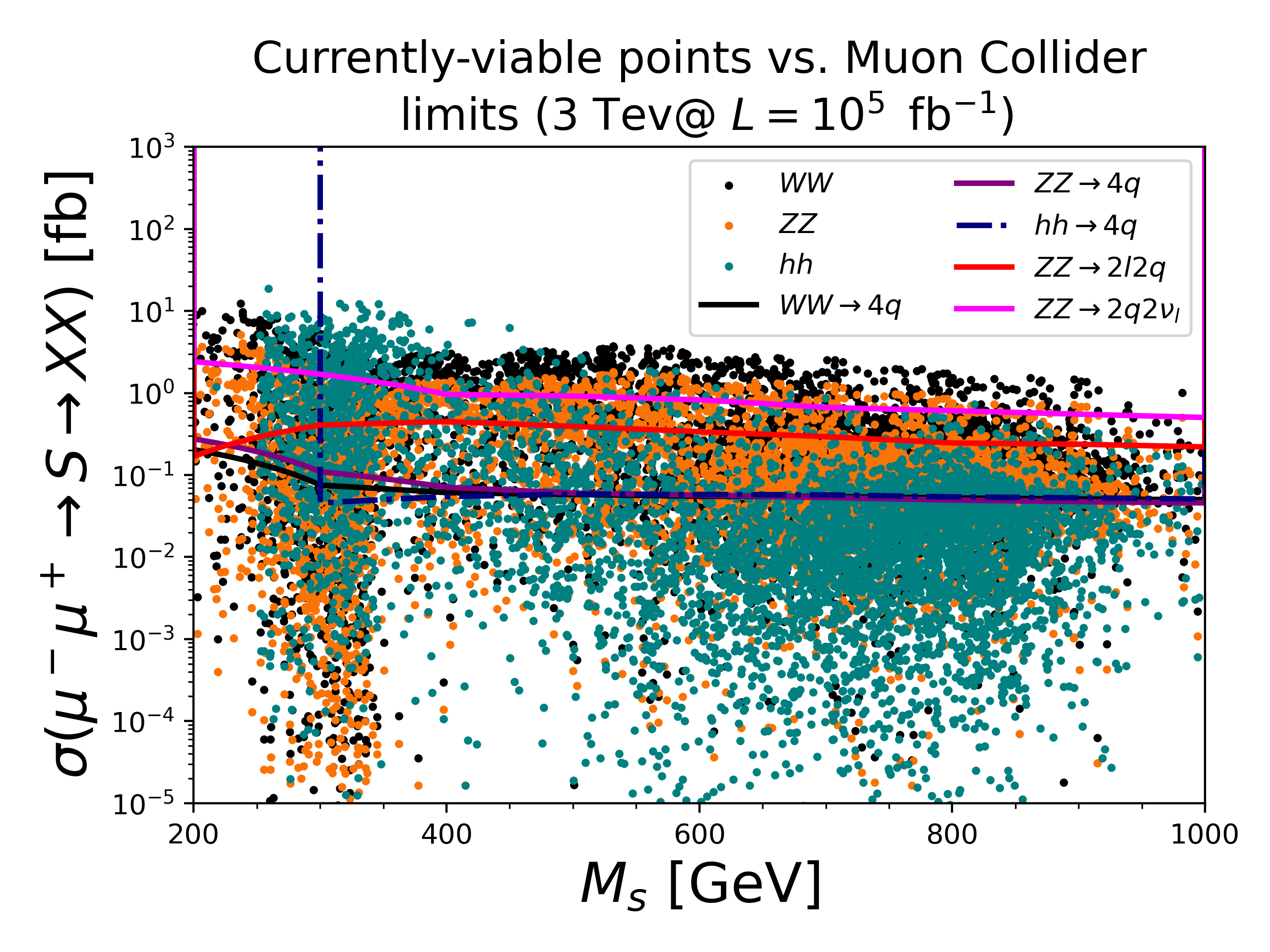}\\
(C)
\caption{(A) shows the combined  $WW,\,ZZ,\,hh  \rightarrow nJ$ final states at a muon collider with $\sqrt{s} = 3$ TeV and $L=10^5$~ab$^{-1}$.  (B) shows the FCC-hh exclusion curves for the same channels  at a much higher energy $\sqrt{s} = 100$, TeV and much smaller luminosity $L=20\, ~\mathrm{fb}^{-1}$. (C) represents the discovery plot for $\mathcal{S} =5$.}
\label{all_exclusion_curves}
\end{figure}

Figure~\ref{exclusion_curves} presents the currently viable parameter space points that satisfy the SFOEWPT conditions for both the liberal and conservative categories discussed in Section 2 in conjunction with the sensitivity plots obtained from the previous subsections. Figure~\ref{exclusion_curves} (A) summarizes all the different final states that could be obtained from $WW$ decay, and Fig.~\ref{exclusion_curves} (B,C) for $ZZ$ and $hh$, respectively, at a luminosity of $10^5~\mathrm{fb}^{-1}$.\footnote{The exclusion curve is proportional to $\frac{1}{\sqrt{L}}$ (see Eq.~\eqref{significance}), so we have chosen an order of magnitude higher luminosity than the value used in the previous exclusion curves in order to explore a larger parameter space volume.} The preferred channels, that is, jet-rich channels, have consequently been isolated in Fig.~\ref{all_exclusion_curves}, as they represent the primary limiting channels in a muon collider. Compared to the FCC-hh potential for searches, depicted in Fig.~\ref{all_exclusion_curves} (B), the muon collider appears to be capable of excluding the majority of points that could be examined at the FCC-hh at significantly lower center-of-mass energies, particularly for $S\rightarrow XX\rightarrow 4q$.

In the event of a positive detection of the real singlet at the proposed future muon collider, it will still be possible to exclude a significant portion of FOEWPT parameter space points. This accounts for a modification of the statistical significance in Eq.\eqref{significance} from $\mathcal{S} = 2$ to $\mathcal{S} = 5$, which results in a consistent and slight upward shift of the sensitivity curves across all channels. This result is illustrated in Fig.~\ref{all_exclusion_curves} (c).
From that discovery plot, we can further estimate, in the event of a positive detection, the anticipated mixing between the two scalars. As discussed in Section~\ref{sec:ewptc}, the potential in Eq.~\eqref{reno} results in mixing between the SM Higgs and the new scalar, which is a desirable characteristic for enhancing the FOEWPT. This mixing will scale all the Higgs couplings by $\cos \theta$~\cite{papaefstathiou2022electro}.
\begin{figure}[htb!]
\centering
\includegraphics[scale=0.5]{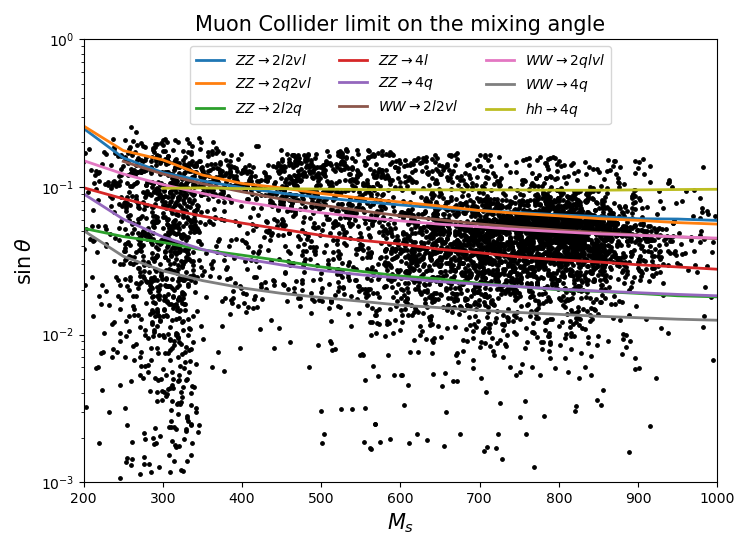}
\caption{The limits in the Higgs-real scalar mixing extracted from the different decay channels that could be probed at the muon collider.}
\label{mixingAngle_limit}
\end{figure}

\noindent
In Ref.~\cite{Han_2021}, the authors discussed the sensitivity of the Higgs boson's properties to new physics, particularly its couplings to weak gauge bosons, and the Higgs trilinear coupling. In this investigation, the source of the `new physics' is the real singlet extension, which inevitably scales all of the Higgs boson's couplings. Of particular importance for this study is the modification of the Higgs trilinear coupling due to this extension, given its role in governing the Higgs potential and, consequently, the type and strength of the EWPT. This work is currently in progress. However, utilizing the discovery plots in Fig.~\ref{all_exclusion_curves} (C), it is possible to extrapolate certain limits on the mixing angle between the two scalars, which may be observed in current and future precision tests of the Higgs properties.
Following ref.~\cite{papaefstathiou2022electro}, the mixing angle between the two scalars can be derived from,
\begin{align}
\sin^4 \theta = \frac{\sigma_{\text{\tiny BSM}}(\mu^+\mu^- \xrightarrow[]{VBF} s \rightarrow f\overline{f} ) }{\sigma_{\text{\tiny SM}}(\mu^+\mu^- \xrightarrow[]{VBF} h \rightarrow f\overline{f} )},
\end{align}
where, $\sigma_{\text{\tiny BSM}}$ represents the expected cross section obtained from our analysis in Section.~\ref{sec:4}, and  $\sigma_{\text{\tiny SM}}$ denotes the SM cross section for each final state topology computed using \texttt{MG5\_aMC}, setting $m_h =m_s$ ($\sin \theta =1$). Through this analysis for the muon collider, it is anticipated that $0.02 \leq \sin \theta \leq 0.15 $ in the event of a positive discovery. This result can already exclude some of the valid parameter space points for FOEWPT, as each set predicts a specific value for the mixing angle, which is illustrated in Fig.\ref{mixingAngle_limit} for both the conservative and liberal categories.

\section{Conclusions}

In this study, we investigated the potential of a high-energy muon collider to explore electroweak phase transitions in the context of the Standard Model extended by a real scalar singlet. Our analysis focused on examining the occurrence of a strong first-order electroweak phase transition, which is essential for explaining baryogenesis, and may result in observable gravitational waves.

As the initial step of our investigation, we identified Lagrangian parameter values for which xSM can produce a strong first-order phase transition that satisfies the sphaleron critical conditions for baryogenesis. These parameter values span a broad parameter space for the xSM. Subsequently, assuming a future muon collider with center-of-mass energy $\sqrt{s} = 3$ TeV, we assessed the sensitivity to a new scalar singlet-like particle through various final states. Our findings indicate that the muon collider can be an effective tool for probing the xSM parameter space if high luminosities can be approached, where the clean environment of a muon collider allows for precise measurements, particularly producing a scalar singlet-like particle via vector boson fusion (VBF). We primarily examined the direct production of the new scalar and its decay into all viable final states, which offers a promising avenue for its discovery or exclusion. We found that decay channels rich with jets in the final states are the most promising for excluding FOEWPT parameter space points at the muon collider. Indirectly, deviations in the Higgs boson couplings from the SM expectations could provide further evidence for the presence of a new scalar singlet-like particle, which would be valuable to explore further in the muon collider. With precision measurements, the zero-temperature part of the effective potential can be reconstructed at the muon collider, whereas gravitational wave observations can provide additional information on thermal dynamics. The scalar singlet-enhanced electroweak phase transition can also generate a signal in gravitational wave detectors, which enables the muon collider to provide complementary information for gravitational wave detectors.

Our findings highlight the unique advantages of a muon collider in exploring electroweak symmetry breaking and potentially new physics beyond the Standard Model. The ability to resolve the nature of the electroweak phase transition in such a collider is of paramount importance, especially for advancing our understanding of early universe dynamics and the mechanisms behind baryogenesis, providing significant advancement in our understanding of electroweak symmetry breaking and its implications in cosmology.

\acknowledgments{The research of C.B. is supported by the Australian Research Council through DP210101636, DP220100643, and LE210100015. A.P. acknowledges support by the National Science Foundation under Grant No.\ PHY 2210161.}

\bibliographystyle{JHEP}
\bibliography{2_refrences.bib}

\providecommand{\href}[2]{#2}\begingroup\raggedright\begin{thebibliography}{10}

\bibitem{kirzhnits1972}
D.A.~Kirzhnits and A.D.~Linde, \emph{Macroscopic consequences of the weinberg
  model}, {\emph{Physics Letters B} {\bfseries 42} (1972) 471}.

\bibitem{coleman1973radiative}
S.~Coleman and E.~Weinberg, \emph{Radiative corrections as the origin of
  spontaneous symmetry breaking}, {\emph{Physical Review D} {\bfseries 7}
  (1973) 1888}.

\bibitem{dolan1974symmetry}
L.~Dolan and R.~Jackiw, \emph{Symmetry behavior at finite temperature},
  {\emph{Physical Review D} {\bfseries 9} (1974) 3320}.

\bibitem{quiros1998}
M.~Quiros, \emph{Finite temperature field theory and phase transitions},
  {\emph{Proceedings, Summer school in high-energy physics and cosmology:
  Trieste, Italy} {\bfseries 1999} (1998) 187}.

\bibitem{jackiw1974functional}
R.~Jackiw, \emph{Functional evaluation of the effective potential},
  {\emph{Physical Review D} {\bfseries 9} (1974) 1686}.

\bibitem{weinberg1973}
S.~Weinberg, \emph{Perturbative calculations of symmetry breaking},
  {\emph{Physical Review D} {\bfseries 7} (1973) 2887}.

\bibitem{dolan1974gauge}
L.~Dolan and R.~Jackiw, \emph{Gauge-invariant signal for gauge-symmetry
  breaking}, {\emph{Physical Review D} {\bfseries 9} (1974) 2904}.

\bibitem{Sakharov:1967dj}
A.D.~Sakharov, \emph{{Violation of CP Invariance, C asymmetry, and baryon
  asymmetry of the universe}},
  \href{https://doi.org/10.1070/PU1991v034n05ABEH002497}{\emph{Pisma Zh. Eksp.
  Teor. Fiz.} {\bfseries 5} (1967) 32}.

\bibitem{dine2003origin}
M.~Dine and A.~Kusenko, \emph{Origin of the matter-antimatter asymmetry},
  {\emph{Reviews of Modern Physics} {\bfseries 76} (2003) 1}.

\bibitem{morrissey2012electroweak}
D.E.~Morrissey and M.J.~Ramsey-Musolf, \emph{Electroweak baryogenesis},
  {\emph{New Journal of Physics} {\bfseries 14} (2012) 125003}.

\bibitem{bhattacharyya2011pedagogical}
G.~Bhattacharyya, \emph{A pedagogical review of electroweak symmetry breaking
  scenarios}, {\emph{Reports on Progress in Physics} {\bfseries 74} (2011)
  026201}.

\bibitem{mclerran1991baryon}
L.~McLerran, M.~Shaposhnikov, N.~Turok and M.~Voloshin, \emph{Why the baryon
  asymmetry of the universe is~ 10- 10}, {\emph{Physics Letters B} {\bfseries
  256} (1991) 477}.

\bibitem{senaha2013overview}
E.~Senaha, \emph{Overview of electroweak baryogenesis}, {\emph{arXiv preprint
  arXiv:1305.1563} (2013) }.

\bibitem{cline2006baryogenesis}
J.M.~Cline, \emph{Baryogenesis}, {\emph{arXiv preprint hep-ph/0609145} (2006)
  }.

\bibitem{braibant1993sphalerons}
S.~Braibant, Y.~Brihaye and J.~Kunz, \emph{Sphalerons at finite temperature},
  {\emph{International Journal of Modern Physics A} {\bfseries 8} (1993) 5563}.

\bibitem{ahriche2007criterion}
A.~Ahriche, \emph{What is the criterion for a strong first order electroweak
  phase transition in singlet models?}, {\emph{Physical Review D—Particles,
  Fields, Gravitation, and Cosmology} {\bfseries 75} (2007) 083522}.

\bibitem{carrington1992effective}
M.~Carrington, \emph{Effective potential at finite temperature in the standard
  model}, {\emph{Physical Review D} {\bfseries 45} (1992) 2933}.

\bibitem{espinosa1993nature}
J.~Espinosa, M.~Quiros and F.~Zwirner, \emph{On the nature of the electroweak
  phase transition}, {\emph{Physics Letters B} {\bfseries 314} (1993) 206}.

\bibitem{arnold1993effective}
P.~Arnold and O.~Espinosa, \emph{Effective potential and first-order phase
  transitions: Beyond leading order}, {\emph{Physical Review D} {\bfseries 47}
  (1993) 3546}.

\bibitem{kajantie1996there}
K.~Kajantie, M.~Laine, K.~Rummukainen and M.~Shaposhnikov, \emph{Is there a hot
  electroweak phase transition at $ m_h \gtrsim m_w $?}, {\emph{arXiv preprint
  hep-ph/9605288} (1996) }.

\bibitem{senaha2020}
E.~Senaha, \emph{Symmetry restoration and breaking at finite temperature: an
  introductory review}, {\emph{Symmetry} {\bfseries 12} (2020) 733}.

\bibitem{espinosa2012strong}
J.R.~Espinosa, T.~Konstandin and F.~Riva, \emph{Strong electroweak phase
  transitions in the standard model with a singlet}, {\emph{Nuclear Physics B}
  {\bfseries 854} (2012) 592}.

\bibitem{choi1993real}
J.~Choi and R.~Volkas, \emph{Real higgs singlet and the electroweak phase
  transition in the standard model}, {\emph{Physics Letters B} {\bfseries 317}
  (1993) 385}.

\bibitem{o2007minimal}
D.~O’Connell, M.J.~Ramsey-Musolf and M.B.~Wise, \emph{Minimal extension of
  the standard model scalar sector}, {\emph{Physical Review D} {\bfseries 75}
  (2007) 037701}.

\bibitem{ham2005electroweak}
S.~Ham, Y.~Jeong and S.~Oh, \emph{Electroweak phase transition in an extension
  of the standard model with a real higgs singlet}, {\emph{Journal of Physics
  G: Nuclear and Particle Physics} {\bfseries 31} (2005) 857}.

\bibitem{ramsey2020electroweak}
M.J.~Ramsey-Musolf, \emph{The electroweak phase transition: a collider target},
  {\emph{Journal of High Energy Physics} {\bfseries 2020} (2020) 1}.

\bibitem{papaefstathiou2022electro}
A.~Papaefstathiou and G.~White, \emph{The electro-weak phase transition at
  colliders: Discovery post-mortem}, {\emph{Journal of High Energy Physics}
  {\bfseries 2022} (2022) 1}.

\bibitem{branco1998electroweak}
G.~Branco, D.~Delepine, D.~Emmanuel-Costa and R.G.~Felipe, \emph{Electroweak
  baryogenesis in the presence of an isosinglet quark}, {\emph{Physics Letters
  B} {\bfseries 442} (1998) 229}.

\bibitem{barger2009complex}
V.~Barger, P.~Langacker, M.~McCaskey, M.~Ramsey-Musolf and G.~Shaughnessy,
  \emph{Complex singlet extension of the standard model}, {\emph{Physical
  Review D—Particles, Fields, Gravitation, and Cosmology} {\bfseries 79}
  (2009) 015018}.

\bibitem{ELLWANGER20101}
U.~Ellwanger, C.~Hugonie and A.M.~Teixeira, \emph{The next-to-minimal
  supersymmetric standard model},
  \href{https://doi.org/https://doi.org/10.1016/j.physrep.2010.07.001}{\emph{Physics
  Reports} {\bfseries 496} (2010) 1}.

\bibitem{carena2020electroweak}
M.~Carena, Z.~Liu and Y.~Wang, \emph{Electroweak phase transition with
  spontaneous z2-breaking}, {\emph{Journal of High Energy Physics} {\bfseries
  2020} (2020) 1}.

\bibitem{no2014probing}
J.M.~No and M.~Ramsey-Musolf, \emph{Probing the higgs portal at the lhc through
  resonant di-higgs production}, {\emph{Physical Review D} {\bfseries 89}
  (2014) 095031}.

\bibitem{huber2008gravitational}
S.J.~Huber and T.~Konstandin, \emph{Gravitational wave production by
  collisions: more bubbles}, {\emph{Journal of Cosmology and Astroparticle
  Physics} {\bfseries 2008} (2008) 022}.

\bibitem{curtin2014testing}
D.~Curtin, P.~Meade and C.-T.~Yu, \emph{Testing electroweak baryogenesis with
  future colliders}, {\emph{Journal of High Energy Physics} {\bfseries 2014}
  (2014) 1}.

\bibitem{al2022muon}
H.~Al~Ali, N.~Arkani-Hamed, I.~Banta, S.~Benevedes, D.~Buttazzo, T.~Cai et~al.,
  \emph{The muon smasher’s guide}, {\emph{Reports on Progress in Physics}
  {\bfseries 85} (2022) 084201}.

\bibitem{liu2021probing}
W.~Liu and K.-P.~Xie, \emph{Probing electroweak phase transition with multi-tev
  muon colliders and gravitational waves}, {\emph{Journal of High Energy
  Physics} {\bfseries 2021} (2021) 1}.

\bibitem{forslund2022high}
M.~Forslund and P.~Meade, \emph{High precision higgs from high energy muon
  colliders}, {\emph{Journal of High Energy Physics} {\bfseries 2022} (2022)
  1}.

\bibitem{profumo2007singlet}
S.~Profumo, M.J.~Ramsey-Musolf and G.~Shaughnessy, \emph{Singlet higgs
  phenomenology and the electroweak phase transition}, {\emph{Journal of High
  Energy Physics} {\bfseries 2007} (2007) 010}.

\bibitem{laine2017basics}
M.~Laine and A.~Vuorinen, \emph{Basics of thermal field theory--a tutorial on
  perturbative computations}, {\emph{arXiv preprint arXiv:1701.01554} (2017) }.

\bibitem{leigh1992infraredeffectsbubblepropagation}
R.G.~Leigh, \emph{Infrared effects and bubble propagation at the electroweak
  phase transition},  1992.

\bibitem{laine2016}
M.~Laine and A.~Vuorinen, \emph{Basics of thermal field theory}, {\emph{Lect.
  Notes Phys} {\bfseries 925} (2016) 1701}.

\bibitem{arnold1992phase}
P.~Arnold, \emph{Phase transition temperatures at next-to-leading order},
  {\emph{Physical Review D} {\bfseries 46} (1992) 2628}.

\bibitem{fendley1987effective}
P.~Fendley, \emph{The effective potential and the coupling constant at high
  temperature}, {\emph{Physics Letters B} {\bfseries 196} (1987) 175}.

\bibitem{fernandez2023nu}
E.~Fern{\'a}ndez-Mart{\'\i}nez, J.~L{\'o}pez-Pav{\'o}n, J.~No, T.~Ota and
  S.~Rosauro-Alcaraz, \emph{$\nu$ electroweak baryogenesis: the scalar singlet
  strikes back}, {\emph{The European Physical Journal C} {\bfseries 83} (2023)
  715}.

\bibitem{Papaefstathiou_2021}
A.~Papaefstathiou and G.~White, \emph{The electro-weak phase transition at
  colliders: confronting theoretical uncertainties and complementary channels},
  \href{https://doi.org/10.1007/jhep05(2021)099}{\emph{Journal of High Energy
  Physics} {\bfseries 2021} (2021) }.

\bibitem{athron2020phasetracer}
P.~Athron, C.~Bal{\'a}zs, A.~Fowlie and Y.~Zhang, \emph{Phasetracer: tracing
  cosmological phases and calculating transition properties}, {\emph{The
  European Physical Journal C} {\bfseries 80} (2020) 1}.

\bibitem{fowlie2018fast}
A.~Fowlie, \emph{A fast c++ implementation of thermal functions},
  {\emph{Computer Physics Communications} {\bfseries 228} (2018) 264}.

\bibitem{wainwright2012cosmotransitions}
C.L.~Wainwright, \emph{Cosmotransitions: computing cosmological phase
  transition temperatures and bubble profiles with multiple fields},
  {\emph{Computer Physics Communications} {\bfseries 183} (2012) 2006}.

\bibitem{lewicki2024impacttheoreticaluncertaintiesmodel}
M.~Lewicki, M.~Merchand, L.~Sagunski, P.~Schicho and D.~Schmitt, \emph{Impact
  of theoretical uncertainties on model parameter reconstruction from gw
  signals sourced by cosmological phase transitions},  2024.

\bibitem{gould2023higherorderscosmologicalphase}
O.~Gould and C.~Xie, \emph{Higher orders for cosmological phase transitions: a
  global study in a yukawa model},  2023.

\bibitem{cms2013search}
C.~collaboration et~al., \emph{Search for a standard-model-like higgs boson
  with a mass in the range 145 to 1000 gev at the lhc}, {\emph{arXiv preprint
  arXiv:1304.0213} (2013) }.

\bibitem{buttazzo2018fusing}
D.~Buttazzo, D.~Redigolo, F.~Sala and A.~Tesi, \emph{Fusing vectors into
  scalars at high energy lepton colliders}, {\emph{Journal of High Energy
  Physics} {\bfseries 2018} (2018) 1}.

\bibitem{costantini2020vector}
A.~Costantini, F.~De~Lillo, F.~Maltoni, L.~Mantani, O.~Mattelaer, R.~Ruiz
  et~al., \emph{Vector boson fusion at multi-tev muon colliders},
  {\emph{Journal of High Energy Physics} {\bfseries 2020} (2020) 1}.

\bibitem{Alwall:2011uj}
J.~Alwall, M.~Herquet, F.~Maltoni, O.~Mattelaer and T.~Stelzer, \emph{{MadGraph
  5 : Going Beyond}},
  \href{https://doi.org/10.1007/JHEP06(2011)128}{\emph{JHEP} {\bfseries 06}
  (2011) 128} [\href{https://arxiv.org/abs/1106.0522}{{\ttfamily 1106.0522}}].

\bibitem{Bahr:2008pv}
M.~Bahr et~al., \emph{{Herwig++ Physics and Manual}},
  \href{https://doi.org/10.1140/epjc/s10052-008-0798-9}{\emph{Eur. Phys. J.}
  {\bfseries C58} (2008) 639}
  [\href{https://arxiv.org/abs/0803.0883}{{\ttfamily 0803.0883}}].

\bibitem{Bellm:2017bvx}
J.~Bellm et~al., \emph{{Herwig 7.1 Release Note}},
  \href{https://arxiv.org/abs/1705.06919}{{\ttfamily 1705.06919}}.

\bibitem{Gieseke:2011na}
S.~Gieseke et~al., \emph{{Herwig++ 2.5 Release Note}},
  \href{https://arxiv.org/abs/1102.1672}{{\ttfamily 1102.1672}}.

\bibitem{Arnold:2012fq}
K.~Arnold et~al., \emph{{Herwig++ 2.6 Release Note}},
  \href{https://arxiv.org/abs/1205.4902}{{\ttfamily 1205.4902}}.

\bibitem{Bellm:2013hwb}
J.~Bellm et~al., \emph{{Herwig++ 2.7 Release Note}},
  \href{https://arxiv.org/abs/1310.6877}{{\ttfamily 1310.6877}}.

\bibitem{Bellm:2019zci}
J.~Bellm et~al., \emph{{Herwig 7.2 release note}},
  \href{https://doi.org/10.1140/epjc/s10052-020-8011-x}{\emph{Eur. Phys. J. C}
  {\bfseries 80} (2020) 452}
  [\href{https://arxiv.org/abs/1912.06509}{{\ttfamily 1912.06509}}].

\bibitem{Bewick:2023tfi}
G.~Bewick et~al., \emph{{Herwig 7.3 Release Note}},
  \href{https://arxiv.org/abs/2312.05175}{{\ttfamily 2312.05175}}.

\bibitem{hwsim}
{Papaefstathiou, Andreas}, ``{The \texttt{HwSim} analysis package for HERWIG
  7}.'' {https://gitlab.com/apapaefs/hwsim}.

\bibitem{Brun:1997pa}
R.~Brun and F.~Rademakers, \emph{{ROOT: An object oriented data analysis
  framework}}, \href{https://doi.org/10.1016/S0168-9002(97)00048-X}{\emph{Nucl.
  Instrum. Meth. A} {\bfseries 389} (1997) 81}.

\bibitem{Cacciari:2011ma}
M.~Cacciari, G.P.~Salam and G.~Soyez, \emph{{FastJet User Manual}},
  \href{https://doi.org/10.1140/epjc/s10052-012-1896-2}{\emph{Eur. Phys. J. C}
  {\bfseries 72} (2012) 1896}
  [\href{https://arxiv.org/abs/1111.6097}{{\ttfamily 1111.6097}}].

\bibitem{Catani:1993hr}
S.~Catani, Y.L.~Dokshitzer, M.H.~Seymour and B.R.~Webber, \emph{{Longitudinally
  invariant $K_t$ clustering algorithms for hadron hadron collisions}},
  \href{https://doi.org/10.1016/0550-3213(93)90166-M}{\emph{Nucl. Phys. B}
  {\bfseries 406} (1993) 187}.

\bibitem{Ellis:1993tq}
S.D.~Ellis and D.E.~Soper, \emph{{Successive combination jet algorithm for
  hadron collisions}},
  \href{https://doi.org/10.1103/PhysRevD.48.3160}{\emph{Phys. Rev. D}
  {\bfseries 48} (1993) 3160}
  [\href{https://arxiv.org/abs/hep-ph/9305266}{{\ttfamily hep-ph/9305266}}].

\bibitem{Cacciari:2008gp}
M.~Cacciari, G.P.~Salam and G.~Soyez, \emph{{The anti-$k_t$ jet clustering
  algorithm}}, \href{https://doi.org/10.1088/1126-6708/2008/04/063}{\emph{JHEP}
  {\bfseries 04} (2008) 063} [\href{https://arxiv.org/abs/0802.1189}{{\ttfamily
  0802.1189}}].

\bibitem{Han_2021}
T.~Han, D.~Liu, I.~Low and X.~Wang, \emph{Electroweak couplings of the higgs
  boson at a multi-tev muon collider},
  \href{https://doi.org/10.1103/physrevd.103.013002}{\emph{Physical Review D}
  {\bfseries 103} (2021) }.

\end{thebibliography}\endgroup

\end{document}